\documentclass[a4paper,12pt]{article}

\usepackage{jheppub}

\usepackage[left]{lineno}

\usepackage{latexsym}
\usepackage{amssymb}
\usepackage{amsmath}
\usepackage{epsfig}
\usepackage{rotating}
\usepackage{graphicx}
\usepackage[utf8]{inputenc}

\numberwithin{equation}{section}

\newcommand{\be}{\begin{equation}}
\newcommand{\ee}{\end{equation}}
\newcommand{\bea}{\begin{eqnarray}}
\newcommand{\eea}{\end{eqnarray}}
\newcommand{\bean}{\begin{eqnarray*}}
\newcommand{\eean}{\end{eqnarray*}}
\newcommand{\nn}{\nonumber}

\newcommand{\bra}{\langle}
\newcommand{\ket}{\rangle}

\newcommand{\tr}{\mbox{tr}}
\newcommand{\eps}{\epsilon}

\newcommand{\id}{1\!\!1}

\newcommand{\half}{\frac{1}{2}}

\newcommand{\e}{e}

\newcommand{\C}{\mathbb{C}}

\newcommand{\R}{\mathbb{R}}

\newcommand{\xv}{{\mathbf x}}

\newcommand{\cH}{{\cal H}}

\newcommand{\cO}{{\cal O}}

\newcommand{\bear}{\begin{eqnarray}}
\newcommand{\eear}{\end{eqnarray}}

\newcommand{\Rep}{{\rm Re\,}}

\newcommand{\one}{\mbox{1\hspace{-.7ex}I}}
\renewcommand{\Im}{{\rm Im\,}}
\renewcommand{\Re}{{\rm Re\,}}
\newcommand{\Tr}{{\rm Tr}}

\newcommand{\arccosh}{\mbox{arccosh}}

\title{Complex Langevin dynamics and zeroes of the fermion determinant}

\author[a]{Gert Aarts,}
\author[b]{Erhard Seiler,}
\author[c]{D\'enes Sexty,}
\author[d]{Ion-Olimpiu Stamatescu}

\affiliation[a]{Department of Physics, College of Science, Swansea University, Swansea, United Kingdom}
\affiliation[b]{Max-Planck-Institut f\"ur Physik (Werner-Heisenberg-Institut), M\"unchen, Germany}
\affiliation[c]{Bergische Universit\"at Wuppertal, Wuppertal, Germany}
\affiliation[d]{Institut f\"ur Theoretische Physik, Universit\"at Heidelberg and FEST, Heidelberg, Germany}

\emailAdd{g.aarts@swan.ac.uk}
\emailAdd{ehs@mppmu.mpg.de}
\emailAdd{sexty@uni-wuppertal.de}
\emailAdd{i.o.stamatescu@thphys.uni-heidelberg.de}

\abstract{
QCD at nonzero baryon chemical potential suffers from the sign problem, due to the complex quark determinant. 
Complex Langevin dynamics can provide a solution, provided certain conditions are met. One of these conditions, holomorphicity of the Langevin drift, is absent in QCD since zeroes of the determinant result in a meromorphic drift. We first derive how poles in the drift affect the formal justification of the approach and then explore the various possibilities in simple models. The lessons from these are subsequently applied to both heavy dense QCD and full QCD, and we find that the results obtained show a consistent picture. We conclude that with careful monitoring, the method can be justified a posteriori, even in the presence of meromorphicity.
}


\arxivnumber{1701.02322}

\begin{document}
\maketitle



\section{Introduction} 
\label{sec:intro}

The determination of the QCD phase diagram in the plane of temperature and baryon chemical potential is one of the outstanding open questions in the theory of the strong interaction, as it is relevant for the early Universe, ongoing heavy-ion collision experiments at the Large Hadron Collider and the Relativistic Heavy Ion Collider, nuclear matter and compact objects such as neutron stars.

Ample progress has been made along (or close to) the temperature axis, where lattice QCD can be used to solve the theory numerically, and in recent years it has been possible to simulate QCD with $2+1$ flavours of light quarks using physical quark masses while taking the continuum limit 
\cite{Borsanyi:2013bia,Bazavov:2014pvz}. This is directly relevant for ultrahigh-energy heavy-ion collisions.
The remainder of the phase diagram has not yet been established from first principles. As is well-known 
\cite{Barbour:1986jf,deForcrand:2010ys},
 at nonzero baryon chemical potential, the quark determinant in the standard representation of the QCD partition function is complex, rather than real and positive, ruling out the immediate use of standard numerical methods based on importance sampling. This is generally referred to as the sign problem.

There are various proposals available to circumvent the sign problem, see e.g.\ the reviews \cite{deForcrand:2010ys,Aarts:2013bla,Gattringer:2014nxa,Sexty:2014dxa,Scorzato:2015qts} and lecture notes \cite{Aarts:2015tyj}.
One approach  which has generated substantial attention in the past years is the 
complex Langevin (CL) method, since it has so far proved to be quite successful 
in simulating systems with a complex action $S$, or complex weight $\rho$, from simple toy models to QCD  
\cite{Aarts:2008rr,Aarts:2008wh,Aarts:2010gr,Aarts:2011zn,Aarts:2012ft,Seiler:2012wz,Sexty:2013ica,Langelage:2014vpa,Aarts:2014bwa,Aarts:2016qrv,Sinclair:2015kva,Sinclair:2016nbg}. While the method was suggested already in the 1980s \cite{Parisi:1984cs,Klauder:1983nn}, recent progress has 
come in several ways: 
the theoretical justification has been provided \cite{Aarts:2009uq,Aarts:2011ax} (see also Refs.\ \cite{Salcedo:2015jxd,Salcedo:2016kyy} for related theoretical developments);
numerical instabilities can be eliminated using adaptive stepsizes \cite{Aarts:2009dg};
explicit demonstrations that the sign problem can be solved in spin models and field theories have been given, even when it is severe \cite{Aarts:2011zn,Aarts:2008wh,Aarts:2009hn}; 
and finally, for nonabelian theories, controlling the dynamics via gauge cooling \cite{Seiler:2012wz}, possibly adaptive \cite{Aarts:2013uxa} (see also Ref.\ \cite{Nagata:2015uga}), has been shown to be necessary and effective, resulting in the first results for full QCD \cite{Sexty:2013ica,Aarts:2014bwa,Sinclair:2015kva,Sinclair:2016nbg,Nagata:2016mmh}.
Promising steps beyond gauge cooling have also been taken \cite{Attanasio:2016mhc}.

There is, however, a serious conceptual problem that has to be faced.
It is by now quite well established that when the weight $\rho\sim \exp(-S)$ is free from zeroes in 
the whole complexified configuration space, the only worry is the 
possibility of slow decay in imaginary directions \cite{Aarts:2009uq,Aarts:2011ax}, which will result in incorrect convergence. However, for theories which include fermions, such as QCD, integrating out the latter  will yield a determinant which will always have zeroes for some complexified configurations. 
These zeroes lead to a meromorphic drift; the formal justification for the CL method 
\cite{Aarts:2009uq,Aarts:2011ax} requires holomorphicity, however (this will be reviewed below), and poles may cause convergence to wrong results. The relevance of this has first 
been pointed out by Mollgaard and Splittorff 
\cite{Mollgaard:2013qra,Mollgaard:2014mga} in the context of a random matrix model and has been further investigated in Refs.\ 
\cite{Greensite:2014cxa,Nishimura:2015pba,Nagata:2016vkn,Ito:2016efb}.
Possible consequences for the behaviour of the spectrum of the Dirac operator \cite{Splittorff:2014zca} have been studied in random matrix theory \cite{Ichihara:2016uld}, as has the interplay with gauge cooling \cite{Nagata:2016alq}.

This problem has both theoretical and practical aspects. Concerning the former, it requires a re-analysis of the derivation and justification of the method, given for the holomorphic case in Refs.\  \cite{Aarts:2009uq,Aarts:2011ax}. To do so is the first aim of this paper and is the topic of Sec.\ \ref{sec:formal}. In practice, it has been observed in a number of papers that a meromorphic  drift  will not necessarily cause convergence to wrong results -- sometimes without this issue being explicitly flagged up (one example being when the meromorphicity is due to the Haar measure). However, this aspect is not yet properly understood; while there is a collection of results for a variety of models, an overall understanding is lacking. In Sec.\ \ref{sec:poles} we address this issue using simple models, in which a detailed understanding can be obtained. Lessons from this analysis are summarised in Sec.\ \ref{sec:lessons}.  In Sec.\ \ref{sec:spin} we then move to a more intricate SU(3) model and see how the lessons apply in that context. Finally, in Sec.\ \ref{sec:lattice} we turn to lattice QCD -- heavy dense QCD and full QCD -- and compare our findings with the understanding developed previously. A discussion of the results obtained in the various models is contained in Sec.\ \ref{sec:disc}. 
We conclude that an overall consistent picture can be extracted, applicable across all models considered, and give guidance on how to tackle this problem in future simulations.
The Appendices contain some additional material, including proposals on how to handle poles in the drift in special cases.
We note that partial results have already been presented  in Refs.\ \cite{Seiler-sign,Aarts:2016mso,Aarts:2016bdr}.

 \section{Formal justification in the presence of poles} 
\label{sec:formal}

We briefly recall the basic principles of the CL method, adapting the results for its justification \cite{Aarts:2009uq,Aarts:2011ax} to include a meromorphic drift, i.e.\ a drift with a pole.

Given a holomorphic action $S$ we denote by $\rho$ the (normalised) complex density 
\be
\rho(x) = \frac{e^{-S(x)}}{Z},
\qquad\qquad 
Z = \int dx\, e^{-S(x)},
\ee
on the original real field space. For simplicity we assume here a flat configuration space, i.e.\ $\R^n$. 
A complex drift $K(x+iy)$ is defined by 
analytic continuation as
\be
K(x+iy)= \frac{\nabla \rho(x+iy)}{\rho(x+iy)} = -\nabla S(x+iy). 
\ee
The CL equation, a stochastic differential equation in the complexified 
field space, with the drift given by the real and imaginary parts of $K$, 
\bea
&& \dot x = K_x + \eta_R, \qquad K_x\equiv\Re K, \qquad  \bra \eta_R(t)\eta_R(t')\ket = 2N_R\delta(t-t'), \\
&& \dot y = K_y + \eta_I,  \qquad \; K_y\equiv\Im K, \qquad \bra \eta_I(t)\eta_I(t')\ket = 2N_I\delta(t-t'),
\eea
leads to the Fokker-Planck equation describing the evolution of the 
(positive) probability density $P(x,y;t)$,
\be 
\dot P(x,y;t)=L^T P(x,y;t),  
\ee 
with 
\bea
L^T &=& \nabla_x\left[N_R\nabla_x-K_x\right] + \nabla_y\left[N_I\nabla_y- K_y\right], \\
L &=& (N_R \nabla_x+K_x)\nabla_x+ (N_I \nabla_y+K_y)\nabla_y, 
\label{fpe}
\eea 
where $N_R-N_I=1$ and $N_I\ge 0$. We used here `complex noise' ($N_I>0$) for presentation purposes; below we specialise to real noise  ($N_I=0$), as advocated earlier \cite{Aarts:2009uq,Aarts:2011ax}.

Averaging over the noise, the 
evolution of holomorphic observables $\cO(x+iy)$ is governed by the 
equation
\be
\dot \cO(x+iy;t)= L\cO(x+iy;t)= \tilde L\cO(x+iy;t),
\label{obsevol}
\ee
with 
\be
\tilde L=\left[\nabla_z-(\nabla_z S(z))\right]  \nabla_z,
\ee
where in the last step we used the Cauchy-Riemann equations, i.e.\
holomorphy of $\cO(x+iy;t)$, and $z=x+iy$.

The consistency of the complex Langevin method with the original problem 
hinges on the quantity 
\be 
F(t,\tau)\equiv \int P(x,y;t-\tau) \cO(x+iy;\tau)\, dxdy, \label{fttau} 
\ee 
which is supposed to interpolate between 
\be 
 F(t,0)=\int P(x,y;t) \cO(x+iy;0) \, dxdy \equiv \bra O\ket_{P(t)}
 \label{ft0} 
\ee 
and 
\be 
 F(t,t)=\int \cO(x;0)\rho(x;t)\, dx  \equiv \bra O\ket_{\rho(t)},
 \label{ftt} 
\ee 
where $\rho(x;t)$ is the complex density evolved according to 
\be 
\dot \rho(x;t)= \nabla_x \left(\nabla_x-K(x)\right) \rho(x;t). 
\ee 
Here it is 
necessary to choose the initial density $\rho(x;0)$ positive, typically a $\delta$-function.

Correctness of the CL method requires that the two quantities $F(t,0)$ and 
$F(t,t)$ are equal, i.e.\ $\bra O\ket_{P(t)} =  \bra O\ket_{\rho(t)}$, at least as $t\to\infty$.
 To show this equality, in Ref.\ \cite{Aarts:2011ax} it was 
argued that
\bea
\frac{\partial}{\partial \tau} F(t,\tau) &=& -\int \left(L^T P(x,y;t-\tau)\right)\cO(x+iy;\tau)\, dxdy \nn \\
&&  + \int  P(x,y;t-\tau) L\cO(x+iy;\tau) \, dxdy = 0;
\label{interpol}
\eea
this required that formal integration by parts, without possible boundary 
terms, is correct. 

For holomorphic actions, this requires care in the imaginary directions, $|y|\to\infty$. Slow decay, for instance power-law decay in polynomial models, does not allow partial integration to be carried out for all holomorphic observables $z^n$ without picking up contributions at the boundary. On the other hand, if the distribution is strictly localised in a strip in the complex configuration space, no boundary terms will appear and the results from the CL simulation can be justified, see for instance Ref.\ \cite{Aarts:2013uza} for an explicit example.

In the case of a meromorphic drift, the topic of this paper, we have to introduce two boundaries: one at large $|y|$ and near the location(s) of the pole(s), which we denote generically as $z_p$. Let us first reconsider Eq.\ (\ref{ftt}): by definition we have
\be
F(t,t)=\int P(x,y;0) \cO(x+iy;t)\, dxdy.
\ee
We may consider a single trajectory starting at $(x,y)=(x_0,0)$, which
means choosing
\be
P(x,y;0)=\delta(x-x_0)\delta(y).
\ee
We then find 
\be 
F(t,t)=\cO(x_0;t).
\ee
This is well defined. Furthermore, provided $x_0\neq z_p$, the time-evolved 
observable $\cO(z;t)$ is holomorphic for $z\neq z_p$. However, according to 
Eq.\ (\ref{obsevol}), we have to expect that $\cO(z;t)$ has an essential 
singularity at $z=z_p$, since formally
\be
\cO(z;t) = \exp(\tilde L t)\cO(z) = \sum_{k=0}^\infty \frac{t^k}{k!}\tilde  L^k\cO(z), 
\ee   
and each term of the series in general will produce a pole of higher order.  This is the first finding.

Now let us look at Eq.\ (\ref{interpol}): For simplicity we assume 
that there is only a single pole at $z=z_p$ and consider a one-dimensional
configuration space. Integration by parts can be used at first only for the domain
\be
G_{\epsilon,Y}\equiv \left\{z=x+iy\,\vert\; |y|<Y; |z-z_p|>\epsilon \right\},
\ee
in which the dynamics is nonsingular; later we have to take the limits $Y\to\infty$ and 
$\epsilon\to 0$. The first integral in Eq.\ (\ref{interpol}) is of the form
\be
\int_{G_{\epsilon,Y}} (\nabla\cdot J) \cO \, dxdy,
\ee
where $J$ is the `probability current'
\be
J={\bf N} \nabla P -  K P,
\ee
with
\be
{\bf N}=
\left(   
\begin{matrix}
\ \ N_R & 0\\
 0&N_I
\end{matrix}
\right).
\ee
Using the divergence theorem (Gauss's theorem) one finds that the first
integral is equal to
\be
\label{eq:gauss}
-\int_{G_{\epsilon,Y}} J \cdot \nabla \cO \, dxdy +
\int_{\partial G_{\epsilon,Y}} n\cdot J \cO \, ds,
\ee
where $ds$ stands for the line element of the boundary $\partial G$ and
$n$ denotes the outer normal.
The boundary has 3 disconnected pieces: two straight lines at $y=\pm Y$
and a circle at $|z-z_p|=\epsilon$.
   We assume now as usual that $P$ has sufficient decay so that the
contributions from $y=\pm Y$ disappear for $Y\to \infty$. Then the
question remains if the circle around $z_p$ gives a nonvanishing
or even divergent contribution.

Numerically it has been found that always
\be   
P(x_p,y_p)=0,
\ee
and furthermore that $P$ vanishes at least linearly with the distance from 
$z_p$, with some angular dependence. But the expected essential singularity 
of the evolved observable $\cO(x+iy;t)$ at $z_p$ could lead to a finite or even
divergent contribution as $\epsilon\to 0$. Numerically, however, 
we never found divergent behaviour, so presumably the boundary terms are 
finite. But they may be nonzero, spoiling the proof of correctness. This is the second finding, the appearance of boundary terms, similar to the ones that may appear at $|y|=Y$.

Let us apply  integration by parts a second time to the bulk integral
\be
-\int_{G_{\epsilon,Y}} J_D \cdot \nabla \cO \, dxdy,
\ee
where $J_D$ denotes the `diffusive current'
\be
J_D\equiv {\bf N}\nabla P.
\ee
The integral above is then
\be
-\int_{G_{\epsilon,Y}} {\bf N}\nabla P  \cdot \nabla \cO \, dxdy.
\ee
Green's first identity (also a consequence of the divergence theorem) says
that this is equal to
\be
\int_{G_{\epsilon,Y}} P \nabla \cdot {\bf N}\nabla \cO \,dxdy
-\int_{\partial G_{\epsilon,Y}} P  n \cdot {\bf N} \nabla \cO \, ds.
\ee
The discussion of the new boundary terms is almost identical to the one above; again what happens depends on the detailed 
behavior of $\cO(x+iy;t)$ near $z_p$.

In practice we found (numerically) no indication of any divergence caused by the existence of an essential singularity of $\cO(x+iy;t)$.\footnote{There are exceptions to the claim that a meromorphic drift will cause an essential singularity in $\cO(x+iy;t)$, but unfortunately they 
are nongeneric. One example is discussed in App.\ \ref{sec:2nd}.}
 The reason for this seems to be that  
both $P(x,y;t)$ and  $\cO(x+iy;t)$ have nontrivial angular dependence. In Sec.\ \ref{sec:u1} we discuss a probably typical situation in which $P(x,y;t)$ vanishes identically in two opposite 
quadrants near the pole. So if $\cO(x+iy;t)$ shows strong growth only in 
those quadrants, the product may well be integrable, i.e.\ the boundary 
terms near the pole remain bounded.   

To summarise, we find that the time-evolved observable will generically have an essential singularity at the pole, which, however, is counteracted by the vanishing distribution. Concerning the justification, partial integration at the boundaries now also includes integration around the pole, which requires the distribution to vanish rapidly enough for partial integration to be possible without picking up boundary terms. In the following section, we will study this first in simple models, focussing on the essential elements.

 \section{Poles: inside or outside the distribution}
\label{sec:poles}

From the formal derivation in the previous section, it is clear that the 
essential question concerns the interplay between the pole (and observables 
evaluated close to the pole) and the equilibrium distribution. Logically 
there are three possibilities: 
\begin{enumerate}
\item poles are outside the distribution; 
\item poles are on the edge of the distribution; 
\item poles are inside the distribution. 
\end{enumerate}
It can be expected that in the first case poles are not dangerous, as they 
are avoided in the Langevin process (possibly after thermalisation). What 
happens in the second and third possibility is not a priori clear.
In this section we will discuss each of these cases using simple 
zero-dimensional models, with the aim of extracting insight that can be 
carried over to more complicated theories, including QCD. 
Some additional remarks on simple models with poles are given in App.\ \ref{sec:2nd} and \ref{sec:real}.

\subsection{One-pole model}
\label{subsec:one}

The simplest model of a system with a pole is given by the density on $\R$,
\be
\label{eq:rho31}
\rho(x)= (x-z_p)^{n_p} \exp(-\beta x^2),
\ee
where we take $\beta$ real. When $z_p$ is real, the weight is real as 
well, but the model has a sign problem for odd $n_p$, while for even $n_p$ 
the zero in the distribution may potentially lead to problems with 
ergodicity. When $z_p$ is complex, the weight is complex of course.

The complex drift appearing in the Langevin process is given by
\be
K(z)=\frac{\rho'(z)}{\rho(z)}= \frac{n_p}{z-z_p}-2\beta z.
\label{drift}
\ee
While the original weight vanishes at $z_p$, the drift diverges and is 
hence meromorphic. We will refer to this model as the ``one-pole model''. 
Special cases (with $n_p=1$) have been considered long ago 
\cite{Salcedo:1993tj,Fujimura:1993cq}, while recently this model has been studied again, in 
particular for a large range of values of $n_p$ \cite{Nishimura:2015pba}.  Our focus 
is somewhat different; we are mostly interested in the interplay between 
the location of the pole and the distribution and, for real $z_p$, the 
difference between $n_p=1$ and $n_p=2$.

This model captures the presence of a meromorphic drift in QCD in a very rudimentary way, as follows.
Consider the QCD partition function for $n_f$ degenerate flavours, 
\be
Z = \int DU\, \det[M(U)]^{n_f} e^{-S_{\rm YM}(U)} = \int DU\, e^{-S_{\rm eff}(U)},
\ee
with 
\be 
S_{\rm eff}(U) = S_{\rm YM}(U) - n_f\ln\det M(U) = S_{\rm YM}(U) -n_f 
\sum_i\ln\lambda_i(U),
\ee
where in the last expression we have written the fermion determinant in terms of the eigenvalues of the Dirac operator,  $\lambda_i(U)$, which depend on the gauge field configuration, as indicated with the $U$ dependence.  
  The drift contributing to the update of link $U$ will now have a contribution from the fermion determinant as
\be
K_{F} \sim n_f \sum_i\frac{D\lambda_i(U)}{\lambda_i(U)},
\ee
where $D$ denotes the derivative.
When $\lambda_i$ goes to zero (and the determinant vanishes), the drift has a pole. In the one-pole model, the complicated dependence of $\lambda_i$ on $U$ is replaced by a simple pole located at $z_p$, i.e.\ 
\be
n_f \sum_i\frac{D\lambda_i(U)}{\lambda_i(U)}  \to \frac{n_p}{z-z_p}.
\ee
In QCD, the links $U$ are of course fluctuating and the dependence is considerably more complicated. The relation between the number of flavours ($n_f$) and the order of the zero ($n_p$) depends on details of the fermion determinant.

\subsubsection{Strips in the complex plane}

To continue, we allow the location $z_p$ of the pole in the drift to be complex in general and take $\beta$ real and positive. The drift has fixed points ($K(z)=0$) at
\be
z_{1,2}=\frac{z_p}{2} \pm \frac{z_p}{2} \sqrt{1 + \frac{2n_p}{\beta  z_p^2}}.
\ee
A fixed point $z_i$ is attractive (repulsive) if ${\rm Re}\, 
K'(z_i)<0\;\; ({\rm Re}\, K'(z_i)>0)$. We find
\be
K'(z_{1,2}) = -2\beta\left( \frac{2\beta}{n_p}z_{1,2}^2+1\right),
\ee
and hence, for real or imaginary $z_p$, this yields
\begin{itemize}
\item [(a)]
$z_p=x_p$ real $\Rightarrow$  both fixed points $z_{1,2}$ are real and attractive;
\item [(b)]
$z_p=iy_p$ imaginary, $y_p^2<2n_p/\beta$ $\Rightarrow$
$z_{1,2}$ complex: both fixed points are attractive;
\item[(c)] $z_p=iy_p$ imaginary, $y_p^2>2n_p/\beta$ $\Rightarrow$
$z_{1,2}$ imaginary: the fixed point closer to the real axis is 
attractive, the other one repulsive.
 \end{itemize}
In order to find where the pole is with respect to the equilibrium distribution $P(x,y)$, and be able to discuss the three cases above (pole is outside, on the edge or inside the distribution), we note the following. The drift in the imaginary direction is given by
\be
 K_y(x,y) = \Im K(x+iy) = -n_p\frac{y-y_p}{(x-x_p)^2+(y-y_p)^2}-2\beta y.
 \ee
 Without loss of generality we take $y_p\geq 0$. Hence it immediately follows that the drift is pointing downwards when $y>y_p$ and upwards when $y<0$. In the case of real noise (which we use from now on), this implies that the equilibrium distribution will be nonzero only in the strip $0 < y < y_p$ \cite{Aarts:2009uq,Aarts:2013uza}. Hence generically in this model the pole will be on the edge of the distribution. Moreover, since the distribution is strictly zero outside the strip, partial integration at $y\to\pm\infty$ is not a problem and therefore this aspect of the justification is under complete control.
 
\begin{figure}[t] 
\centerline{\includegraphics[width=.7\columnwidth] 
{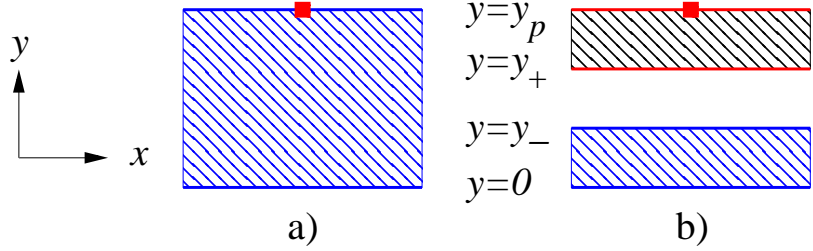}}
\caption{One-pole model: strips where the equilibrium distribution 
$P(x,y)$ is nonzero. The pole is located at $z_p=x_p+iy_p$, with $y_p>0$ 
(red square). Left: $y_p^2<2n_p/\beta$: $P(x,y)> 0$ when $0<y<y_p$ and the 
pole is on the edge. Right: $y_p^2>2n_p/\beta$: $P(x,y)> 0$ when $0<y<y_-$ 
and the pole is outside the distribution.  The strip $y_+<y<y_p$ can be 
visited during the Langevin process, provided that the process is 
initialised at $y>y_+$, but will eventually be abandoned (transient).
 }
\label{fig:strips}
\end{figure}

 Following the analysis of Refs.\  \cite{Aarts:2009uq,Aarts:2013uza}, we can in fact derive a stronger result. It follows from the FPE that the equilibrium distribution has to satisfy the condition
 \be
 \int_{-\infty}^\infty dx\, K_y(x,y)P(x,y) = 0.
 \ee
Since $P(x,y)\geq 0$, it follows that if $K_y(x,y)$ has a definite sign as a function of $x$ for given $y$,  $P(x,y)$ has to vanish for this $y$ value. Following exactly the same steps as in Sec.\ 4.2 of Ref.\  \cite{Aarts:2013uza}, we find the following. 
As a function of $x$, $K_y(x,y)$ has an extremum at $x=x_p$ and the value at the extremum is given by
\be
F(y) = -\frac{n_p}{y-y_p}-2\beta y.
\ee
The zeroes of $F(y)$, at
\be
y_\pm = \frac{y_p}{2}\pm\frac{y_p}{2}\sqrt{1-\frac{2n_p}{\beta y_p^2}},
\ee 
determine the presence of additional boundaries at $y_\pm$, provided they are real \cite{Aarts:2013uza}. We find that 
\begin{itemize}
\item[a)] $y_p^2<2n_p/\beta$: no additional boundaries;
\item[b)] $y_p^2>2n_p/\beta$: additional boundaries at $y_\pm$, $P(x,y)=0$ when 
$y_-<y< y_+$. 
\end{itemize}
This situation is sketched in Fig.\ \ref{fig:strips}.
 
In the latter case, no conclusion from this argument can be drawn regarding the 
strips $0<y<y_-$ and $y_+<y<y_p$. However, an additional analysis of the classical 
flow pattern shows that the strip $0<y<y_-$ is an attractor, while the strip 
$y_+<y<y_p$ can only be visited when the process starts at $y>y_+$. The drift 
inside this strip is pointing mostly towards $y=y_-$ and hence this region will 
eventually be abandoned. It will therefore at most be present as a transient.

We conclude that in this model the pole is either on the edge of (case a) or outside (case b) the 
distribution. In the following we address each of these possibilities.

\subsubsection{Ergodicity and bottlenecks for real poles}
\label{sec:ergo}

\begin{figure}[t]
\centerline{
\includegraphics[width=.45\columnwidth]{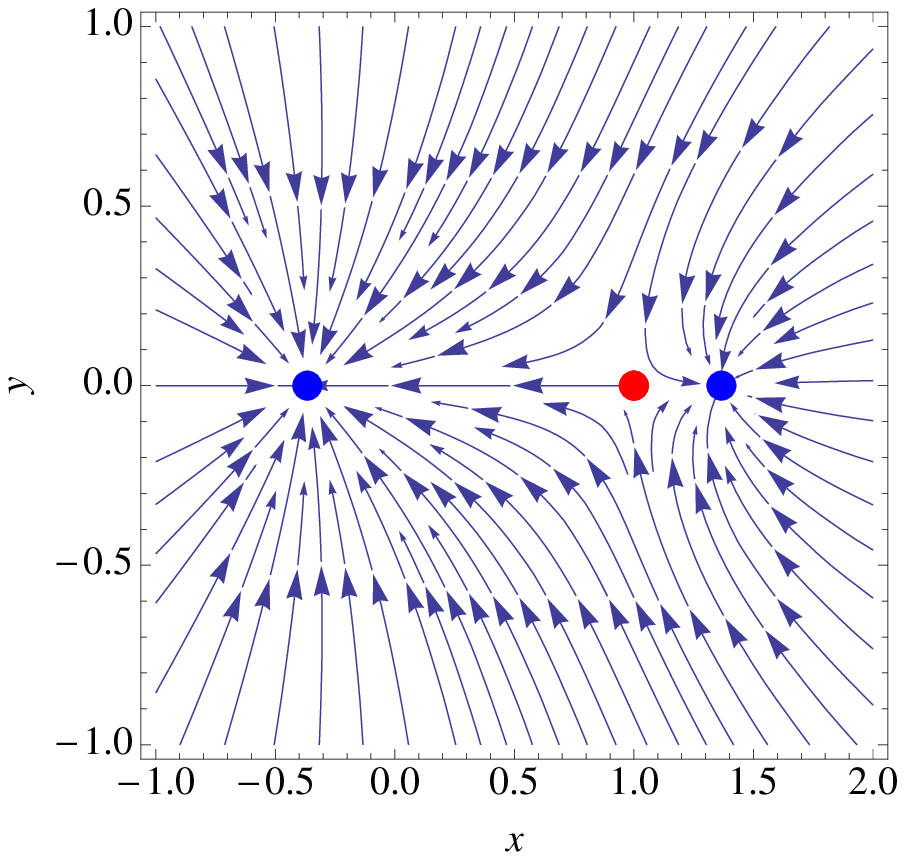} 
\includegraphics[width=.45\columnwidth]{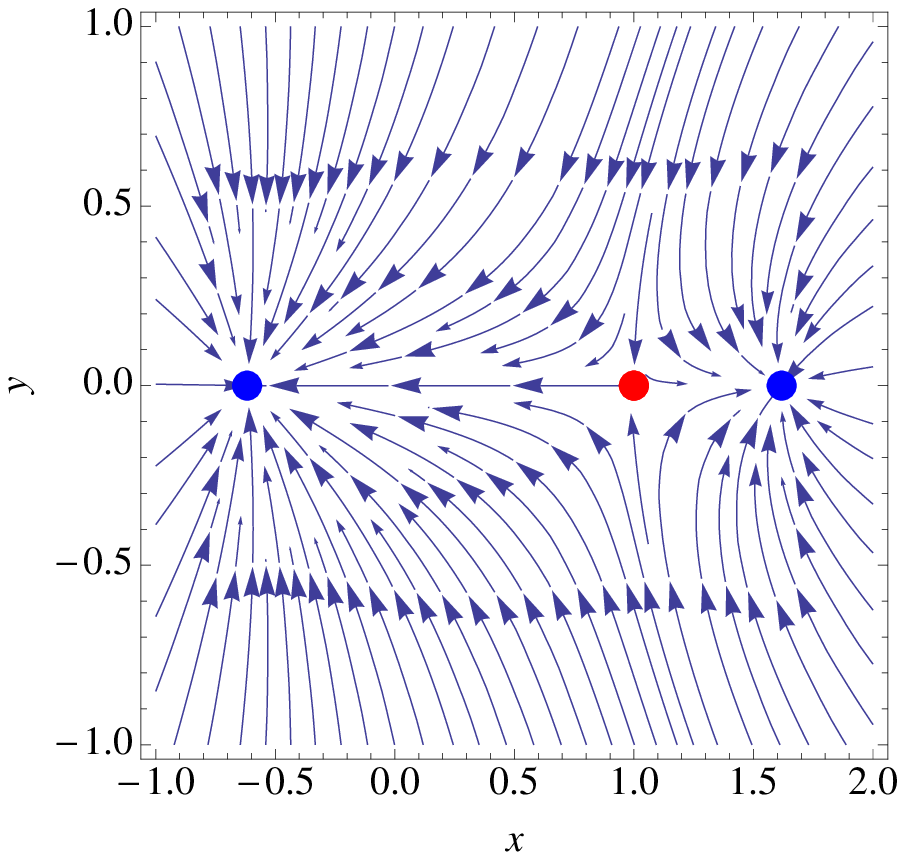}}
\caption{Classical flow patterns for $\beta=1$ and $z_p=1$, with $n_p=1$ (left) and 
$n_p=2$ (right). The blue (red) circles indicate the fixed points (pole). The real 
axis is an attractor.
}
\label{flowreal}
\end{figure}

We first discuss the real case, with $z_p=x_p$, since this allows us to 
introduce the concept of a `bottleneck', which will turn out also be relevant for 
the complex case. In this case, the distribution $\rho(x)$ is real, but with a sign problem for odd $n_p$. As follows from the analysis above, both fixed points are attractive and the equilibrium distribution lies on the real axis. This is illustrated in Fig.\ \ref{flowreal} for $\beta=z_p=1$ and $n_p=1,2$. We note that close to the pole, the drift is repulsive along the real direction and attractive along the imaginary direction; it is easy to see that this is true in general.

In the limit of continuous Langevin time, trajectories of a real Langevin process will not cross the poles \cite{naga}. This leads to a 
`separation phenomenon', a point made some time ago \cite{flower}. In an actual simulation, because of the finite 
step size, crossing of the poles may happen (depending on the step size) \cite{Fujimura:1993cq}. It is instructive to 
look at the corresponding stationary Fokker-Planck equation (on the real axis)
\be
\partial_x(\partial_x-K(x))P(x)=0, \quad\quad\quad  K(x)=\frac{\rho'(x)}{\rho(x)}. 
\ee
Clearly $P(x)\sim\rho(x)$ is a solution, but wherever there is a sign problem, it cannot be the stationary probability distribution, since $P(x)$ should be nonnegative. Instead we find two linearly independent, nonnegative solutions:
\be
P_+(x)=\rho(x)\theta(\rho(x)), \quad\quad\quad P_-(x)=-\rho(x)\theta(-\rho(x));
\ee
any linear combination of $P_+$ and $P_-$ with nonnegative coefficients
is likewise a possible long time average. Hence the Fokker-Planck Hamiltonian 
has two ground states.

If the simulation manages to slip through the barrier sufficiently easily, we expect to get
\be
P_q(x)= P_+(x)+P_-(x) = |\rho(x)|,
\ee
i.e.\ the phase-quenched model, as already found in Ref.\ \cite{Fujimura:1993cq}. We have 
verified this numerically for $n_p=1$.
One way to cross the bottleneck and facilitate tunneling through the pole is by adding a 
small amount of imaginary noise. However, the drift (\ref{drift}) is insensitive to 
sign changes in $\rho$ and the phase-quenched result is recovered. We conclude that the Langevin process cannot give correct results for odd $n_p$.

For even $n_p>0$, there is no sign problem, but the lack of ergodicity 
exists as well. In this case, because of the stronger repulsion away from 
the pole, our simulations typically do {\it not} cross the pole, and hence
produce incorrect results when started on one side of the pole. 
In this case, adding a small imaginary noise term does facilitate the crossing and 
leads to correct results.\footnote{For the special case $z_p=0$ the symmetry $x \to -x$ allows one 
to start the process with equal probability on either side of the pole and obtain correct results as well.}

In conclusion, we find that zeroes in the distribution lead to a bottleneck and hence ergodicity problems. Whether this zero is crossed depends on the order of the zero: the higher the order, the more difficult the crossing is.  We will see that the same is true in the complex case, even though it is easier to go around the pole in the complex plane in that case.

\subsubsection{Poles outside the distribution}
\label{subsec:out}

\begin{figure}[t]
\centerline{
\includegraphics[width=.45\columnwidth]{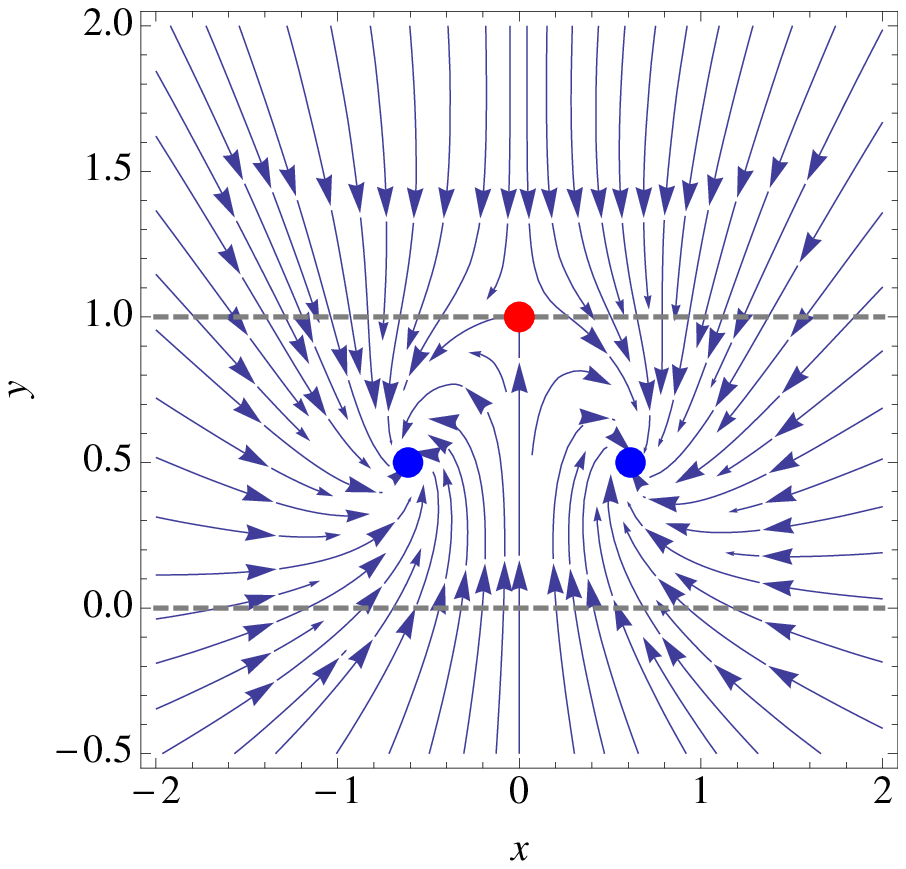}
\includegraphics[width=.45\columnwidth]{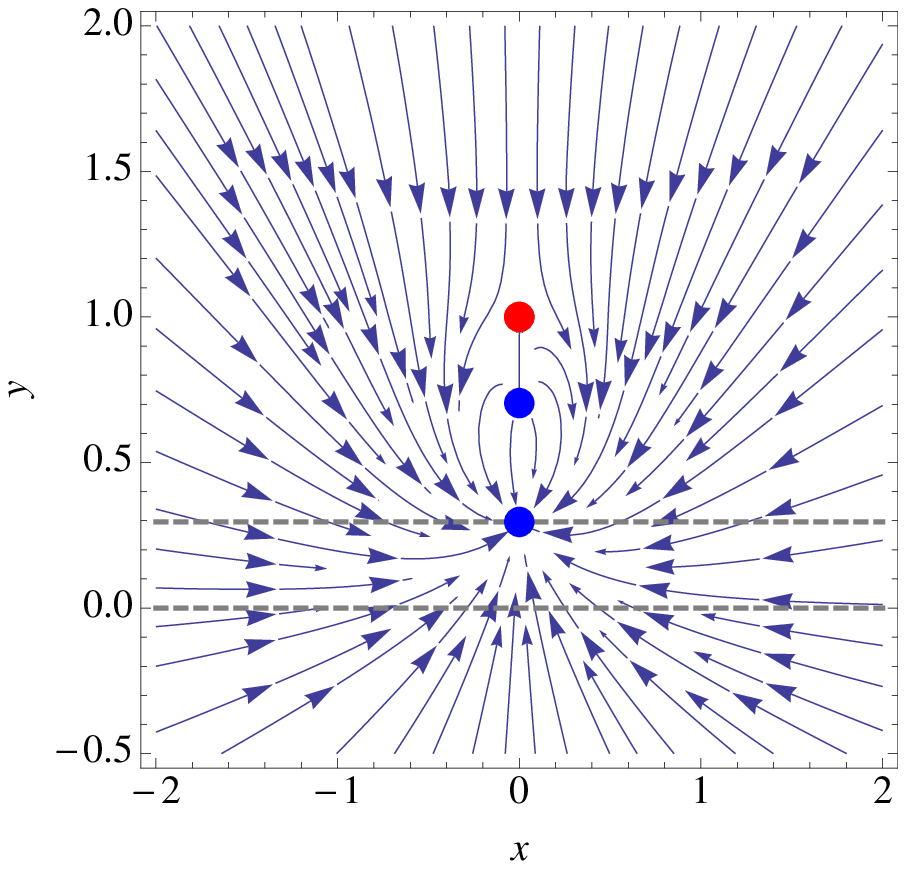}
}
\caption{Classical flow diagrams for $z_p=i$, $n_p=2$ and $\beta=1.6$ (left) and 
$\beta=4.8$ (right).
The blue (red) circles are fixed points (pole) and the equilibrium 
distribution is contained between the dashed horizontal lines.
}
\label{flowcompl}
\end{figure}

We now consider the complex case and take $n_p=2, z_p=i$ ($y_p=1$) and three 
$\beta$ values: $\beta=1.6, 3.2, 4.8$. The relevant parameter determining the 
distribution is $2n_p/\beta y_p^2$, which takes the values 5/2, 5/4 and 5/6  respectively.  Hence for $\beta=4.8$ the distribution is confined to the strip $0<y<y_-\approx 0.296$ and the pole is outside the strip, while for the other $\beta$ values the distribution touches the pole and $0<y<y_p=1$.

\begin{figure}[t]
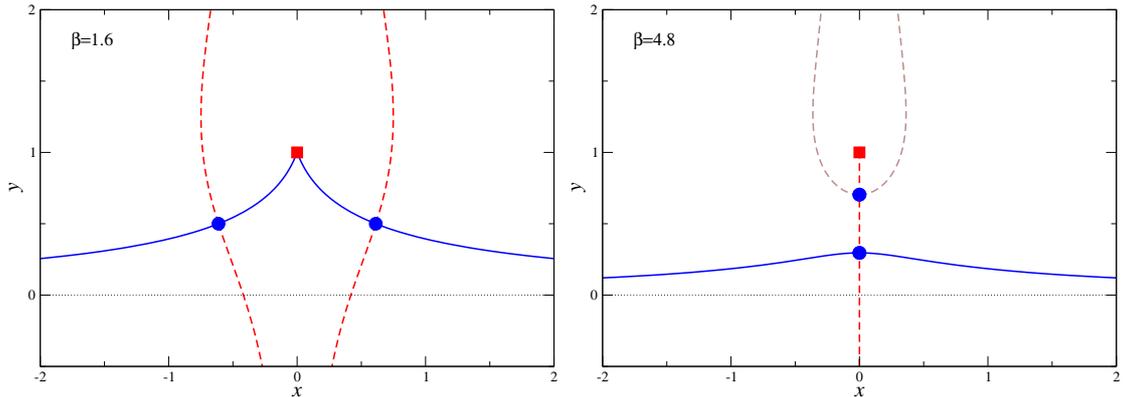

\centerline{
\includegraphics[width=.48\columnwidth]{figs-POLES/plot-thimble-beta1.6.eps}
\includegraphics[width=.48\columnwidth]{figs-POLES/plot-thimble-beta4.8.eps}
}
\caption{Thimbles corresponding to Fig.\ \ref{flowcompl}. See text for details.
}
\label{thimbles}
\end{figure}

The classical flow diagrams are shown in Fig.\ \ref{flowcompl}, for $\beta=1.6$ and $4.8$. It is easy to see from the flow patterns that the general conclusions apply. For completeness, the corresponding thimbles\footnote{In short, (stable) thimbles correspond to deformations of the original integral:  they emerge from the classical fixed points and along the thimbles the imaginary part of the weight is constant \cite{Cristoforetti:2012su}. Thimbles may end at singularities of the drift \cite{Aarts:2014nxa}.} are shown in  Fig.\ \ref{thimbles}.
Here the full (blue) lines are the stable, contributing thimbles and the dashed lines are the unstable, noncontributing thimbles. We note that at $\beta=4.8$ the unstable thimble for the lower fixed point is the stable thimble for the upper fixed point. At the lower $\beta$ value the thimbles meet at the pole, while at the higher value the stable thimble avoids the  pole, consistent with the Langevin analysis.

\begin{table}[t]
\begin{center}
\begin{tabular}{c | c | c l}
$\beta$ & $n$ & complex Langevin& exact\\
\hline
$1.6$ & 1	& $-0.0029(80) + i 0.5223(12)$		& $i 0.909091$   \\ 
&  2 	& $0.4193(25) - i 0.0043(68)$ 		& $0.0284091$  \\
&  3 	& $0.0053(71) + i 0.7605(30)$ 		&  $i0.852273$ \\
&  4 	& $0.2226(96) - i 0.001(12)$  		& $-0.239702$  \\
\hline
$3.2 $ &  1 	& $0.0013(31) + i 0.36985(58)$ 	& $i 0.37037$  \\
&  2 	& $0.0994(14) - i 0.0001(20)$ 		& $0.0983796$\\
&  3 	& $0.0029(11) + i 0.17439(76)$  	& $i 0.173611$ \\
&  4 	& $0.0192(10) - i 0.0018(15)$  		& $0.0189887$ \\
\hline
$4.8$ & 1 	& $0.00052(54) + i 0.23256(5)$  	& $i 0.232558$   \\
&  2 	& $0.07993(19) + i 0.00027(22)$	& $0.0799419$ \\
&  3 	& $-0.00019(16) + i 0.07266(9)$ 	& $i 0.0726744$ \\
&  4 	& $0.01743 (12) +  i 0.00007(14)$ 	& $0.0174116$ 
 \end{tabular}
\caption{Results  for $\bra z^n\ket$  using complex Langevin simulations for the weight (\ref{eq:rho31}), with $n_p=2$, 
$z_p=i$ and various $\beta$ values, compared to the exact result.
}
\label{onepole}
\end{center}
\end{table}

\begin{figure}[t]
\centerline{
\includegraphics[width=.49\columnwidth]{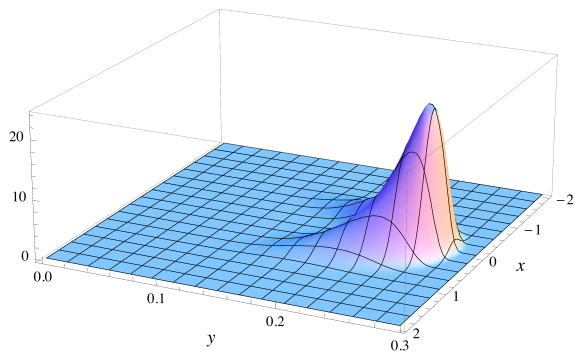}
}
\caption{Histogram $P(x,y)$ for  $z_p=i$, $n_p=2$ and $\beta=4.8$.} 
\label{beta48} 
\end{figure}

We first consider the case $\beta=4.8$. The histogram for $P(x,y)$ is shown in Fig.\ \ref{beta48} and is confined between $0<y<y_-\simeq 0.296$, as it should be. The results for the observables $\bra z^n\ket$ ($n=1,2,3,4$) from a complex Langevin simulation are shown in Table \ref{onepole}; we observe excellent agreement with the exact result. It is clear that this is in line with the formal derivation.
We hence state the following 

{\it Proposition:} If the drift is such that the equilibrium distribution 
is confined to a simply connected region not containing any poles of the 
drift, the complex Langevin process converges to the exact results.

\subsubsection{Poles on the edge of the distribution}
\label{subsec:edge}

\begin{figure}[t]
\centerline{
\includegraphics[width=.49\columnwidth]{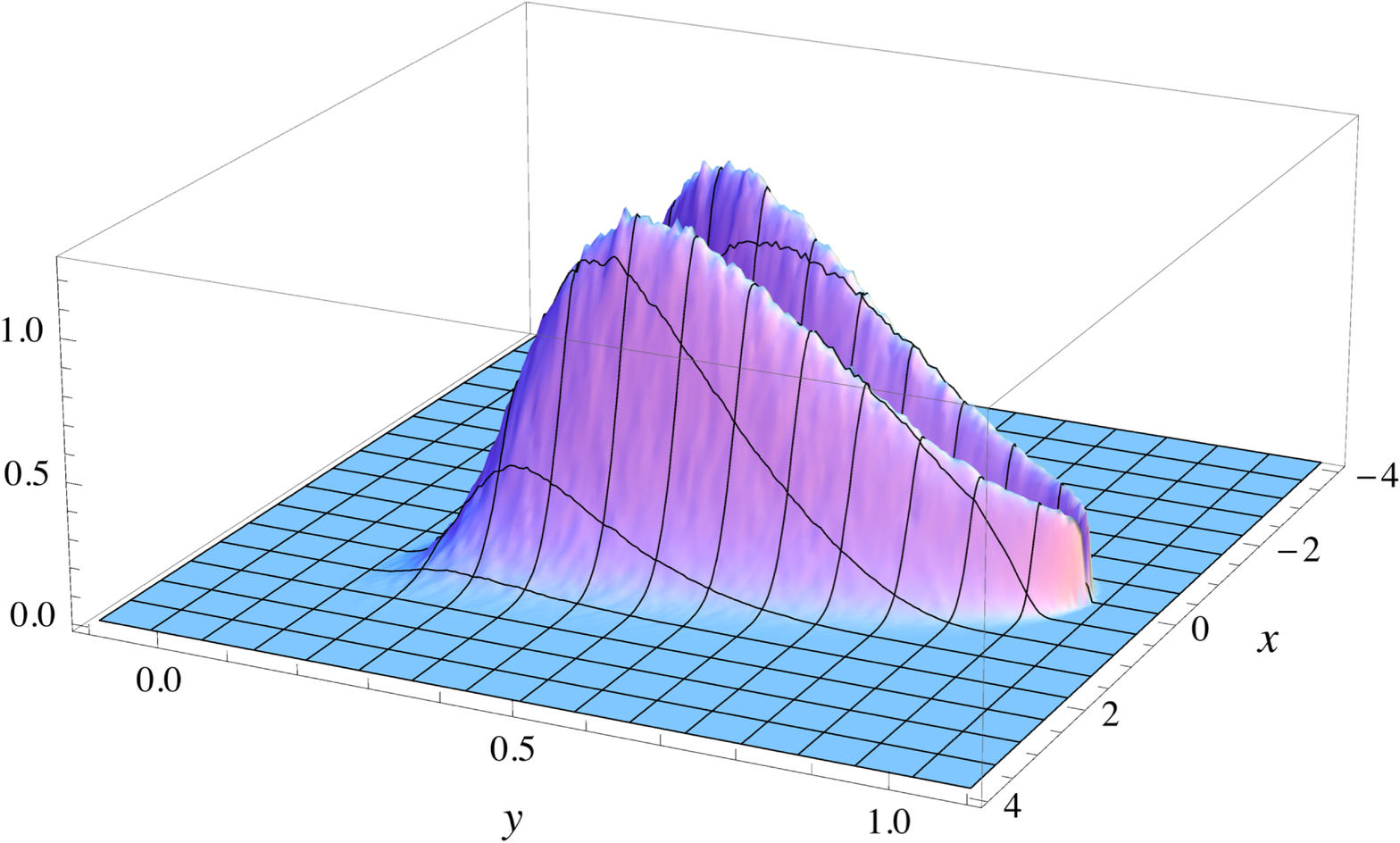}
\includegraphics[width=.49\columnwidth]{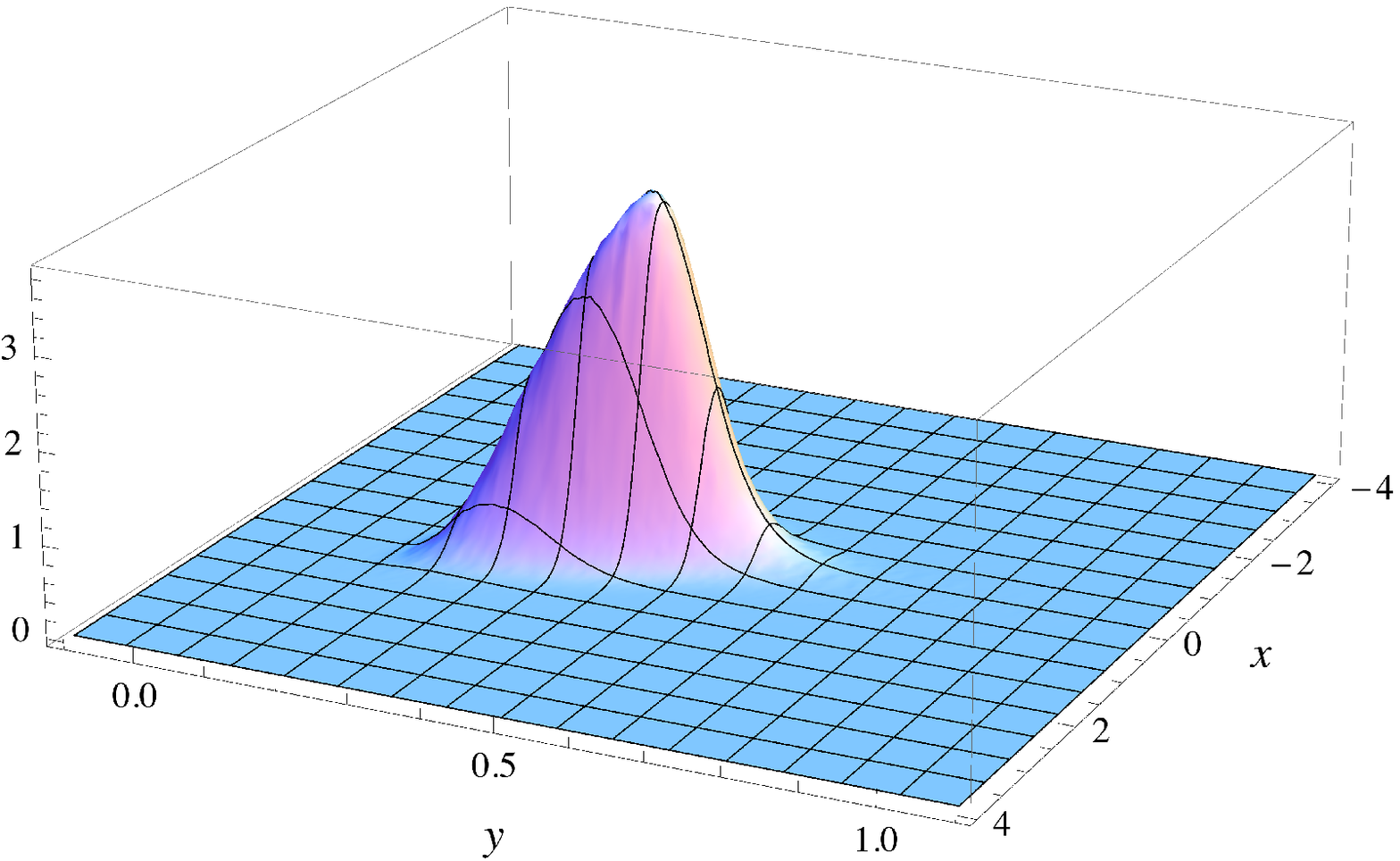}
}
\caption{Histogram $P(x,y)$ for  $z_p=i$, $n_p=2$, $\beta=1.6$ (left) and 
$\beta=3.2$ (right).
}
\label{beta16-32} 
\end{figure}

We now turn to $\beta=1.6$ and $3.2$, with the pole at the edge of the distribution. The corresponding histograms for $P(x,y)$ are shown in Fig.\ \ref{beta16-32} and the results for $\bra z^n\ket$ are listed in Table \ref{onepole}.
Here we note that the Langevin results for $\beta=1.6$ are wrong, while the results for $\beta=3.2$  appear to be correct (within the error).
To understand this better we employ two methods.

\begin{figure}[t]
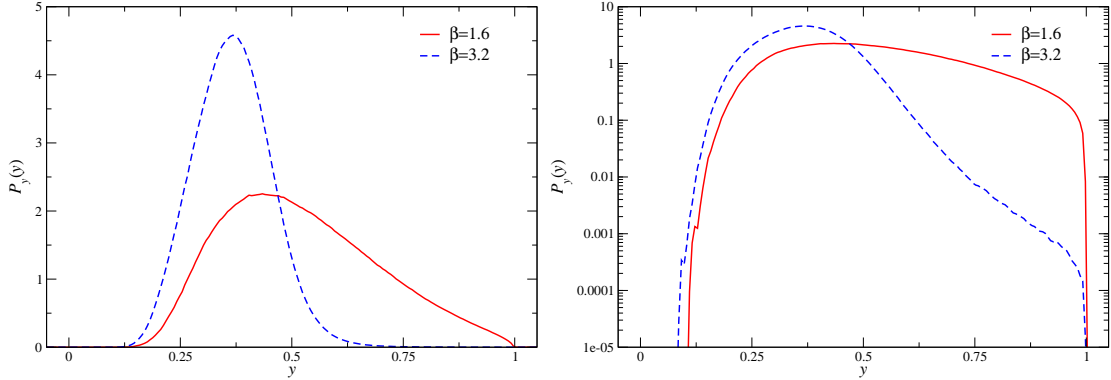

\centerline{
\includegraphics[height=.22\textheight]{figs-POLES/plot-hist-y-beta-nf2.eps}
\includegraphics[height=.22\textheight]{figs-POLES/plot-hist-y-beta-nf2-log.eps}
}
\caption{Partially integrated distributions $P_y(y)$ on a linear scale (left) and logarithmic scale (right) for $\beta=1.6, 3.2$, other parameters as above.
}
\label{fig:py} 
\end{figure}

First we note that the histograms look quite different. At $\beta=1.6$ the distribution is nonzero very close to the pole, which one expects yields boundary terms in the formal justification, which invalidate the outcome. On the other hand, at $\beta=3.2$ the distribution is peaked predominantly away from $y=1$ and the pole appears to be avoided. We make this more precise by computing the partially integrated distribution
\be
P_y(y) = \int_{-\infty}^\infty dx\, P(x,y).
\ee
The results are shown in Fig.\ \ref{fig:py} on a linear scale (left) and on a logarithmic scale (right). On a linear scale it is easy to see that at $\beta=1.6$, $P_y(y)$ is nonzero up to $y=1$ and goes to zero linearly (at the pole the distribution is zero of course). Based on the formal justification, we conclude that this slow decay invalidates the applicability of the approach. On the other hand, at $\beta=3.2$ the distribution appears to drop exponentially in an extended interval $0.5<y\lesssim 1$, possibly with two exponentials.  Hence expectation values of polynomials $\bra z^n\ket$ can be computed safely, as illustrated in Table  \ref{onepole}. 

For $\beta=1.6$, it can be seen that there is a nonvanishing 
boundary term around the pole. Instead of a small circle surrounding the 
pole at $z=i$ we may consider a horizontal line $y=1-\epsilon$ approaching 
the pole for $\epsilon\to 0$. Then the boundary term in Eq.\ \ref{eq:gauss} 
becomes (for $N_I=0$)
\bea
&& \lim_{\epsilon\to 0}\int K_y(x,1-\eps) P(x,1-\epsilon) \cO 
(x+i-i\epsilon) \,dx =
\nn \\
&&
\lim_{\epsilon\to 0}\int \left(n_p\frac{\epsilon}{x^2+\epsilon^2}-2\beta 
(1-\epsilon)\right)
P(x,1-\epsilon) \cO(x+i-i\epsilon) \,dx.
\eea
The smooth terms can be replaced by their values for $\epsilon=0$ and 
a boundary term arises because 
\be
\lim_{\epsilon\to 0} \int K_y(x,1-\eps)P(x,1-\epsilon)\, dx \neq 0.
\ee

\begin{figure}[t]
\centerline{
\includegraphics[width=.32\columnwidth]{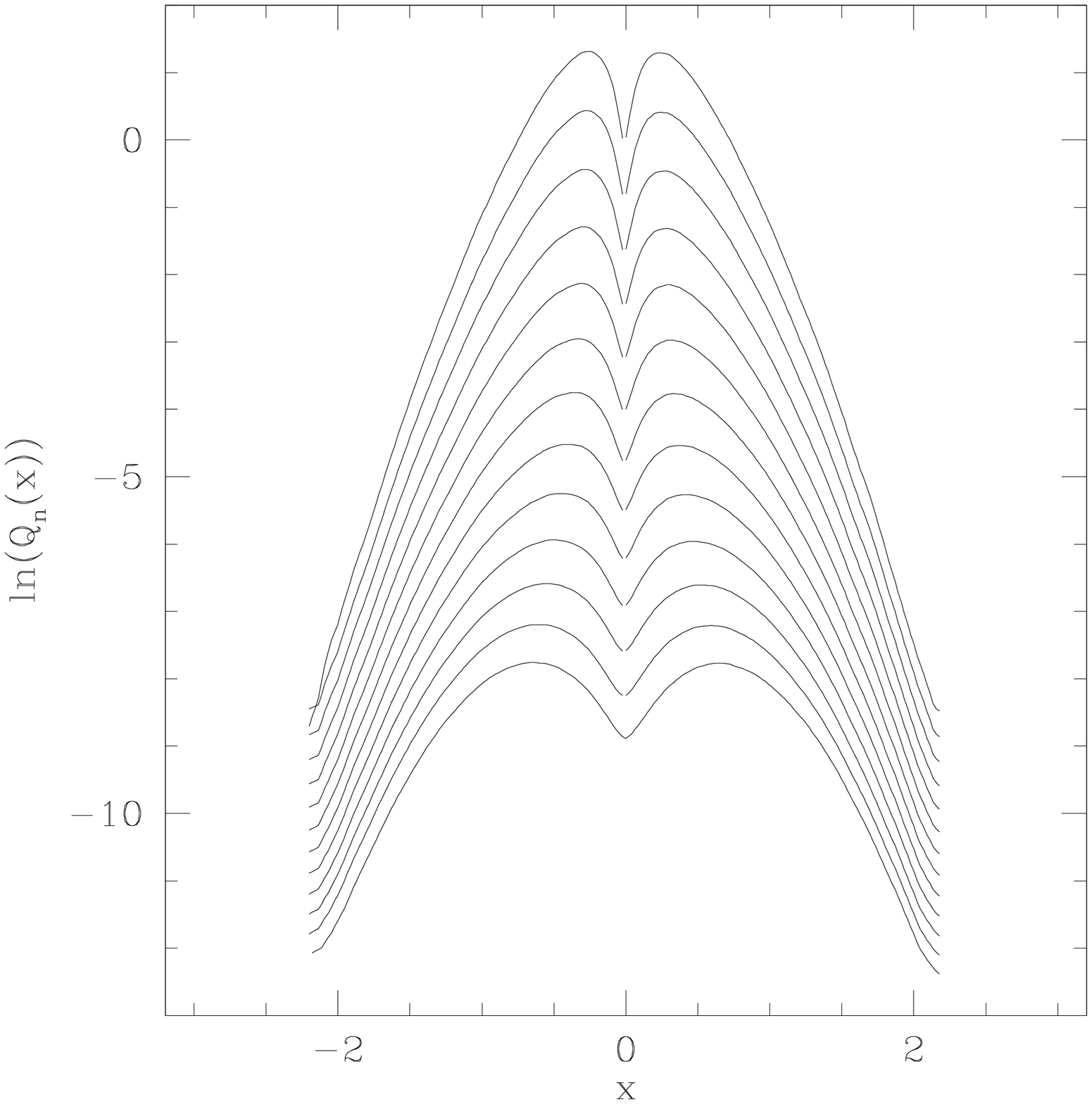}
\includegraphics[width=.32\columnwidth]{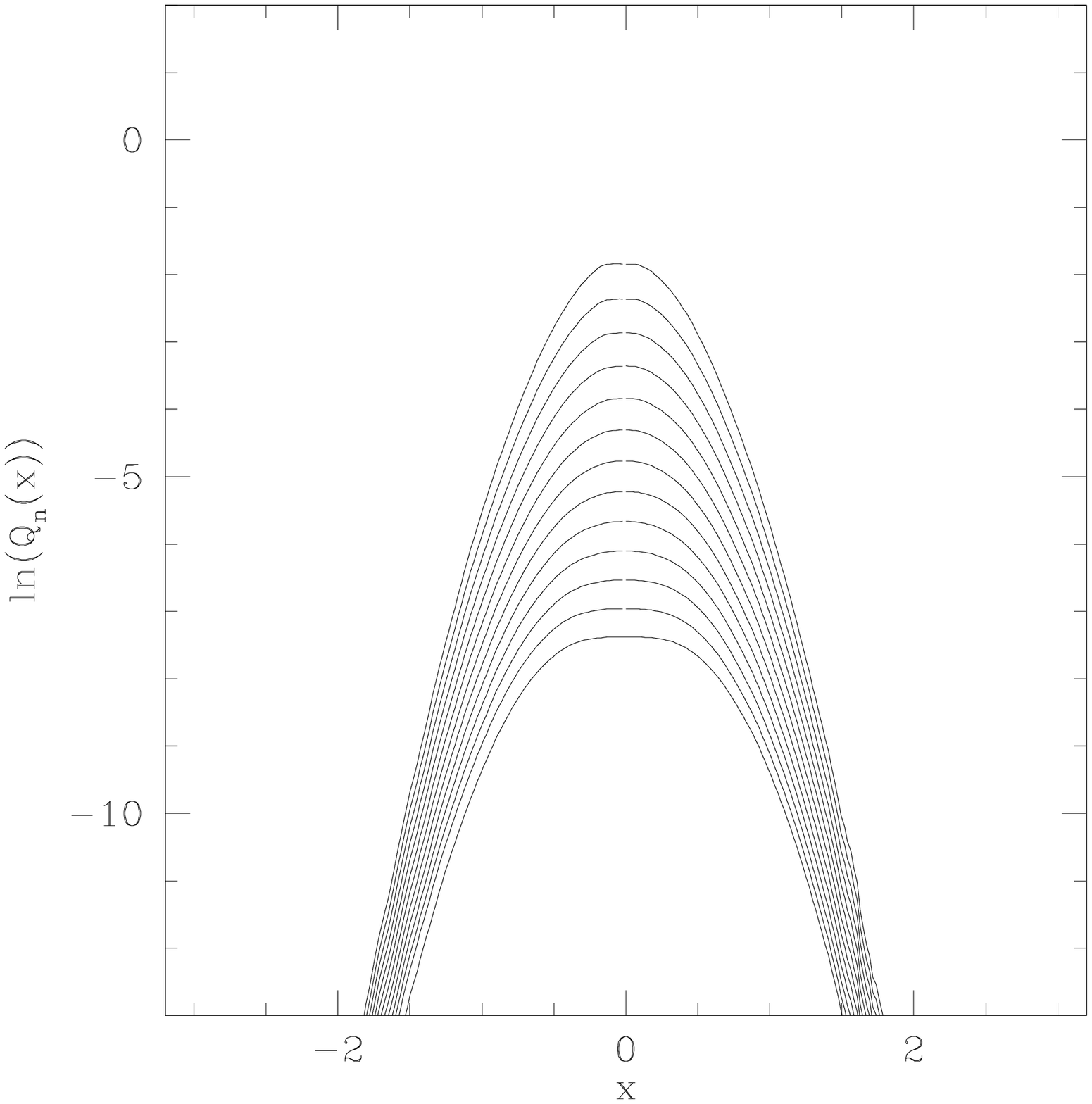}
\includegraphics[width=.32\columnwidth]{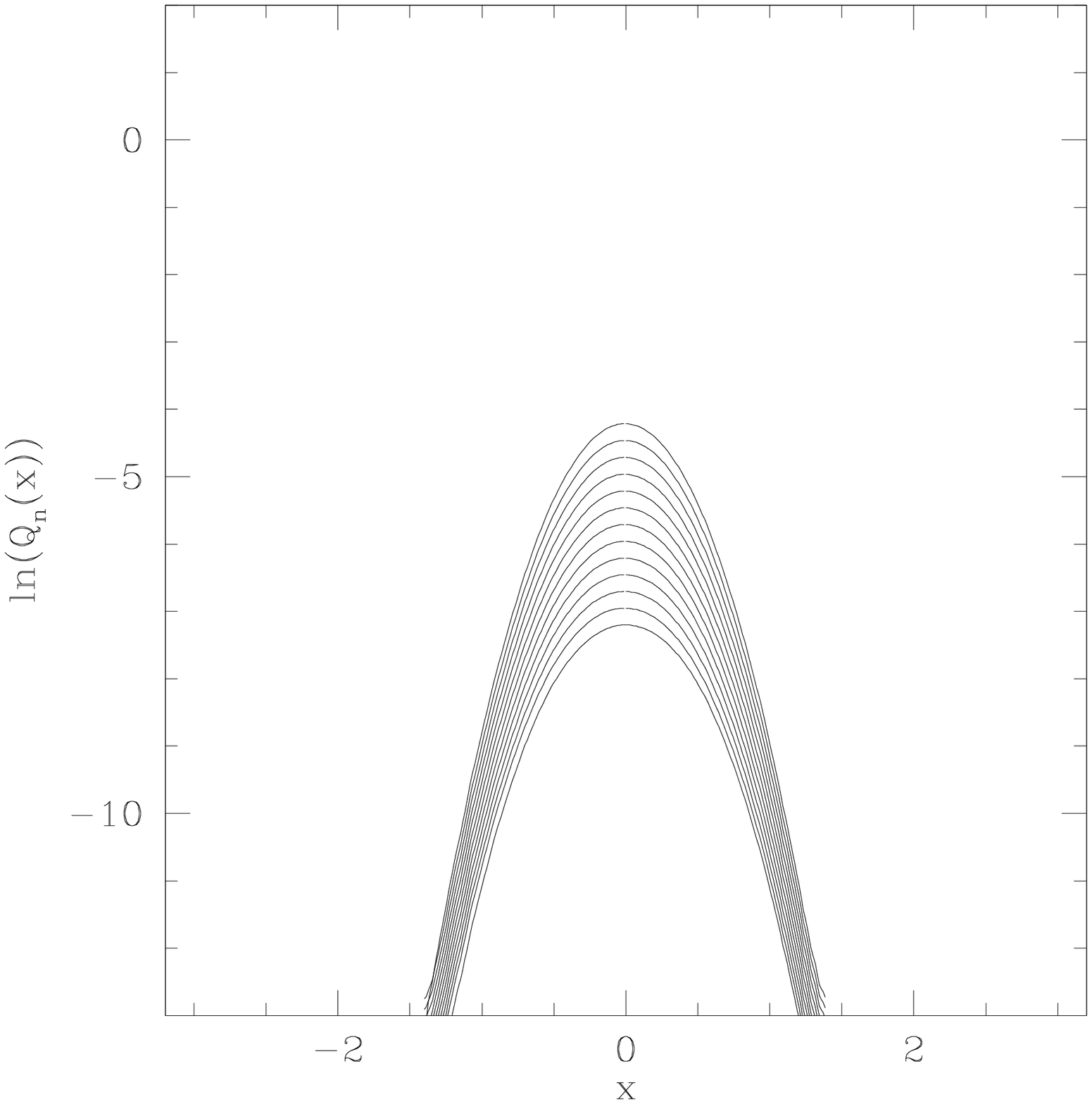}
}
\caption{$Q_k(x)$, see Eq.\ (\ref{qndef}), for $z_p=i$, $n_p=2$ and $\beta=1.6$ 
(left), 3.2 (middle) and 4.8 (right), for $k=0, -1, \ldots, -12$. In each figure the top 
(bottom) curve corresponds to $k=0$ ($-12$).
}
\label{qnx}
\end{figure}

Next we try to elucidate in more detail what is causing success or 
failure. For this purpose let us remember that in Ref.\ \cite{Aarts:2011ax} we 
established a criterion for correctness that went as follows: if the 
consistency conditions $\bra \tilde L\cO \ket=0$ hold for `all' 
observables {\it and} a bound of the form
\be
|\bra \cO\ket|< \mbox{const}\; \max_{x\in\R}|O(x)| 
\label{bound} 
\ee
holds, then the process produces correct results. Since for 
the original complex integral such a bound obviously holds, it is a 
{\it necessary} condition for correctness. Now the consistency condition 
simply expresses the fact that we have reached convergence, so it should 
be satisfied; the bound Eq.\ (\ref{bound}), however, may fail. We can see 
from the CL simulation that Eq.\ (\ref{bound}) apparently fails for 
$\beta=1.6$, but not for the other two values. In order to see this, define 
\be
Q_k(x)\equiv \int _\infty ^\infty dy\, P(x,y) \e^{-ky}  = \bra e^{-ky}\ket_y.
\label{qndef}
\ee
These functions are related to the expectation values of $\exp(ikz)$ by
\be
\bra \exp(ik(x+iy)) \ket =\int dx\, Q_k(x) \e^{ikx}. 
\ee
In Fig.\  \ref{qnx} we show the functions  $\log Q_k(x)$ for integer values  
$k=0,-1,\ldots, -12$. In all three cases the shape of the functions seems 
to stabilise with growing $k$, whereas there is approximately constant 
shift upwards with $k$. This suggests the following asymptotic behaviour,
\be
Q_k(x)\sim \exp(ck) f(x)\,,
\ee
with some constant $c>0$, so 
\be
\bra \exp(ik(x+iy))\ket \sim \exp(ck) \int dx\, f(x) \e^{ikx} = \exp(ck)\hat f(k).
\ee
How can this remain bounded for $k\to\infty$? The only possibility is that 
the Fourier transform $\hat f(k)$ decays exponentially; this will be the 
case if $f(x+iy)$ is analytic in a strip $|y|<const$. In particular $f(x)$ 
has to be smooth. Looking at Fig.\ \ref{qnx} one can see clearly that for 
$\beta=1.6$ $f(x)$ is developing a kink, wheres in the other two cases it 
at least appears to be smooth and the effect of the pole appears to be negligible. 
Hence we may conclude that the incorrect convergence is due to the failure of the bound (\ref{bound}).

To summarise the findings in the one-pole model, we conclude that if close to the pole the distribution drops to 
zero fast enough, e.g.\ exponentially in the case considered here, the 
meromorphicity of the Langevin drift is not necessary an obstacle and 
correct results can still be obtained.
When on the other hand the distribution is not falling rapidly at the pole, incorrect convergence is observed.

\subsection{U(1) one-link model}
\label{sec:u1}

In order to analyse what happens when poles are inside the distribution, we 
switch to the following U(1) integral with a complex weight, 
\be
Z=\int_{-\pi}^\pi dx\, \rho(x),   \qquad\quad 
\rho(x) = \left[1+\kappa \cos(x-i\mu)\right]^{n_p}\exp[\beta \cos(x)].
\ee
This model was introduced in Ref.\ \cite{Aarts:2008rr} (for $n_p=1$) as a toy 
model for QCD, with a complex `fermion determinant'
\be
D(x;\mu) =  1+\kappa \cos(x-i\mu), 
\ee
satisfying $[D(x;\mu)]^* = D(x;-\mu^*)$.  Complex Langevin dynamics was 
studied extensively in Ref.\ \cite{Aarts:2008rr} for $\kappa<1$, while problems 
for $\kappa>1$ were first reported in Ref.\ \cite{Mollgaard:2013qra}.
Subsequently thimbles were analysed in Ref.\cite{Aarts:2014nxa}.

When $\kappa<1$ the weight is positive when $\mu=0$, while for $\kappa>1$ 
there is already a sign problem at $\mu=0$. Concerning Langevin dynamics, 
we note that good results are obtained when $\kappa<1$ (and $k$ not too large and negative), while problems 
emerge for $\kappa>1$ and $\beta$ not too large \cite{Mollgaard:2013qra,Aarts:2014nxa}. 
It should be noted that in view of the later sections even values of $n_p\ge 2$ 
can be physical as the QCD determinant has double
zeroes when the Wilson fermion formulation is used.

\begin{figure}[t]
\centerline{ 
  \includegraphics[width=.90\columnwidth]{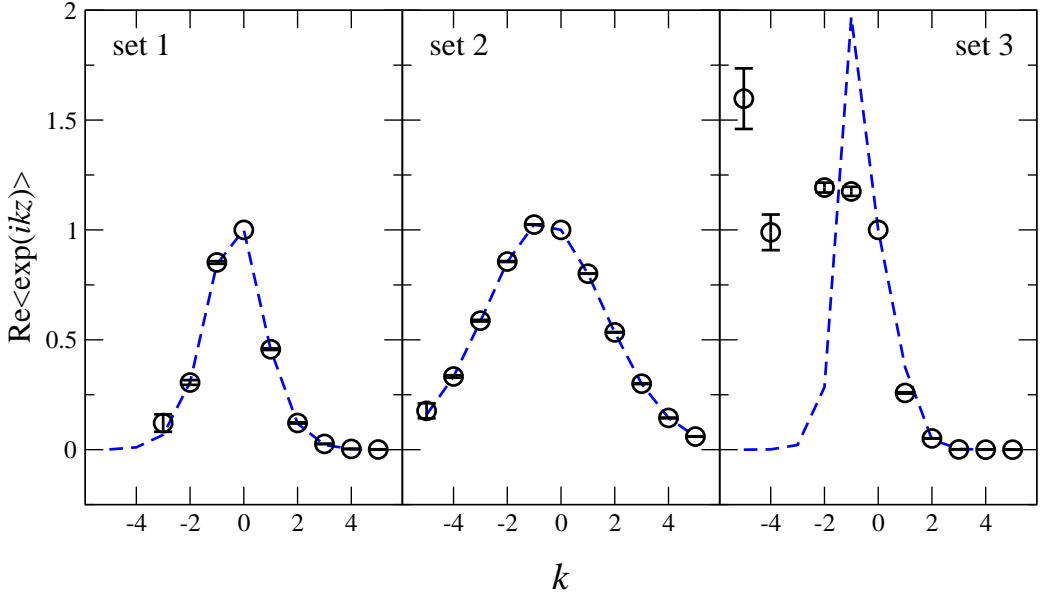} 
 }
\caption{Re $\bra\exp{ikz}\ket$ for $k=-5, -4, \ldots, 4,5$ vs $k$, for parameter sets (1,2,3), all with $n_p=1$ (the imaginary parts are negligible). The lines are the exact results.
}
\label{u1_plot}
\end{figure}

The complex drift reads
\be
K(z)=-\beta \sin(z)-\frac{n_p\kappa \sin(z-i\mu)}{1+\kappa \cos(z-i\mu)}.
\ee
When $\kappa<1$ there is an attractive fixed point at $x=0$ and repulsive fixed 
points at $x=\pm\pi$, with poles located at $z_p=\pm\pi+iy_p$, where $\cosh(y_p-\mu)=1/\kappa$. When $\kappa>1$, poles are at $z_p=x_p+i\mu$, with $\cos x_p=-1/\kappa$. 
We start with a brief discussion of three sets of parameters, all with $n_p=1$:
\begin{itemize}
\item[(1)] $\kappa=0.5, \beta=1, \mu=1$: pole at  $x_p=\pm\pi, y_p=\mu+\arccosh(1/\kappa)$;
\item[(2)] $\kappa=2, \beta=5, \mu=1$: poles at $x_p=\pm\frac{2}{3}\pi, y_p=\mu$;
\item[(3)] $\kappa=2,\beta=0.3,\mu=1$: poles at  $x_p=\pm\frac{2}{3}\pi, y_p=\mu$.
\end{itemize}
Results of CL dynamics for the observables $\bra e^{ikz}\ket$ ($k=\pm1, \ldots,\pm 5$) are shown in Fig.\ \ref{u1_plot}. For set (1), we observe good results, except when $k$ is large and negative, $k=-4,-5$. For those values, fluctuations are large and increasing the simulation time does not improve this, a sign of non or poor convergence. For set (2), excellent agreement with exact results is obtained. For set (3), we observe agreement for large and positive $k$, but increasingly worse behaviour as $k$ is reduced. The results for $k=-4,-5$ have larger errors, but the values of the averages are robust as the Langevin time is increased, hence here we find incorrect convergence.
Since for our choice of parameters the poles are located at $y_p>0$, we note that exponentials with $k>0$ ($k<0$) will be less (more) sensitive to the presence of the poles, as a suppression (enhancement) with $e^{-ky}$ ($e^{ky}$) arises naturally. This is indeed supported by the data.

In the following we focus on the case where $\kappa>1$ and $\beta\lesssim 1$, since this is where complex Langevin dynamics converges, but possibly to an incorrect result. 
Moreover, we will compare $n_p=1,2$ and 4.

\subsubsection{Poles inside the distribution}
\label{subsec:in}

\begin{figure}[t] 
\includegraphics[width=.48\columnwidth]
{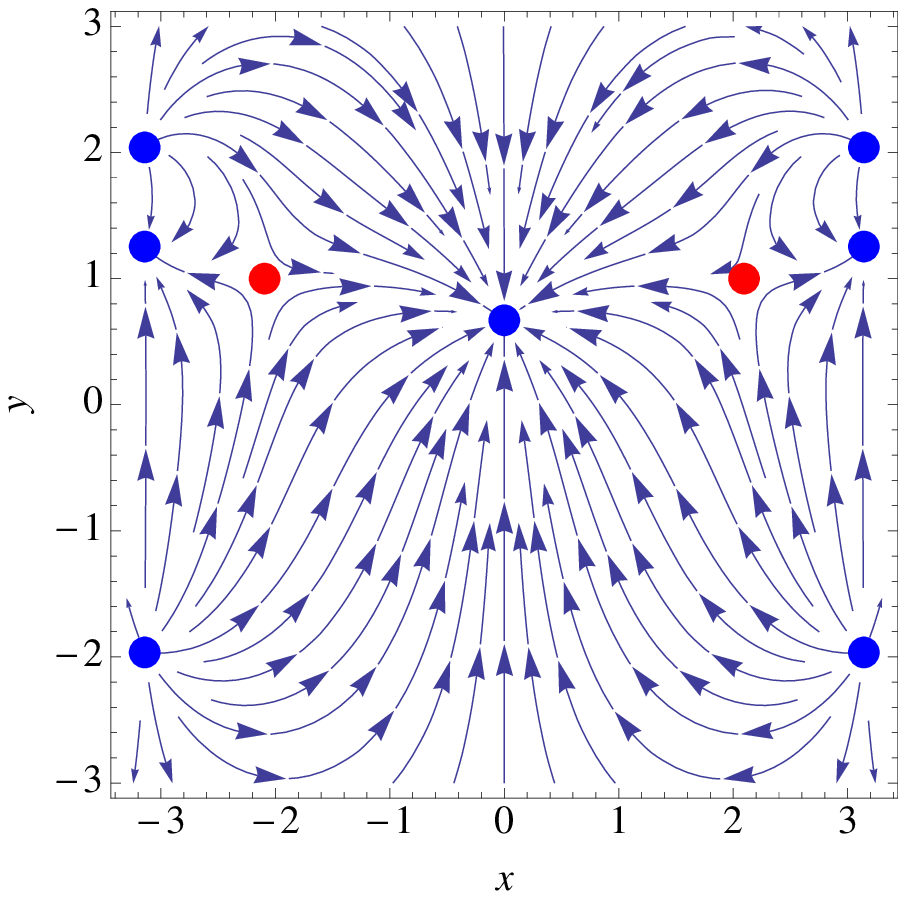} 
\includegraphics[width=.48\columnwidth]
{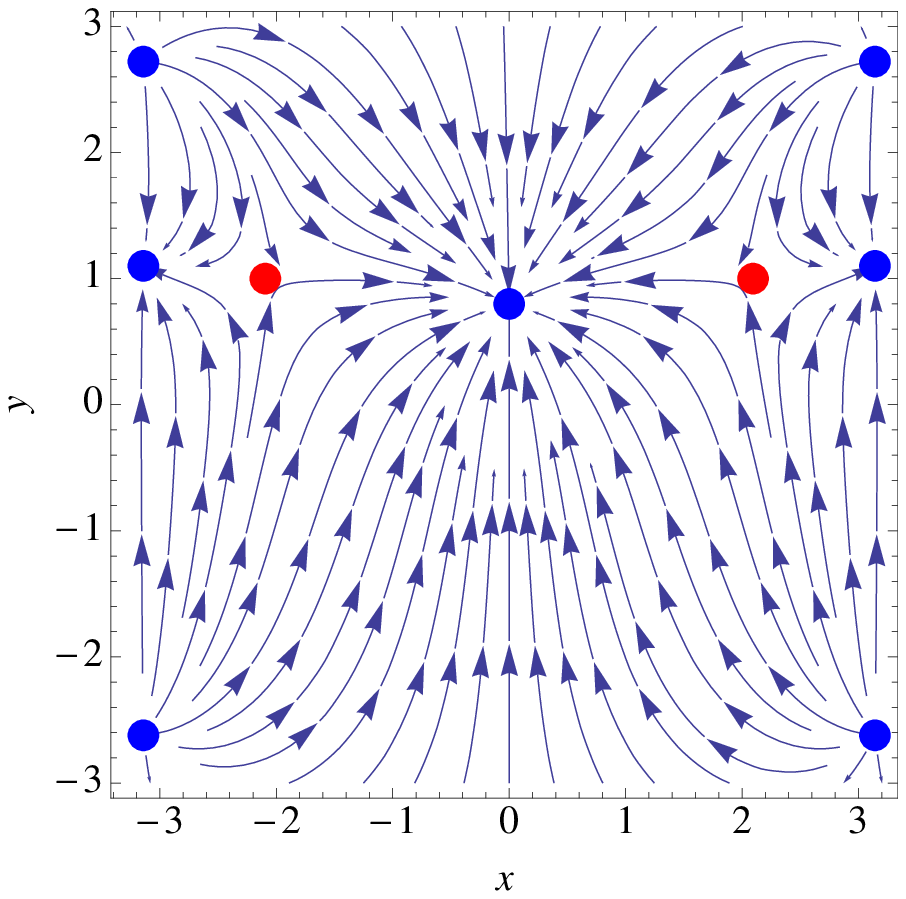} 
\caption{Classical flow diagrams in the U(1) model with $\beta=0.3, \kappa=2, 
\mu=1$,  $ n_p=1$ (left) and $n_p=2$  (right).  The blue (red) circles 
are fixed points (poles). }
\label{flow-u1}
\end{figure}

We consider parameter set (3), with $\beta=0.3, \kappa=2, \mu=1$ 
and $n_p=1,2$ and 4. 
Classical flow diagrams are given in Fig.\ \ref{flow-u1} for $n_p=1, 2$
(note the periodicity in $x$). Besides the attractive `perturbative' fixed
point at $x=0$, there is an additional attractive fixed point at
$x=\pm\pi$. The other two fixed points are repulsive. It is clear to see  
from the flow diagrams, and can be confirmed following a similar analysis
as above, that the equilibrium distribution will be contained in a
horizontal strip between the two attractive fixed points. Finally, the
pole is attractive in the imaginary direction and repulsive in the real
direction (as always), making the pole an approximate bottleneck, just as
in the real case considered in Sec.\ \ref{sec:ergo}. Hence, as the 
attractive fixed points move closer together in the imaginary direction, 
the distribution gets narrower and narrower.

\begin{figure}[t] 
\includegraphics
[width=.48\columnwidth]{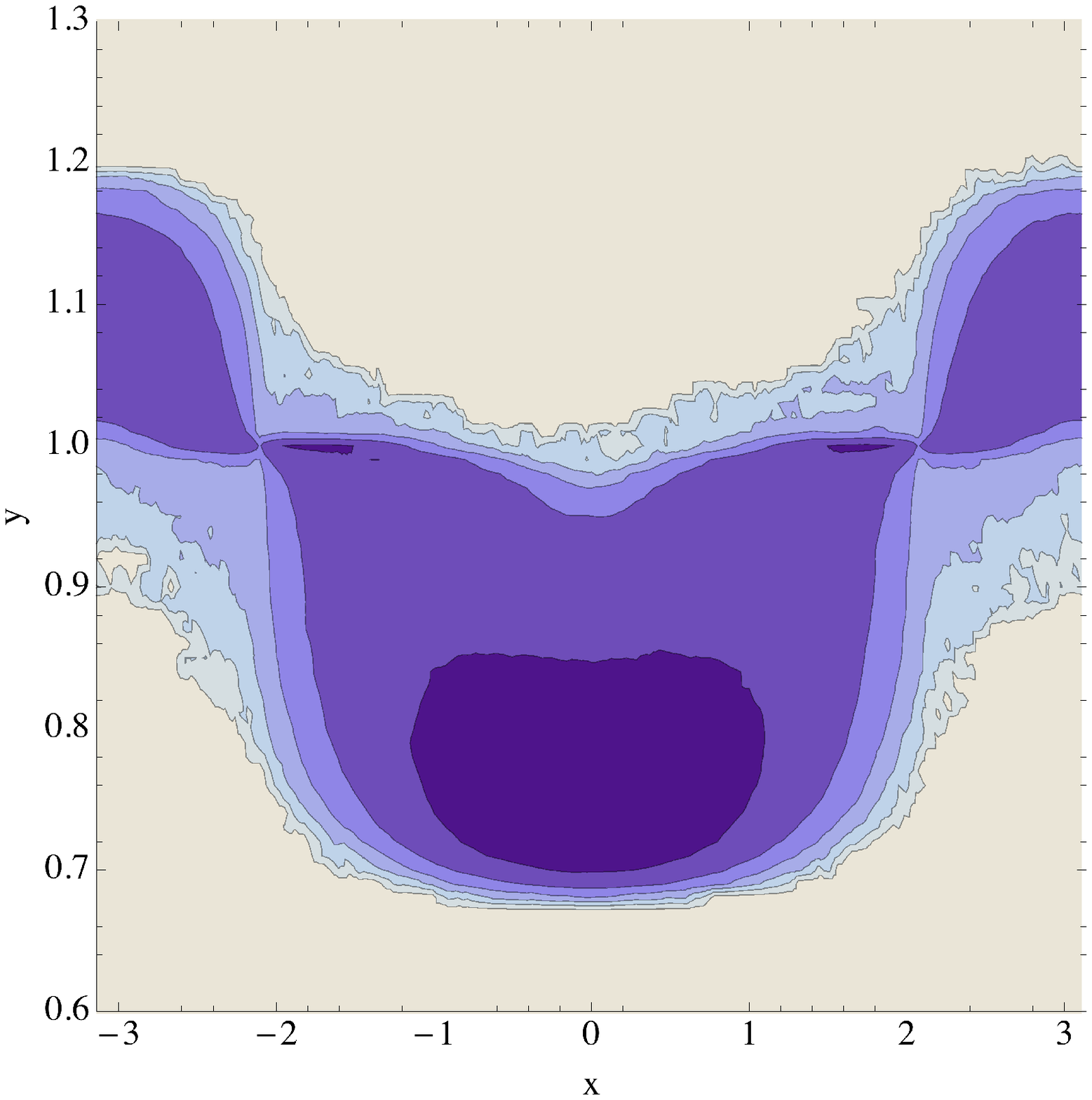} 
\includegraphics
[width=.48\columnwidth]{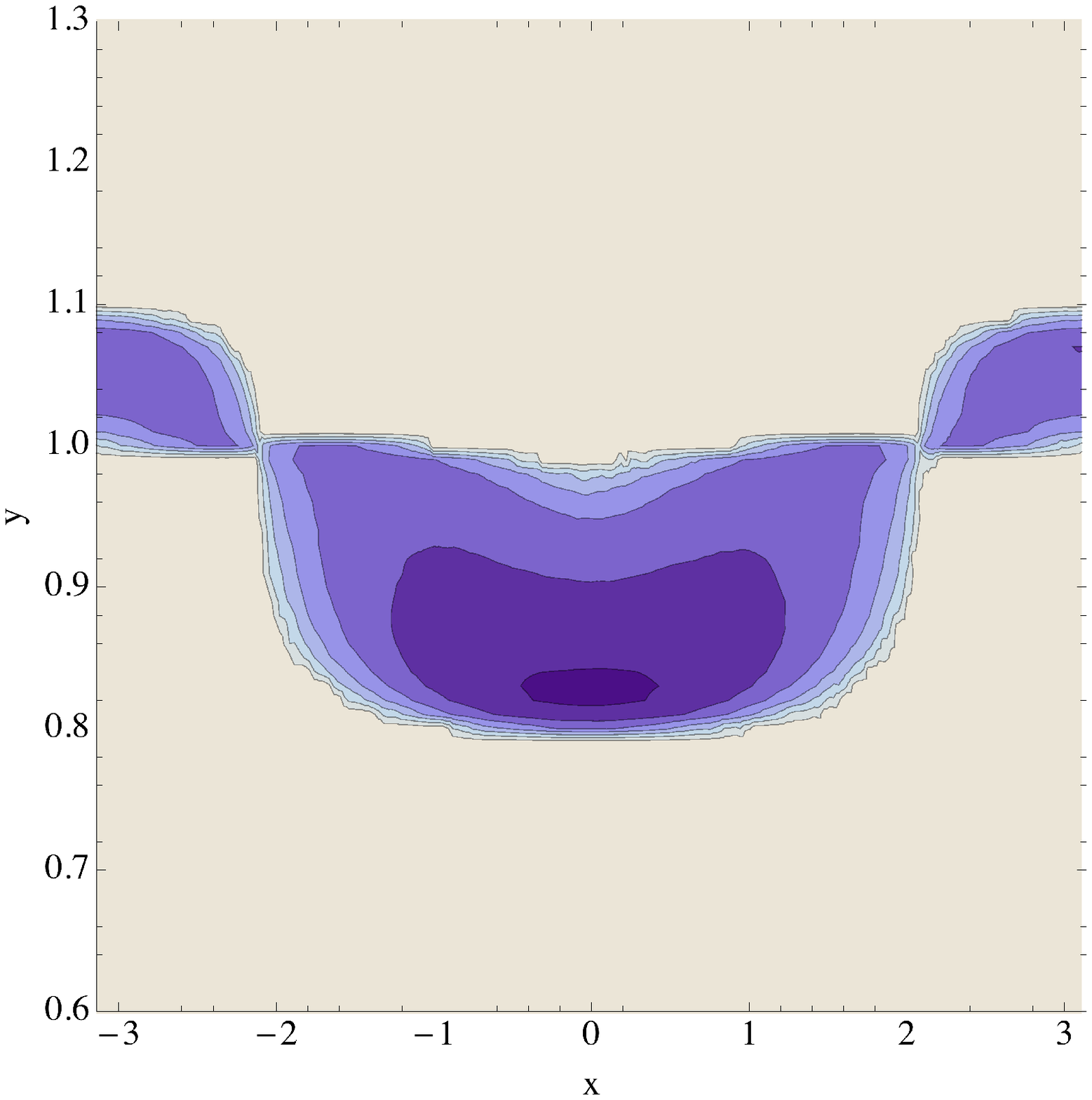} 
\caption{Logarithmic contour plots of the distribution in the $xy$ plane, for 
$\beta=0.3, \kappa=2, \mu=1$,  $n_p=1$ (left) and $n_p=2$  (right).  
 }
\label{cont-xy}
\end{figure}

 In Fig.\ \ref{cont-xy} we show logarithmic contour plots of the 
equilibrium distribution sampled during the CL process in the complex plane for $n_p=1$ (left) and 2 
(right). Note that the darker colours correspond to the most frequently 
visited regions. The position of the pole can clearly be identified as the 
place where the distribution is pinched, resulting in a bottleneck; this 
effect gets stronger with increasing $n_p$. The distribution is strictly zero outside the strip set by the attractive fixed points.

To better understand this structure, we note that the approximately disconnected regions (i.e. the `head' and the `ears' in Fig. \ref{cont-xy}) are characterised by the sign of the real part of the determinant, 
\be
\Re D = 1+\kappa \cos(x)\cosh(y-\mu),
\ee
and hence we will refer to them as $G_\pm$,
\be
G_+=\left\{ (x,y) \; | \;  \Re D>0 \right\}, 
\qquad\qquad
G_-=\left\{ (x,y) \; | \;  \Re D<0 \right\},
\ee
with $G_+$ the `head' and $G_-$ the `ears'.
For $n_p=1$, we observed frequent crossings between the two regions. For $n_p=2$, the crossings are rarer but still frequent enough such that both regions are visited during long runs. This might, however, be due to the finite time step. In the 
continuous time limit it is possible that the two regions that are not connected by the process, i.e.\ the process might not be ergodic. 
Of course rare crossings make it hard to collect good statistics.
For $n_p=4$ (not shown) no crossings were observed and the distribution only has support in $G_+$.
 
\begin{figure}[t] 
\includegraphics[width=.48\columnwidth]
{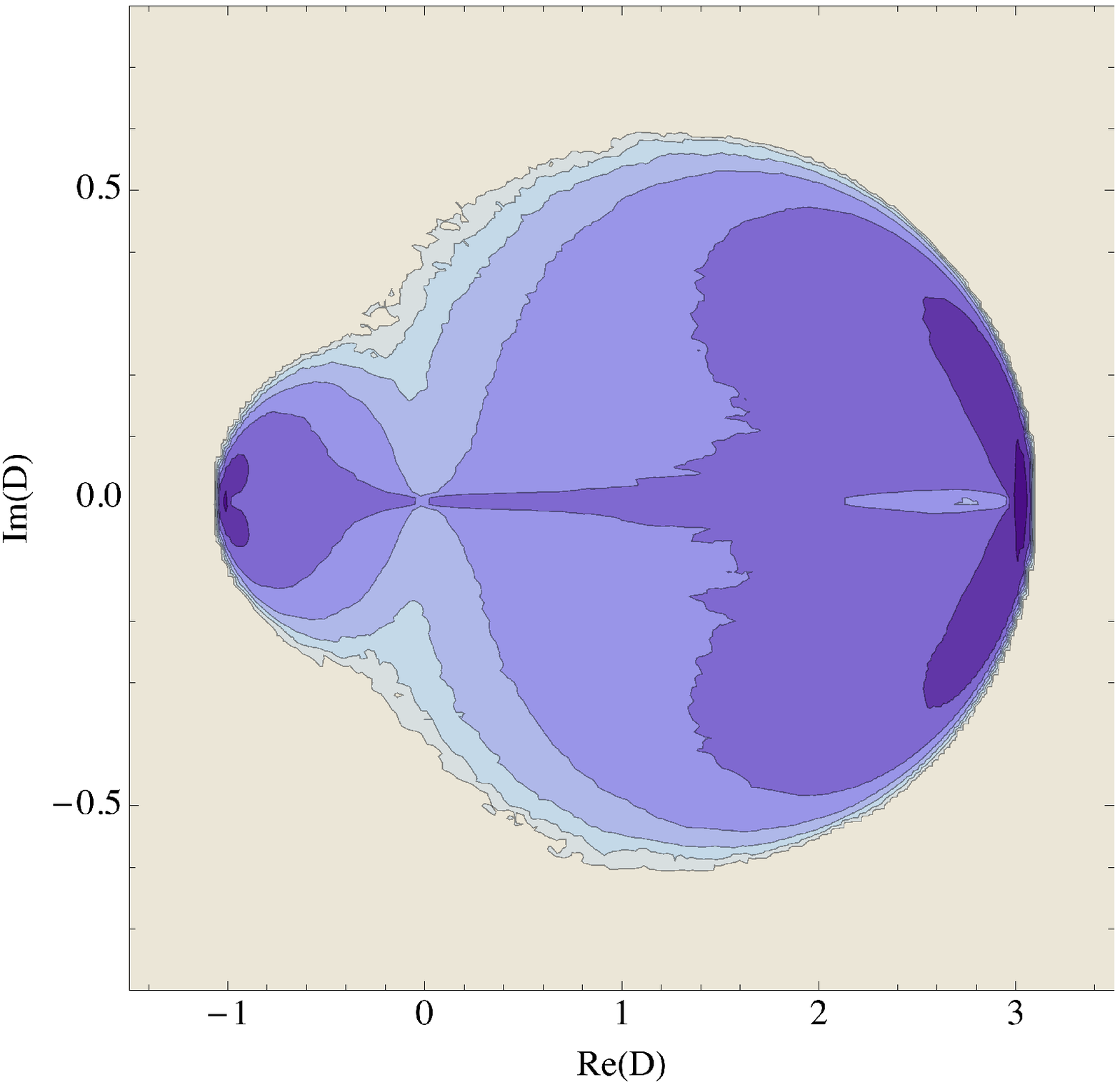} 
\includegraphics[width=.48\columnwidth]
{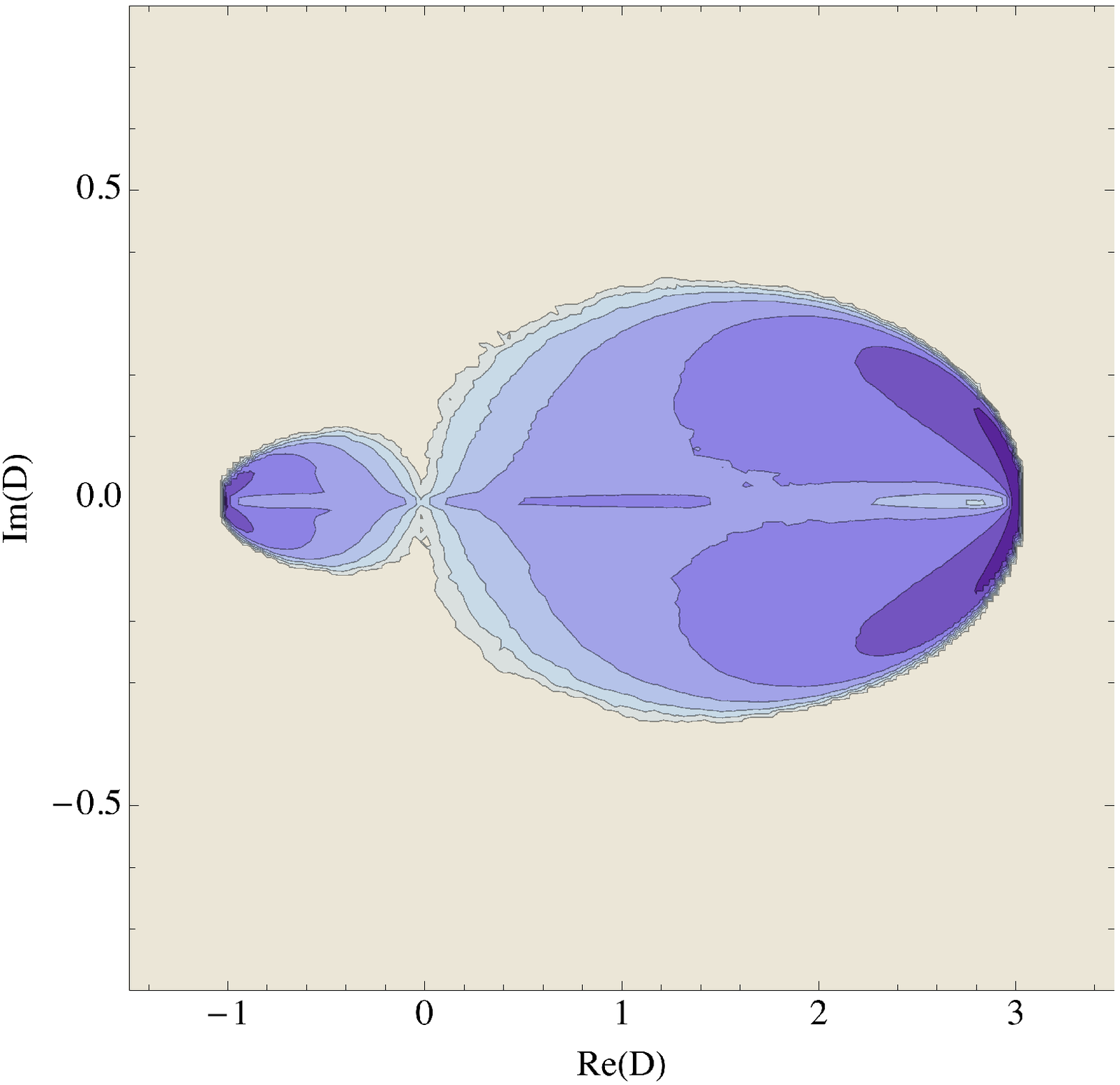} 
\caption{Logarithmic contour plots of the distribution of the complex 
determinant $D$, for $\beta=0.3, \kappa=2, \mu=1$,  $n_p=1$ (left) and 
$n_p=2$  (right). 
}
\label{cont-D}
\end{figure}

To translate these findings to an observable easily accessible also in more complicated models and lattice theories, we consider the complex determinant. Logarithmic  contour plots of $D$ are shown in  Fig.\ \ref{cont-D}. We observe a similar structure, with the zero of $D$ acting as the bottleneck. We will use this diagnostics in the more complicated models discussed below.

\begin{figure}[t]
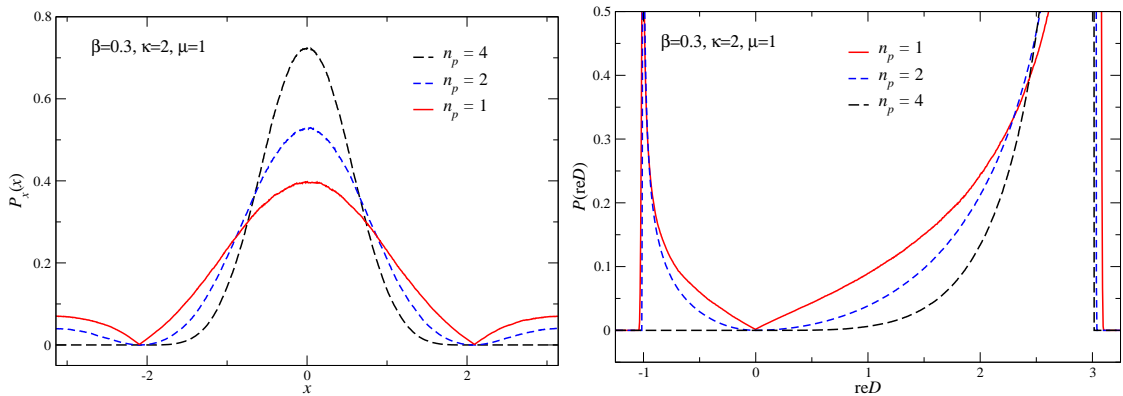
 
\includegraphics 
[width=.48\columnwidth]{figs-POLES/plot-U1-b03-k2-mu1-nf124-Px.eps} 
\includegraphics 
[width=.48\columnwidth]{figs-POLES/plot-U1-b03-k2-mu1-nf124-PreD.eps} 
\caption{Partially integrated distributions $P_x(x)$ (left),  
distributions of Re $D$ (right) for $n_p=1,2,4$.}
\label{histou1}
\end{figure}

\begin{figure}[t]
\centerline{ 
  \includegraphics[width=.70\columnwidth]{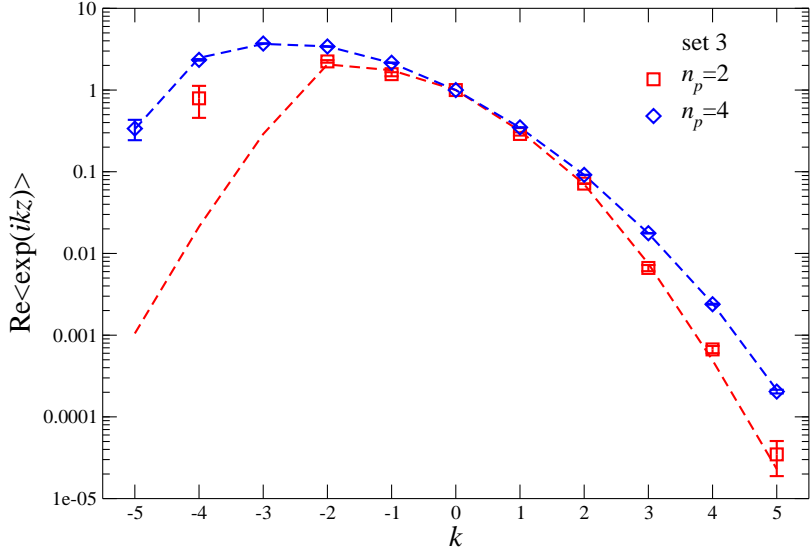}
 }
\caption{Re$\bra\exp{ikz}\ket$ for $k=-5, \ldots, 5$ vs $k$, on  a logarithmic scale, for parameter set (3), with $n_p=2,4$.
The lines are the exact results.
}
\label{fig:set3-24}
\end{figure}

In view of the formal justification, see Sec.\ \ref{sec:formal}, it is important to know the rate at which the distribution goes to zero at the pole. 
 This is shown in Fig.\ \ref{histou1} for the partially integrated distribution $P_x(x)$ (left) and the real part of $D$ (right). For $n_p=1$ we observe a linear decrease at the pole (recall that $x_p=\pm 2\pi/3$), while for $n_p=2$ the decay is faster. For $n_p=4$, the pole is not crossed and the entire dynamics takes places in $G_+$. Since the pole does not negatively influence the dynamics in this case, we expect good agreements with the exact results, although there may be problems with ergodicity, similar to the real case.
   This is demonstrated in Fig.\ \ref{fig:set3-24}, where Re $\bra e^{ikz}\ket$ is shown on a logarithmic scale, for 10 values of $k$. For $n_p=2$ we find approximate agreement, especially for $k$ close to 0. For $n_p=4$, good agreement is seen for all $k$ values considered. 
 This is consistent with the formal derivation: for $n_p=4$ the pole is avoided and only the region sufficiently far from $D=0$ is relevant.
 
Finally, we stress once more that further support for the validity of the formal arguments comes from the observed interplay of the observables and the pole:  it is possible that for some observables good agreement is found, while for others it is not. This crucially depends on the region in configuration space most relevant for the observable under consideration, as exemplified in this model by the observables $\bra e^{ikz}\ket$, with $k\gtrless 0$.

\subsubsection{What does the CL simulation actually compute?}
\label{sec:what}

In order to further understand the relevance of the contributions from the nearly disconnected regions $G_\pm$, we have analysed the results from Langevin simulations for $\bra e^{ikz}\ket$ by separating  the trajectories based on the sign of the real part of $D$. The results are summarised in Table \ref{compG} in the columns labeled CL[$G_\pm$]. We note that the results obtained when restricted to $G_+$ are close to the exact results, listed in the first column, but not quite equal.

We can understand this as follows: first we shift the contour of integration of the original integrals
to go through the zeroes of $\rho(z)$. For set (3) this means Im $z=\mu$. Next we split the integration into two contributions 
coming from the two inequivalent paths connecting the zeroes, one living in $G_+$ and the other in $G_-$, and define
\be
Z_{\pm}\equiv \int_{x\in G_\pm} dx\,  \rho(x+i\mu),
\ee
and similarly
\be
\bra \cO \ket_\pm \equiv \frac{1}{Z_\pm} \int_{x\in G_\pm} dx\, \cO(x+i\mu) \rho(x+i\mu).
\ee
The exact results, restricted to $G_\pm$, are shown in Table \ref{compG} in the columns labeled exact[$G_\pm$]. 
The agreement between the restricted Langevin and exact results  is convincing. This should not be surprising, since the 
formal proof of correctness provided earlier is directly applicable to the model restricted to $G_+$ or $G_-$.

\begin{table}[t]
 \begin{center}
 \begin{tabular}{r |  c cc cc }
  $k$ & exact & CL[$G_+$] & exact[$G_+$] & CL[$G_-$] & exact[$G_-$]   \\
  \hline
 $-2$  & 2.05781 & 1.9589(29)  & 1.94847     & 5.9554(90)   &  5.94936  \\
 $-1$  &  1.74691 &1.87106(43) & 1.87036    & $-$2.6473(11)    & $-$2.64655  \\
 $1$   & 0.316378 & 0.33450(11) & 0.334309 & $-$0.32181(6)   & $-$0.321777  \\
 $2$   & 0.0702397  & 0.07013(9)  & 0.0697774 & 0.08675(10) & 0.0866928 
 \end{tabular}
\caption{Re$\bra e^{ikz}\ket$ for several values of $k$, when restricted to $G_\pm$ (Re $D\gtrless 0$), for 
 $\beta = 0.3, \kappa = 2, \mu = 1$, with $n_p=2$, comparing complex Langevin (CL) and exact results. 
}
  \label{compG}
\end{center}
\end{table}

Since the exact values for the full model can be obtained as
\be
\bra \cO \ket_{\rm exact} = \frac{Z_+ \bra \cO \ket_+ + Z_-\bra \cO \ket_-}{Z_+ +Z_-},
\ee 
a way to obtain the correct results would be to combine the restricted simulation results with the weights  
\be
w_\pm\equiv \frac{Z_\pm}{Z_+ +Z_-}\,. 
\ee
Note that since $Z_-/Z_+ \simeq 0.0281\ll 1$, the deviation between full results and those restricted to $G_+$ is on the order of a few percent as well, as illustrated in Table \ref{compG}.
The problem with this prescription is of course that in realistic models 
the weights are not known. However, below we will see that typically $w_-$ 
is  tiny and can be approximated by zero; here it is nonnegligible because we chose the rather extreme value $\kappa=2$. 

In the case of $n_p=4$, the process never crosses into $G_-$, which indicates a lack of ergodicity, similar to what was found in Sec.\ \ref{sec:ergo}. The process simulates a version of the original integral restricted to a path running between 
the zeroes, which is not quite equal to the full integral.  This causes a tiny systematic error which is, however, not visible in the data since it is highly suppressed; for $n_p=4$, $Z_-/Z_+ \simeq 0.00302\ll 1$.

 \section{Lessons from simple models}
\label{sec:lessons}

The following lessons can be learned from the simple one-variable models.

{\bf Lesson 1:} It has been suggested \cite{Mollgaard:2013qra, 
Mollgaard:2014mga} that the {\it winding} of the Langevin paths around the 
pole is the source of the problem, because the pole corresponds to a logarithmic 
branch point in the action. However, in the one-pole model of Sec.~\ref{subsec:one} we have demonstrated explicitly
that no such winding occurs,  since the pole lies either on the edge or outside the distribution.
 Nevertheless wrong results can be encountered. Further 
indication that it is not the winding which matters has been given in 
Ref.\ \cite{Nishimura:2015pba}, see  also Sec.\ \ref{sec:lattice} for the case of full QCD.

{\bf Lesson 2:} It has been said (see for instance Ref.\ \cite{Nishimura:2015pba}) that it is 
sufficient for correctness if the distribution $P$ is `practically zero' at the pole. Using again evidence from the one-pole model,
we note that this is not correct in general: for small $\beta=1.6$ wrong results are obtained, but $P(x,y)$ vanishes at the pole.
On the other hand, we have shown (and demonstrated numerically for $\beta=4.8$) that it is 
sufficient for $P$ to be nonzero only in a simply connected region whose 
closure does not contain the pole(s). The intermediate case $\beta=3.2$ 
seems to have at least a distribution $P$ vanishing at very high (maybe 
infinite) order at the pole, also leading to good results. All this can be 
understood in the light of the fact discussed in Sec.\ \ref{sec:formal}: the 
observables evolving according to Eq.\ (\ref{obsevol})
typically develop an essential singularity at the location of the pole of the drift.

{\bf Lesson 3:}
A strong attractive fixed point sufficiently far from any poles of the
drift leads to correct results. This almost obvious fact has been
observed already earlier, e.g.\ in QCD with static quarks \cite{Seiler:2012wz}.

{\bf Lesson 4:} The existence of a `bottleneck' between two regions $G_+$ 
and $G_-$, such as in the  U(1) one-link model of Sec.\ \ref{sec:u1},
is a signal for potential trouble. The best variable to analyse this is the determinant 
$D$ (not raised to any power), because it 
can also be used in more complicated lattice  models, as we will see below.

{\bf Lesson 5:} It is possible that the relative weight of one of the two regions is suppressed, i.e.\ $w_-\ll w_+$. 
Then a modification of the process which includes only trajectories with Re $\det D>0$, i.e.\ those contained in $G_+$, 
or using long runs such that the weight of runs in $G_-$ is naturally suppressed, seems to 
produce reasonably good results.  On closer inspection, however, it only gives approximate results, since only one part of the original complex 
integral is represented, namely the part contained in $G_+$. However,  if  indeed $w_-\ll w_+$, this may give a numerically accurate approximation to the complete problem.

{\bf Lesson 6:} The effect of increasing the strength of the pole by increasing 
$n_p$ is twofold: On the one hand the `pull' in the imaginary directions towards 
the pole is increased, which is bad; on the other hand the `push' in the real 
directions away from the pole is strengthened, which is good. 

In the one-pole model,  with the pole on the imaginary axis, the first effect dominates: hence increasing $n_p$ makes the 
situation worse. We have checked that for $n_p=2$, to obtain correct results,  a larger value of
$\beta$ is needed than for $n_p=1$.

For parameter set (3) in the $U(1)$ model the second effect 
dominates: increasing $n_p$ makes the bottleneck between the two regions $G_+$ and 
$G_-$  narrower and inhibits transitions 
between the two regions; furthermore it reduces the relative  weight of $G_-$. For 
$n_p=1$ this bottleneck does not prevent the process from moving between the two 
regions; for $n_p=2$, transitions are already rarer and it seems that each of the regions around the 
two attractive fixed points supports an invariant measure by itself; for 
$n_p=4$ no transitions are observed even for extremely long runs. It should be 
noted  that in lattice QCD with $n_f$ flavours of Wilson fermions the degrees of freedom make $n_p$ at least $2n_f$. 

{\bf Lesson 7:} The interplay between an observable and the distribution 
determines how close the expectation value of the former is to the correct one: if the 
observable is naturally suppressed/enhanced near the pole, it is possible to obtain, 
within the numerical error,  correct/manifestly incorrect results. 
This explains why one can encounter both apparently correctly and 
manifestly incorrectly determined expectation values in a single analysis.

We will now take these lessons and see how they apply to more realistic models.

 \section{Effective SU(3) one-link model}
\label{sec:spin}

In the following section we investigate the role of the zeroes and 
the ensuing lessons in a system with more degrees of freedom, which 
is however still exactly solvable, namely an effective SU(3) one-link 
model. 
Versions of this model have been considered before, 
see e.g.\ Refs.\ \cite{Aarts:2008rr,Aarts:2012ft}.
Here, the form of the model and the choice of parameters 
is motivated by QCD with heavy quarks (HDQCD), to be discussed in
 Sec.\ \ref{sec:lattice}.

The starting point is QCD with $N_f$ flavours of Wilson fermions.
At leading order in the 
hopping expansion, the fermion determinant can be expressed as 
a product of factors involving Polyakov loops at each 
spatial site, see  Sec.\ \ref{sec:lattice} below, 
\be
\label{eq:P}
 \det M = \prod_\xv 
		\det\left(1 + C  \mathcal{P}_\xv \right)^{2N_f} 
		\det\left( 1 + \tilde C  \mathcal{P}^{-1}_\xv \right)^{2N_f},
\end{equation}
where the remaining determinant is in colour space 
only\footnote{Note that these expressions are valid for
 SU($N_c$) and   SL($N_c,\mathbb{C}$).} and $\mathcal{P}_\xv^{(-1)}$ are the 
 (inverse) Polyakov loops, 
\be
  \mathcal{P}_\xv = \prod_{\tau = 0}^{N_\tau - 1} U_{(\xv,\tau),4}		
		\qquad \qquad 
  \mathcal{P}^{-1}_\xv = \prod^0_{\tau=N_\tau-1} U^{-1}_{(\xv,\tau), 4},
\ee
with $N_\tau$ the number of time slices in the temporal direction. 
The parameters $C, \tilde C$ arise from the hopping expansion and read
\be
\label{eq:C}
 C = \left( 2\kappa e^{ \mu}\right)^{N_t},
 \qquad\qquad\qquad
 \tilde  C = \left( 2\kappa e^{-\mu}\right)^{N_t}.
 \ee 
Employing the temporal gauge we can see that the product of local
factors is equivalent to having only one temporal link in each factor.
Using standard relations, again valid for both SU($N_c$) and  SL($N_c,\mathbb{C}$), 
the remaining determinants can be expressed in terms of the traced Polyakov loops, 
\be
P_\xv= \frac{1}{N_c}\tr  \mathcal{P}_\xv, \qquad\qquad  P_\xv'=\frac{1}{N_c}\tr  
\mathcal{P}_\xv^{-1}.
\ee
Explicitly, for $N_c=2$ this gives
\be
 \det \left(1 + C  \mathcal{P}_\xv \right) = 1+2C P_\xv +C^2, \quad\quad
 \det \left(1 + \tilde C  \mathcal{P}_\xv^{-1} \right) = 
 1+2\tilde C P_\xv' + \tilde C^2,
\ee
and for $N_c=3$,
\bea
\det \left(1 + C  \mathcal{P}_\xv \right) &=&  1+3C P_\xv 
+3 C^2 P'_\xv +C^3, \\
\det \left(1 + \tilde C  \mathcal{P}_\xv^{-1} \right) &=&  
1+3\tilde C P'_\xv +3 \tilde C^2 P_\xv + \tilde C^3.
\eea
For larger $N_c$ the relations become more complicated but the determinant
always includes a $C^{N_c}$ term which dominates at large $\mu$
(making  
the sign problem increasingly  harmless toward the saturation regime). Notice that 
for SU($N_c$), $|P_\xv|, |P'_\xv| \leq 1$. In the following we 
concentrate on the $N_c=3$ case.

\subsection{Effective one-link model for HDQCD}
\label{sec:subhd}

\begin{figure}[t]
\begin{center}
\epsfig{file=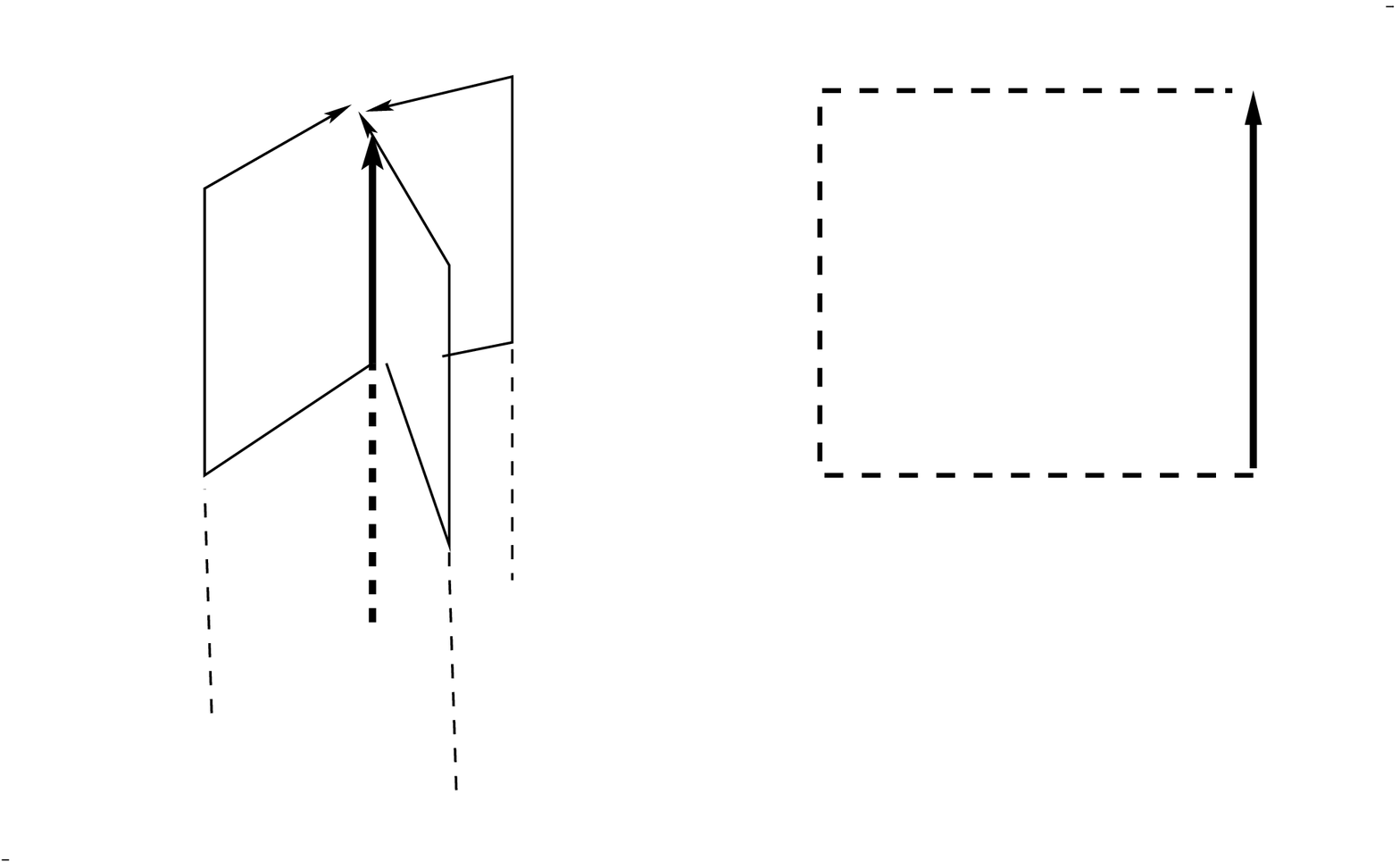, width=8.cm}
 \caption{Effective one-link model  for the  Polyakov line in temporal gauge 
 in the field of its neighbours.}
\label{fig:efmod}
\end{center}
\end{figure}

To define an effective model for HDQCD in four dimensions 
we consider the resulting fermion determinant on a single spatial lattice site, 
such that $P=\tr\, U/3$ and $P'=\tr\, U^{-1}/3$ are the only degrees of freedom. 
Here $U$ is the remaining temporal link in the temporal gauge. To approximate the  
Yang-Mills integration of the lattice model   we consider the temporal link $U$ 
surrounded by its neighbours,  
see Fig.\ \ref{fig:efmod},  and replace the contributions from the staples
 connected to $U$ by a single matrix $A$, such that 
\be 
S_{\rm YM}(U) =  -\frac{\beta}{6}\left( \tr A U + \tr A^{-1}U^{-1} \right).
\ee
For an ordered lattice $A = A^{-1} = 6 \one$, while for a disordered lattice
 $A\in$  GL(3,$\mathbb{C}$) in general. 

There are various ways to proceed \cite{Aarts:2012ft}. Here we diagonalise $U$, 
with eigenvalues $e^{iw_k}$ ($\sum_k w_k=0$, $k=1,2,3$). The group integral then
 includes the reduced Haar measure
\be
H = \sin^2 \frac{w_2 - w_3}{2} \sin^2 \frac{w_3 - w_1}{2} \sin^2 \frac{w_1 - w_2}{2}.
\ee
The complete one-link action to consider now takes the form
\be
\label{eq:Ssu}
S =  -\beta \sum_k \left( e^{\alpha_k+iw_k} +  e^{-\alpha_k-iw_k} \right) 
-\ln\det M -\ln H,
\ee
where the diagonal elements of $A$ are represented by the ${\alpha_k}$'s (where
we took out a factor of $6$).
The Langevin drift is determined by $K= - \nabla S$ and complex Langevin dynamics can be implemented for all
 three $w_k$'s or after eliminating the constraint $\sum_k w_k=0$ \cite{Aarts:2012ft}. 
Zeroes in the Haar measure also lead to poles in the drift, but these generally do not
 lead to problems and, in fact, stabilise the dynamics. This has been discussed 
 in Ref.\ \cite{Aarts:2012ft}. 
More details concerning the distribution of zeroes of $\det M$ are given
in App.\ \ref{app-zeroes}.

As observables we consider
\be
O_n = \tr\left(U^n\right) = \sum_k \e^{in w_k}.
\ee
Exact results are obtained by numerically integrating over the angles $w_k$. 
When the action is real, $\bra O_{-n}\ket=\bra O_n\ket$.

In order to determine reasonable parameter values, relevant for HDQCD, we write 
the fermion determinant as
\be
 \det M = D^{2N_f} {\tilde D}^{2N_f},
\ee
 where
 \bea
 \label{eq:D}
 D &=& 1+3C P +3 C^2 P' +C^3 = \left(1+C^3\right)\left(1+a P + b P'\right), \\
 \label{eq:Dtilde}
 \tilde D &=& 1+3\tilde C P' +3 \tilde C^2 P =  
 \left(1+\tilde C^3\right)\left(1+\tilde a P' + \tilde b P\right),
\eea
with
\be
a= \frac{3 C}{1+C^3},\qquad
b= Ca, \qquad 
{\tilde a}= \frac{3 {\tilde C}}{1+{\tilde C}^3},  \qquad
{\tilde b}= {\tilde C}{\tilde a}.
\ee
Notice that $a, b$ have maxima at $C=2^{-1/3}$ and $2^{1/3}$, respectively,
with the same value $2^{2/3}$ independently on $C$. While the behaviour of the model 
does not depend on how $C, {\tilde C}$ are parametrised, the interpretation in terms 
of physical lattice parameters does.

\begin{figure}[t]
\begin{center}
\epsfig{file=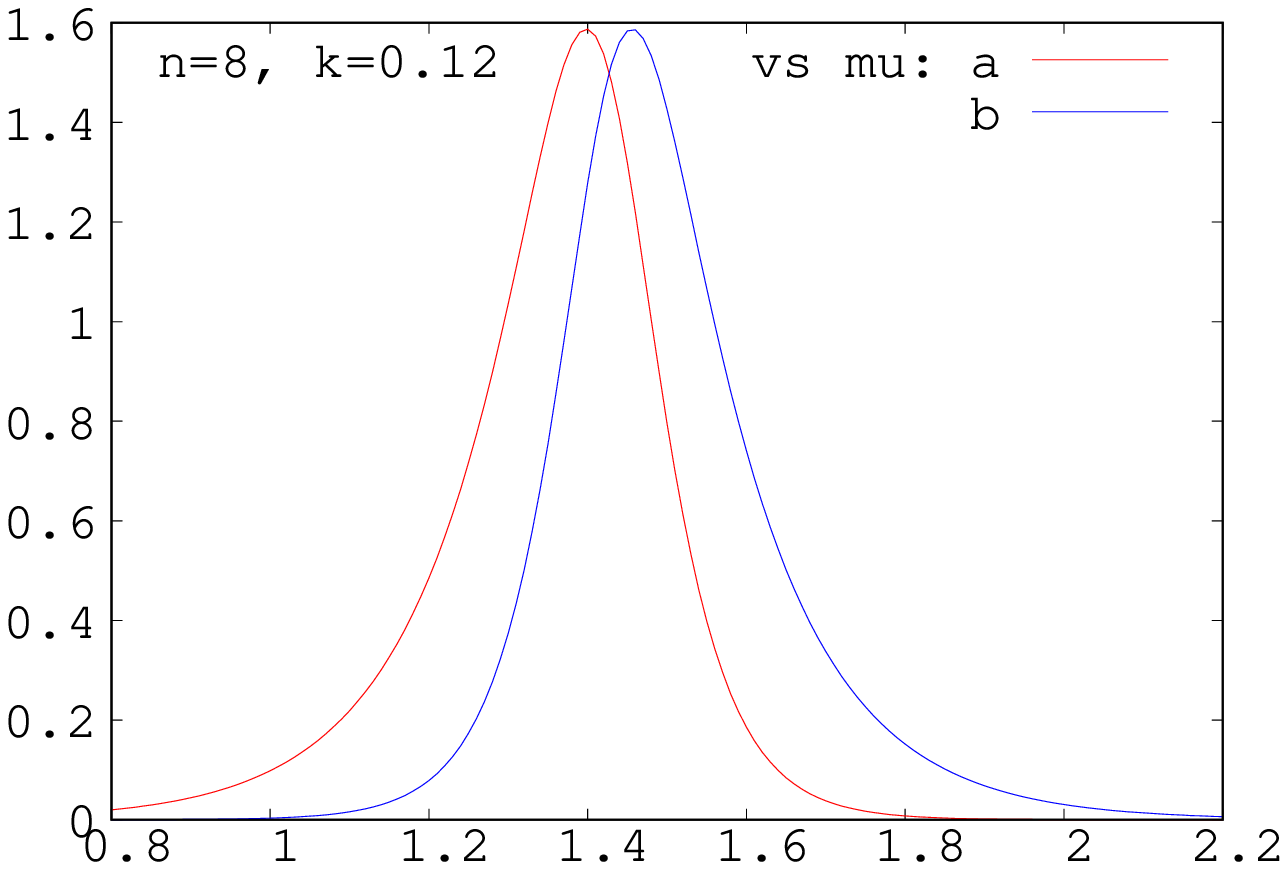, width=0.48\textwidth}
\epsfig{file=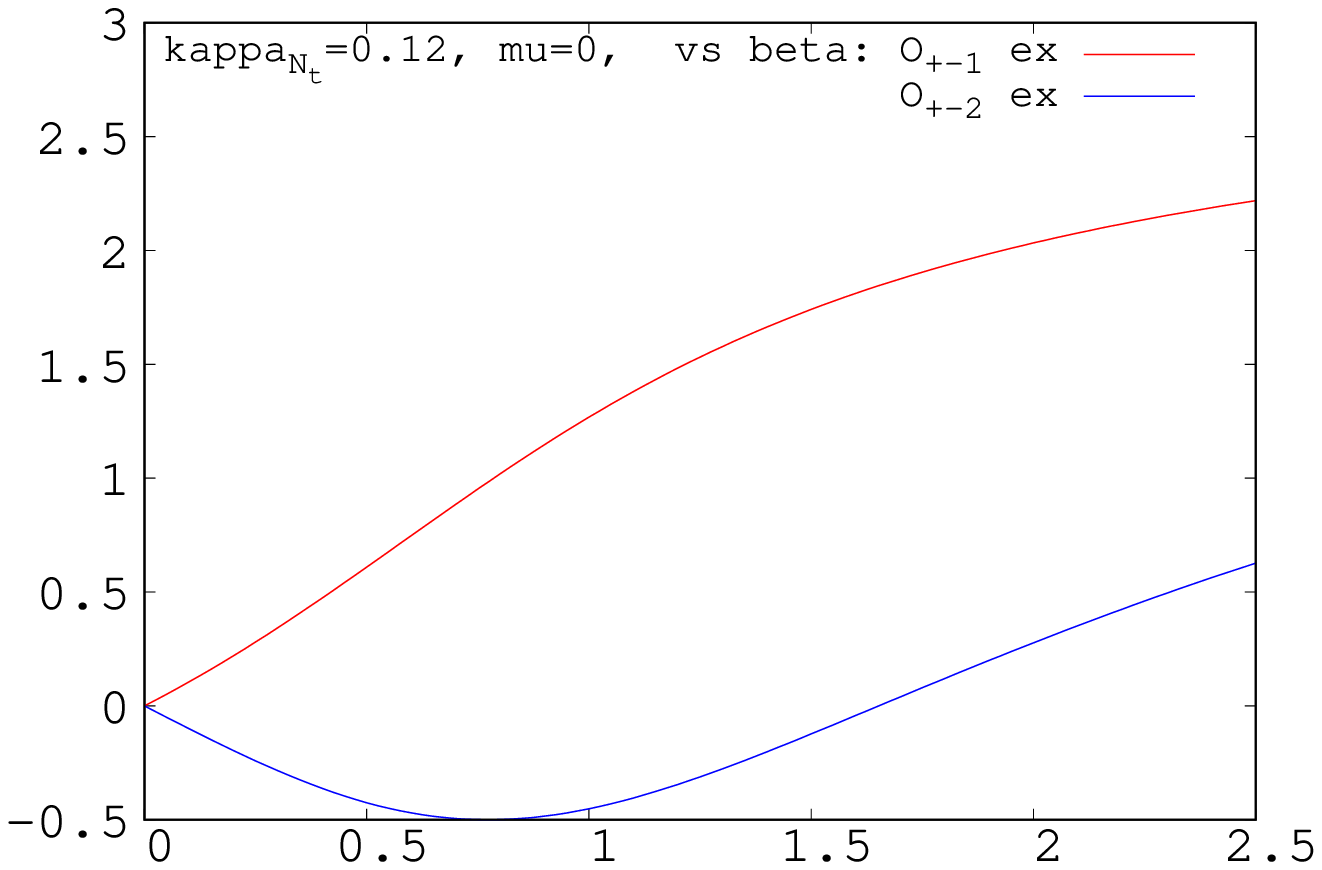, width=0.48\textwidth}
 \caption{ 
Left: Coefficients $a, b$ vs $\mu$ for $N_{\tau}=8$, $\kappa=0.12$.
Right: Observables $\bra O_{\pm 1}\ket , \bra O_{\pm 2}\ket$ vs $\beta$ at $N_{\tau}=8, 
\kappa=0.12, \mu=0$.
 }
\label{fig:ab}
\end{center}
\end{figure}

From Eq.\ (\ref{eq:C}) it follows that the interesting values of $\mu$ are 
around 
\be
\label{eq:muc0}
\mu_c^0 = -\ln(2\kappa),
\ee
the critical chemical potential for onset at zero temperature, 
i.e.\ the chemical potential at which the density changes from zero to nonzero 
\cite{Aarts:2016qrv}. This is illustrated in Fig.\ \ref{fig:ab} (left), where 
the $\mu$ dependence of $a$ and $b$ is shown for given $N_\tau$ and $\kappa$. 
With increasing $\kappa$, $\mu_c^0$ decreases and the peaks shift to the left, 
while with increasing $N_\tau$ the peaks become narrower. We also note that 
the (anti-quark) contribution $\tilde D$ becomes increasingly irrelevant as 
$N_\tau$ increases, as $\tilde C$ becomes exponentially small. Hence we will
usually neglect $\tilde D$.
In the following we use $N_{\tau}=8$, unless stated otherwise, and $N_f=1$.
Since in the one-link model there is no transition as $\beta$ is varied at $\mu=0$, 
see Fig.\ \ref{fig:ab} (right), 
we choose to work at $\beta = 0.25$, but we have also studied larger $\beta$ values. 
We considered two  $\kappa$ values
\be
\kappa=0.120, \quad \mu_c^0 = 1.427,
\qquad\mbox{and} \qquad
\kappa=0.145, \quad \mu_c^0 = 1.238,
\ee
where $\mu_c^0$ is the corresponding critical $\mu$ value (\ref{eq:muc0}),
corresponding to $C=1$.
The sign problem is (nearly) absent exactly at onset, where $C=1$,  $a=b=3/2$, 
and $D$ in Eq.\ (\ref{eq:D}) is real ($\tilde D$ is exponentially close to 1). This 
will explain some of the results below and has been noted before \cite{Seiler-unp,
Rindlisbacher:2015pea}. The behaviour for the two $\kappa$ values is rather
similar therefore we shall only show the results for $\kappa=0.120$.

Finally, to study the effect of the neighbouring links, represented by $A$, we consider
 two cases:
\begin{enumerate}
\item ordered lattice: $\alpha_k =0$;
\item (strongly) disordered lattice: $\{\alpha_k\} = (0.2+1.5i, -0.2+3.1i,  0.2-0.7i)$.
\end{enumerate}

\subsection{Ordered lattice}

 \begin{figure}[t]
\begin{center}
\epsfig{file=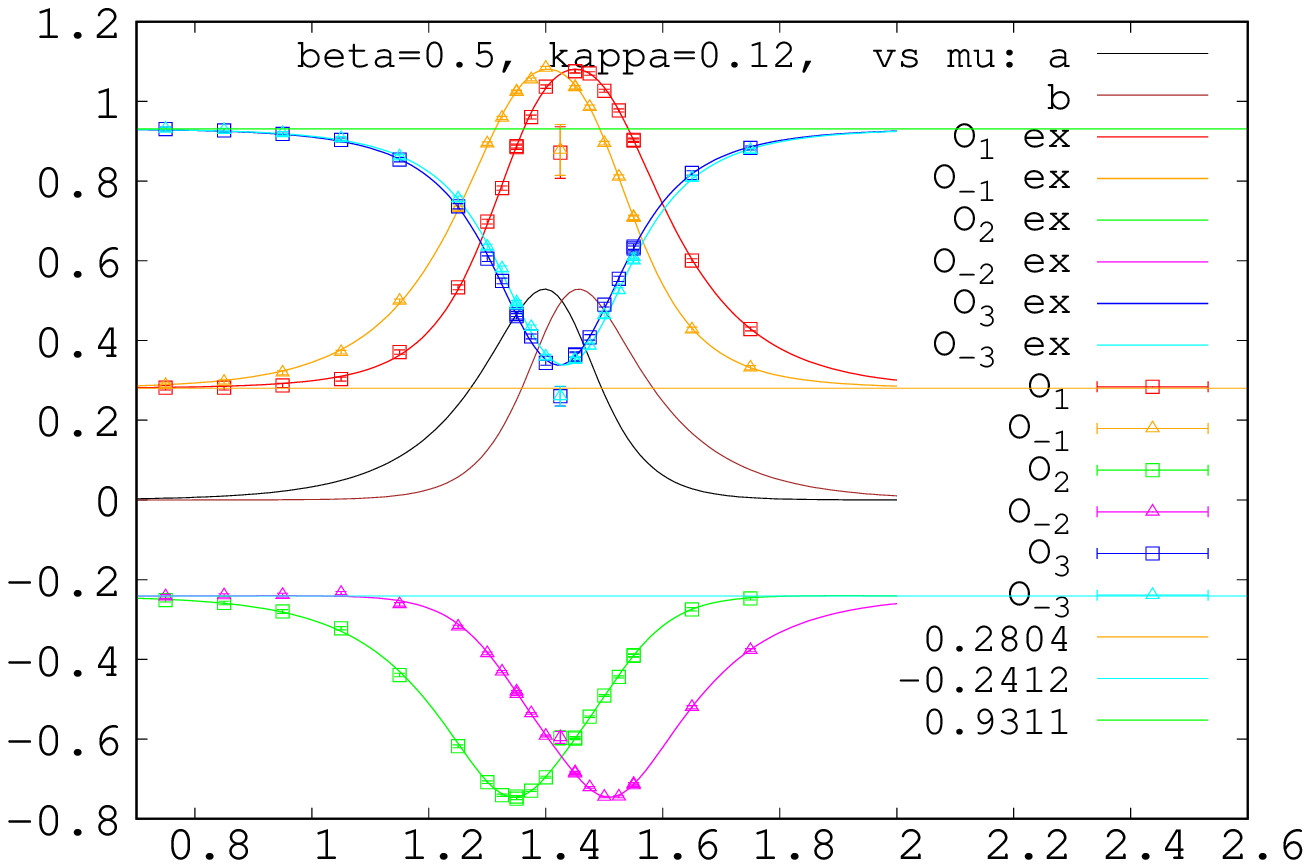, width=0.48\textwidth}
\epsfig{file=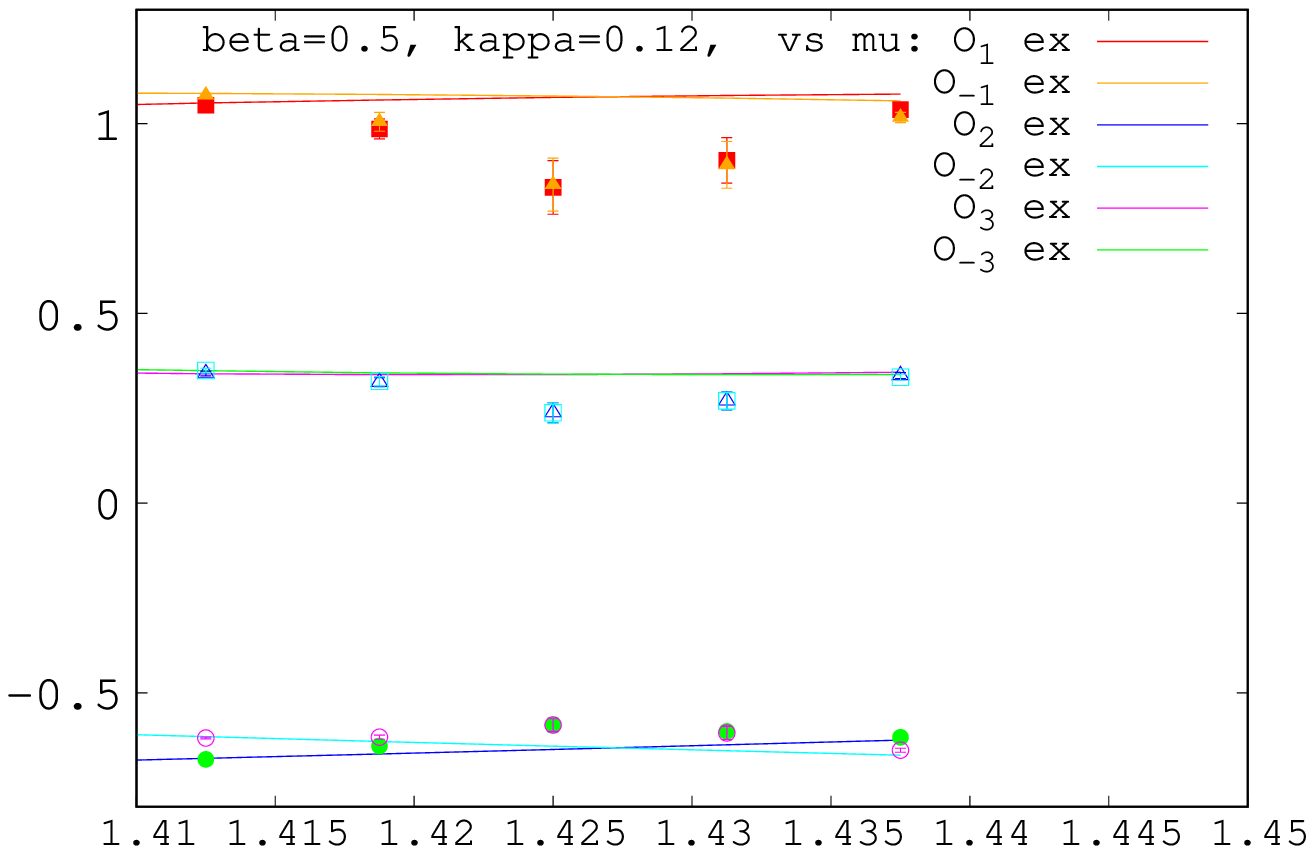, width=0.48\textwidth}
 \caption{
 Observables $\bra O_{\pm n}\ket$ ($n=1,2,3$) vs $\mu$ for $\beta=0.25$, 
 $N_{\tau}=8$ and  $\kappa=0.12$ for the ordered lattice, short runs. 
 Exact results are given by the lines. The figure on the right  
 shows a blow-up around $\mu_c^0= 1.425$.
  }
\label{fig:fsmu_E2u3}
\end{center}
\end{figure}

We first consider the ordered lattice ($\alpha_k =0$).
Fig.\ \ref{fig:fsmu_E2u3} contains results for the observables 
$\bra O_{\pm n}\ket$ ($n=1,2,3$), averaged over 100 trajectories,
 using random starting points.
The runs are relatively short: the total Langevin time is around 130, 
with 20\% thermalisation. Note that $\bra O_{+n}\ket$ and $\bra O_{-n}\ket$
 are typically
 rather close together.
We see very good agreement, except  around $\mu \simeq \mu_c^0= 1.425$. 
The same behaviour is found at the larger $\kappa=0.145$. We hence focus on 
three $\mu$ values: $\mu=$1.375 (below onset, CL fine), 1.425 (close to onset, 
CL problematic), 1.475 (above onset, CL fine).

\begin{figure}
\begin{center}
\epsfig{file=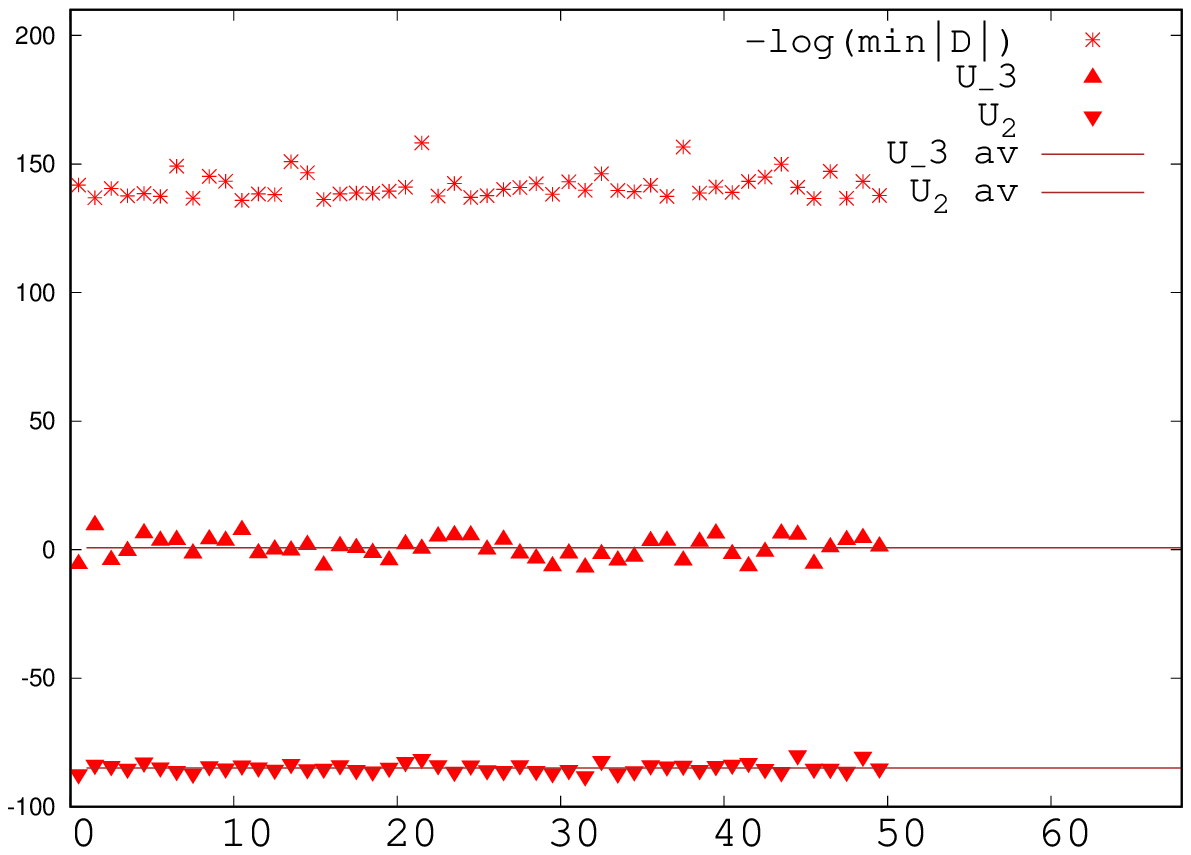, width=0.48\textwidth}
\epsfig{file=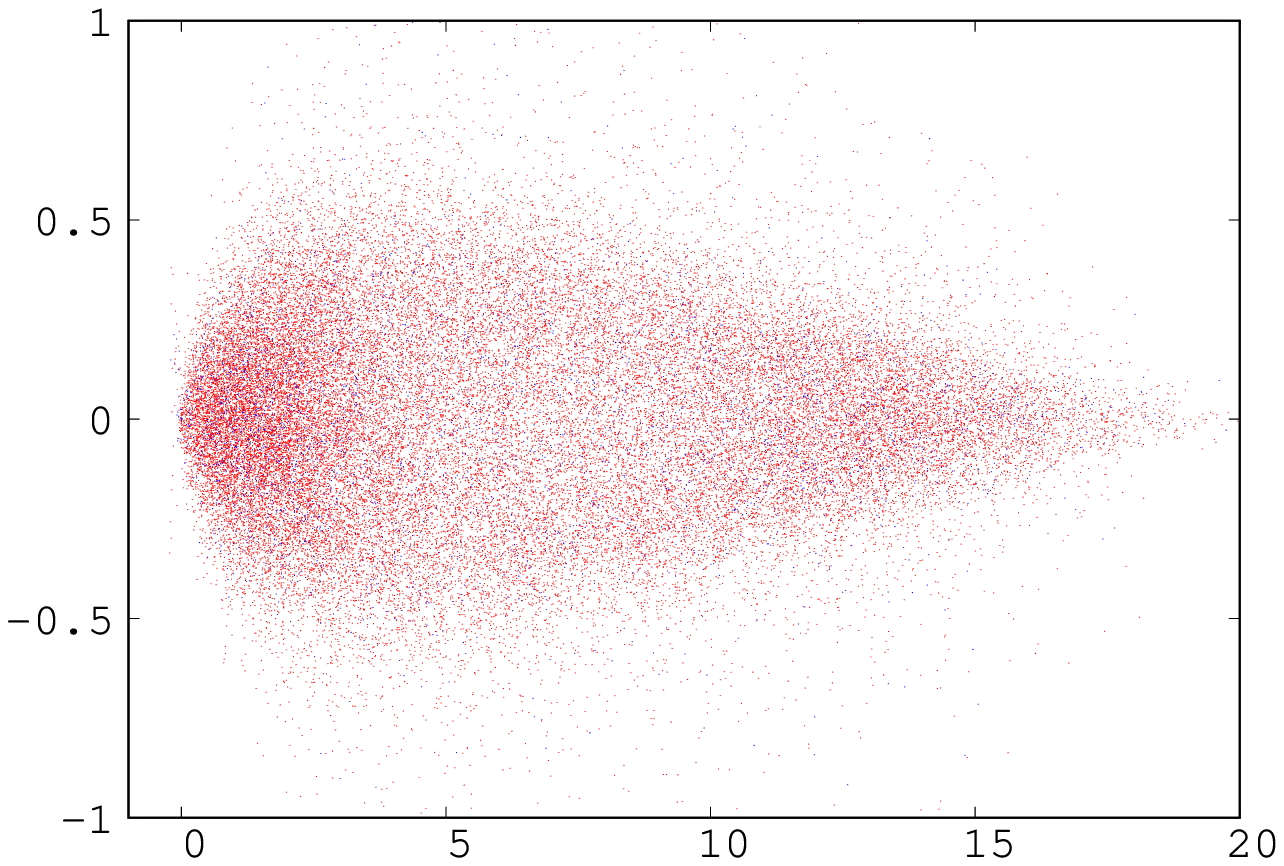, width=0.48\textwidth}
\epsfig{file=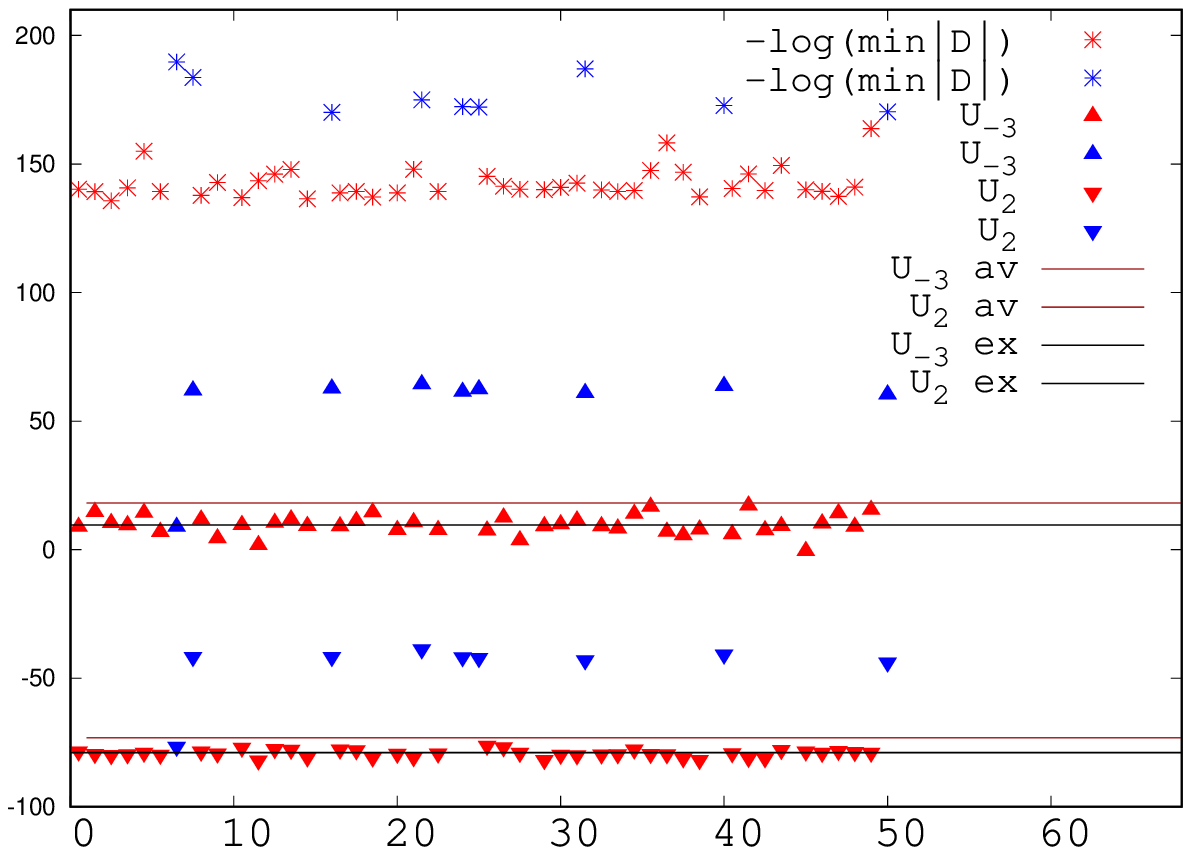, width=0.48\textwidth}
\epsfig{file=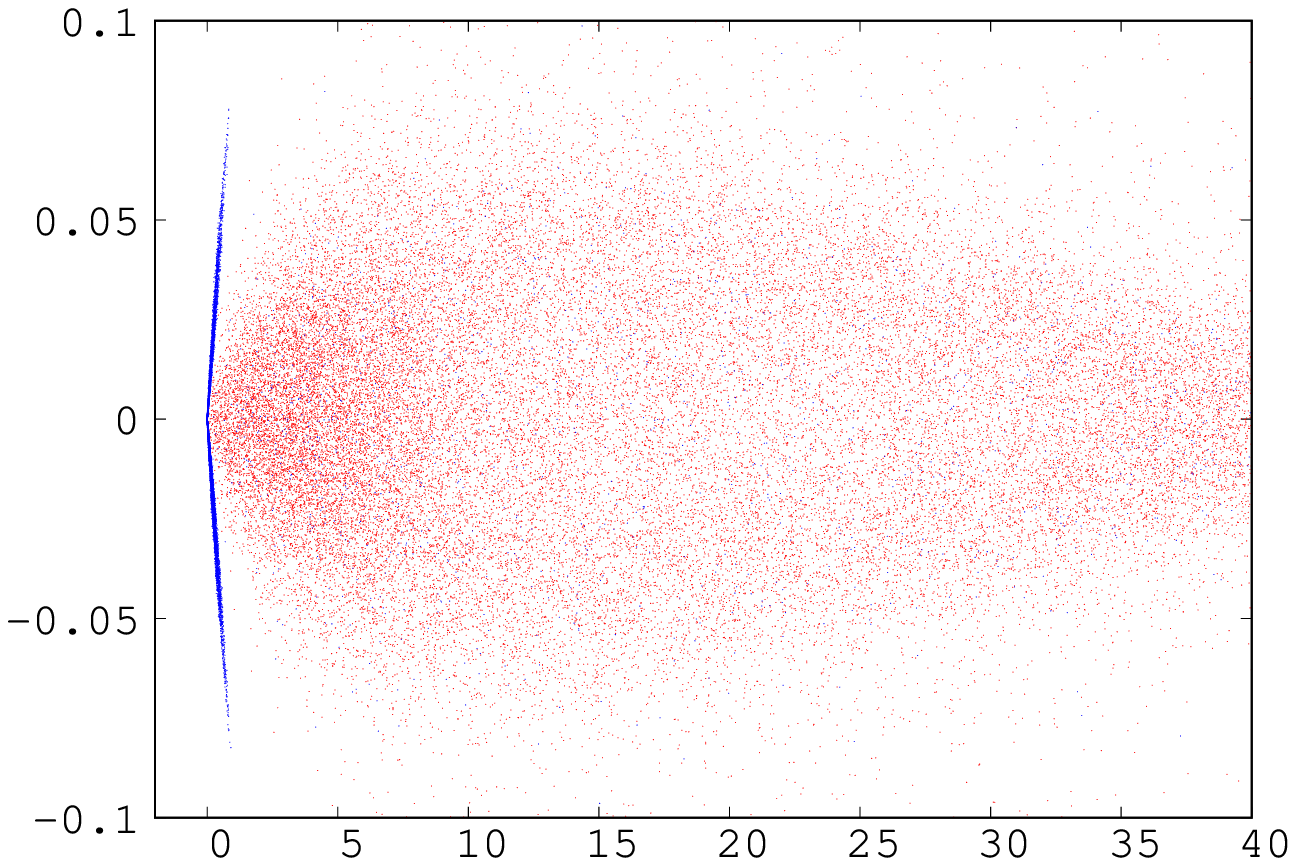, width=0.48\textwidth}
\epsfig{file=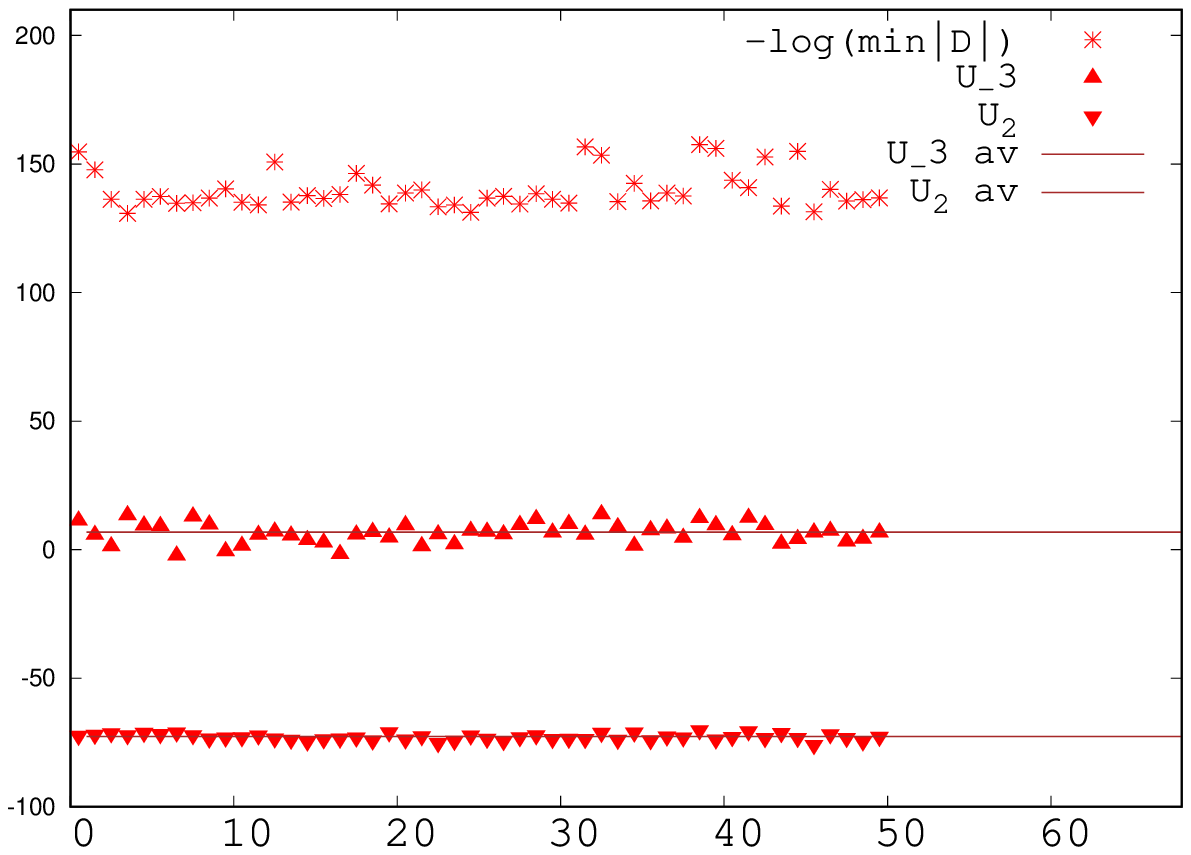, width=0.48\textwidth}
\epsfig{file=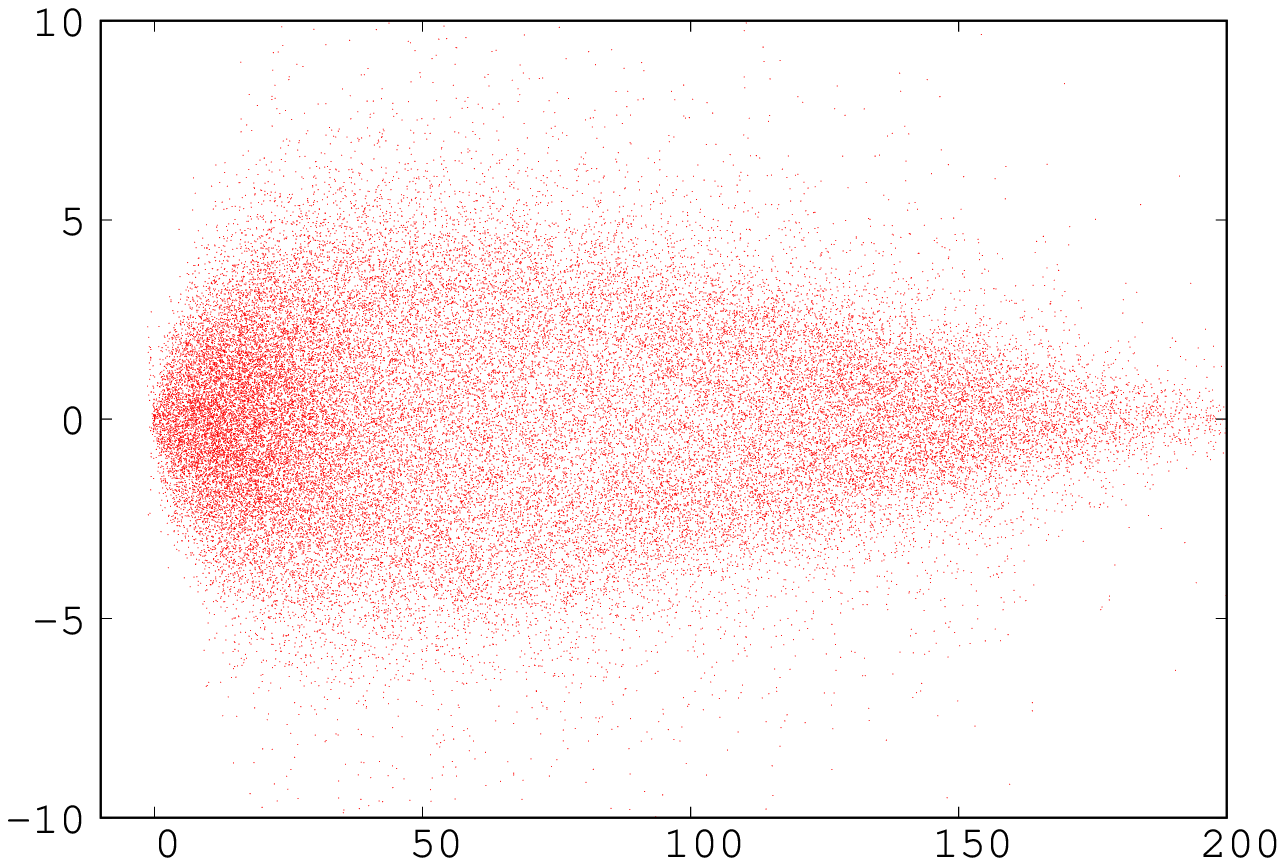, width=0.48\textwidth}
 \caption{Results at $\beta=0.25, \kappa=0.12$ and 
  $\mu=1.375$ (top),  $\mu=1.425$ (middle), $\mu=1.475$ (bottom),
   using short runs, with Langevin time $t \lesssim 130$. 
 Left: Correlation between $d_{\rm min}=\min |\det M |$ (upper data points) 
 and trajectory averages of observables $\bra O_n\ket$ for 50 trajectories 
 (numbers are shifted for clarity).  Right: Scatter plots for $\det M$. 
 The contributions from trajectories with 
 $d_{\rm min}>10^{-6}$ ($d_{\rm min}<10^{-6}$) are shown in red (blue). 
 See text for further details. 
  }
\label{fig:fcor_C13}
\end{center}
\end{figure}

In Fig.\ \ref{fig:fcor_C13} we show results for each of those $\mu$ values, 
using 50 independent, relatively short, trajectories.
 The figures on the left show the observables against trajectory index.  When CL is fine, all trajectories fluctuate around 
the exact result.
However, when CL is problematic (middle figure), the trajectories appear to split in 
two groups, indicated by the red and blue symbols. We identify those using the minimal 
absolute value of the determinant on the trajectory, 
\be
d_{\rm min} = \min_{\rm trajectory} |\det M|.
\ee
Trajectories with $d_{\rm min} > d_c \simeq 10^{-5} - 10^{-8}$ always 
appear to 
lead to correct results, 
while the trajectories having $d_{\rm min}<d_c$ lead to a wrong result 
(for definiteness we  take $d_c =10^{-6}$ in the following).

\begin{figure}[t]
\begin{center}
\epsfig{file=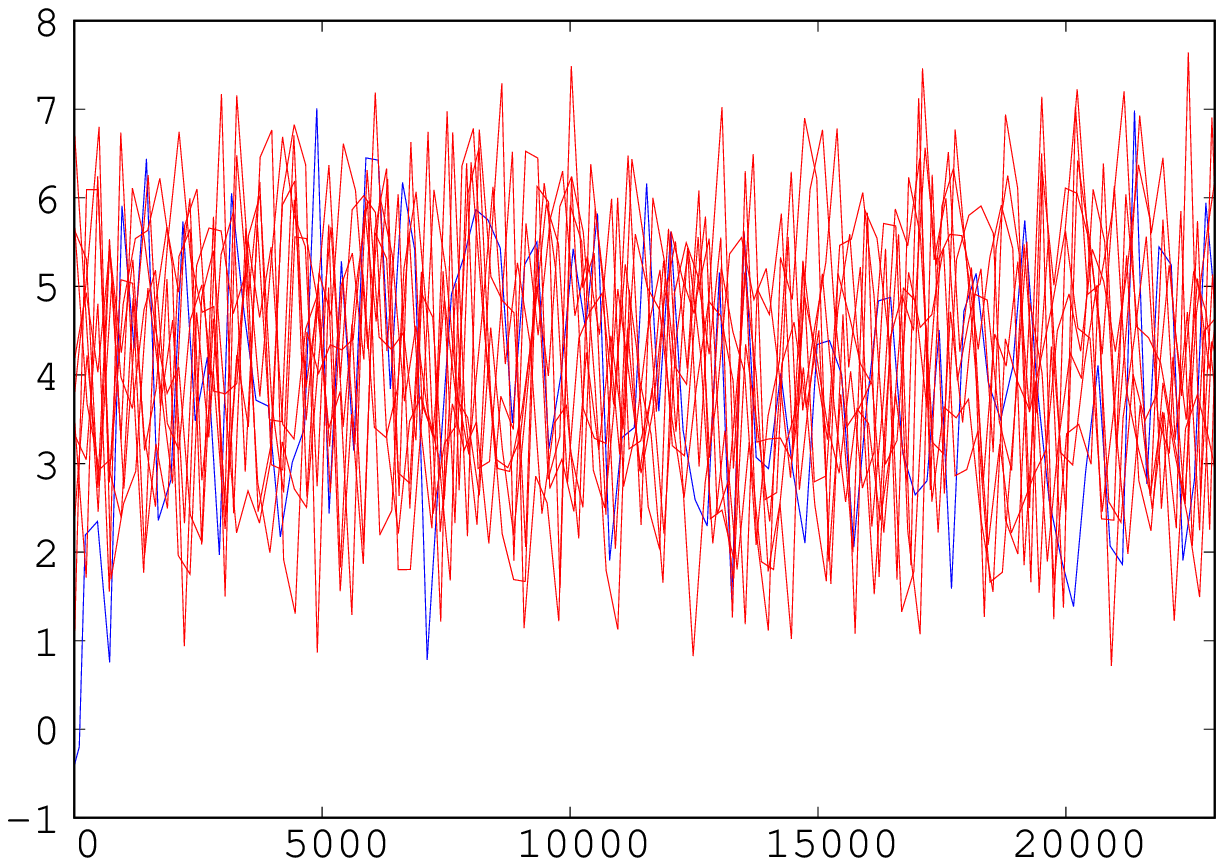, width=0.48\textwidth}\\
\epsfig{file=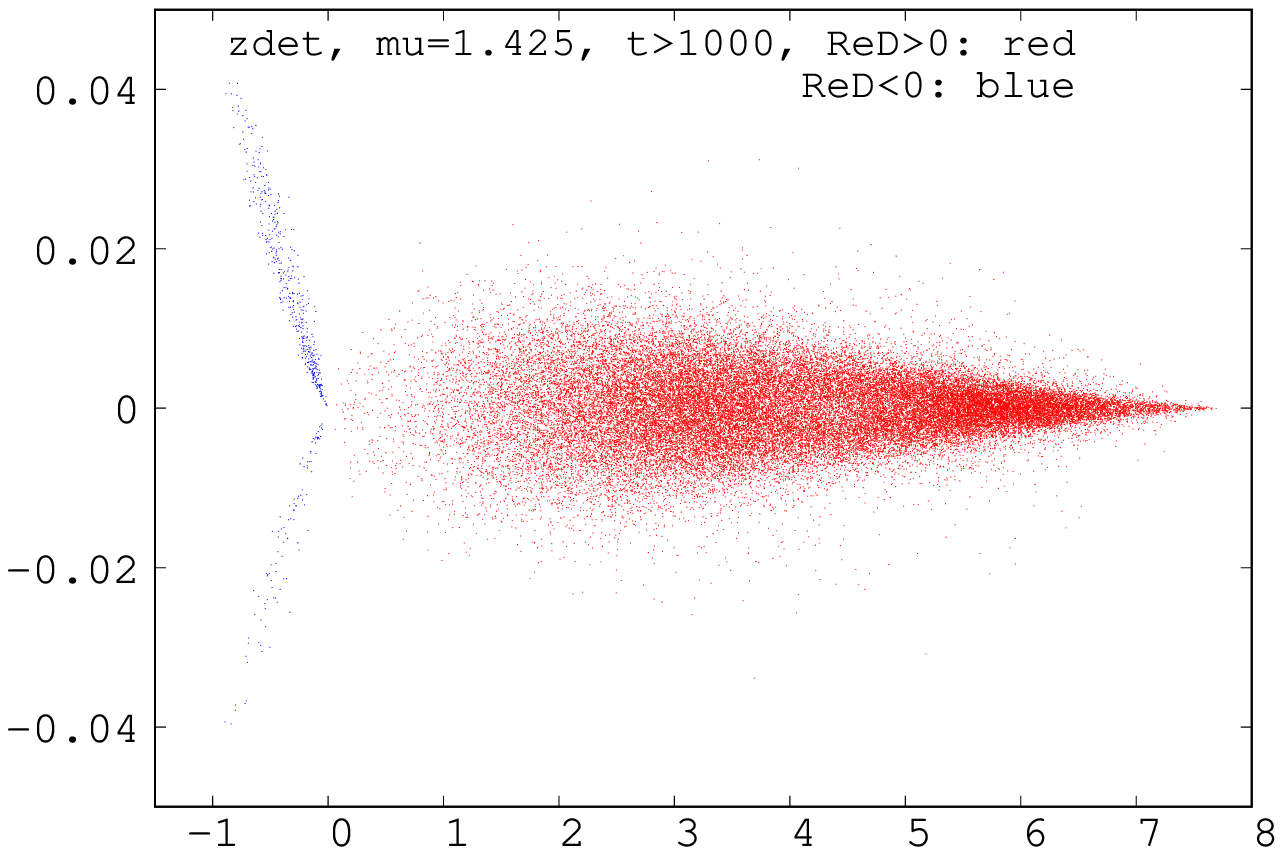, width=0.48\textwidth}
\epsfig{file=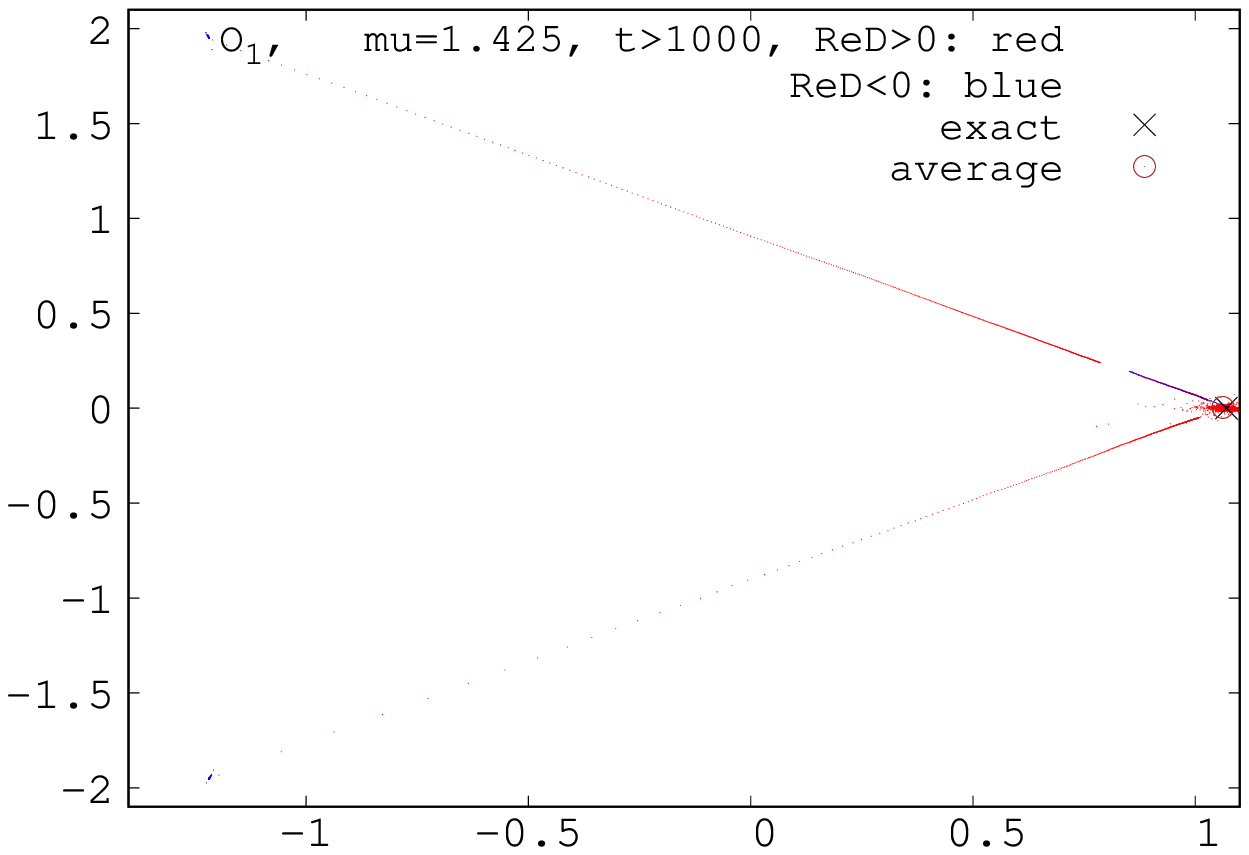, width=0.48\textwidth}
\caption{
Top: Histories of 10 long trajectories at $\mu=1.425$ 
for the ordered case, $\Re D$ vs Langevin time, no thermalization. Trajectories depicted in blue {\em started} in the 
$\Re D<0$ ``blue whiskers''  region (they 
 typically soon switch to the  $\Re D>0$ ``red fish'' region).
Bottom left: Scatter plot for the unsquared determinant 
$D\tilde D \simeq D$ using 50 long trajectories 
with Langevin time 5600. Blue points in the ``red fish'' region come from 
trajectories which started in 
 the ``blue whiskers'' and switched to the former (diluted in the figure). 
 Right: Scatter plot of the observable $O_1$. Here red (blue) points correspond 
 to configurations with $\Re D>0$ ($\Re D<0$).  Other parameters as above. 
}
\label{f.ftr_NAL_2}
\end{center}
\end{figure}

To investigate these two types of trajectories further, we show in 
Fig.\ \ref{fig:fcor_C13} (right) the corresponding scatter plots for the determinant.
When CL works well, the points from all trajectories appear similarly distributed,
 even when $d_{\rm min} $ gets very small.
 At the middle $\mu$ value of  $\mu=1.425$ a different picture appears: the
  trajectories of the first group 
 ($d_{\rm min} > d_c$) give a similar picture as at the lower and higher $\mu$ values 
  (the ``red fish''), 
while the second group  ($d_{\rm min} < d_c$)  yields a very peculiar structure (the 
``blue whiskers'').
The appearance of two essentially disjoint contributions in the determinant is 
very similar to what was observed in the U(1) one-link model. A red/blue code for
identifying the disjoint (``regular''/``deviant'') contributions is used in 
Figs.\  \ref{fig:fcor_C13}, \ref{f.ftr_NAL_2}, \ref{f.trjsc1}, and explained in the captions.

In the scatter plot we showed results for the determinant $\det M = (D\tilde D)^2$ 
which enters in the determination of the drift.  
More information, however, is provided by the unsquared factors 
$D {\tilde D}\simeq D$. 
In Fig.\ \ref{f.ftr_NAL_2} (bottom left)  we show the scatter plot of 
the unsquared factors $D {\tilde D}\simeq D$ at  $\mu = 1.425$.
The ``blue whiskers'' have $\Re D<0$ and come from trajectories which
 approach the pole (zero of the determinant) with $d_{\rm min}<d_c$. 
 As in the simple models, a bottleneck separates them from the region with  $\Re D > 0$. 
 Moreover,  the contributions have a very small weight.
 Depending on the starting configuration, trajectories may run
 for a while in the region with negative Re $D$, before switching to the positive side.
 The contributions from the ``whiskers'' therefore
 practically fades out after enough thermalization: already after
  $t \simeq 1000$ all configurations 
 appear in
the region $\Re D > 0$, see Fig.\ \ref{f.ftr_NAL_2} (top). 
 Some of the results are summarised in Table~\ref{table3}.
 
 Similar as in Sec.\ \ref{sec:what}, we defined here partition functions and weights restricted to subsectors, namely
\be
Z_\pm = \int DU\, \theta(\pm \Re\det M) \rho(U), \qquad\quad w_\pm=\frac{Z_\pm}{Z_++Z_-},
\ee
where $\rho(U)$ is the original complex distribution.
Table 3 also contains an estimate of how much time is spent in the region with $\Re D<0$, which is denoted with $p_-$.
It should be noted that $p_-$ depends on the details of transient behaviour and crossings, and is hence not immediately related to $w_\pm = Z_\pm/Z$.
 

\begin{table}[t]
\begin{center} 
\begin{tabular}{l|r|r|r|r|c|l}
1.375 	& $\bra\tr\,U\ket$ & $\bra\tr\,U^{-1}\ket$ & $\bra\tr\,U^2\ket$ 
			& $\bra\tr\,U^{-2}\ket$ & $t(10^3)$ & $w_-$ ($p_-$)	\\ 
\hline
CL 		& 0.972& 1.064& $-$0.729& $-$0.537& .1+.6	&	\\ 
CLD   	& 0.972& 1.065& $-$0.729& $-$0.537& .1+.6 	&	\\ 
CL$+$   	& 0.972& 1.064& $-$0.730& $-$0.538& 1.+5.6 	&	\\ 
\hline
CL  		& 0.970& 1.063& $-$0.730& $-$0.538& 2.+22.	&$6.\times 10^{-4}$\\ 
CLD    	& 0.970& 1.063& $-$0.730& $-$0.538& 2.+22. 	&	\\ 
CL$+$    	& 0.972& 1.063& $-$0.731& $-$0.538& 2.+22. 	&	\\ 
\hline
exact 	& 0.972& 1.065& $-$0.730& $-$0.537&		&	\\  
ex.$+$ 	& 0.972& 1.065& $-$0.730& $-$0.537&	 	&	\\  
ex.$-$ 	&$-$1.906&$-$0.575& 0.038& $-$0.724&	 	&$-1.09\times 10^{-4}$\\ 
ex.pq	& 1.003& 1.003& $-$0.628& $-$0.628&	 	&	\\ 
\hline 
 &&&&&&\\
1.425 	& $\bra\tr\,U\ket$ & $\bra\tr\,U^{-1}\ket$ & $\bra\tr\,U^2\ket$ 
			& $\bra\tr\,U^{-2}\ket$ & $t(10^3)$ & $w_-$ ($p_-$)	\\ 
\hline
CL 		& 0.885& 0.892& $-$0.597& $-$0.595& .6+2.8	&	\\ 
CLD   	& 1.069& 1.073& $-$0.648& $-$0.640& .6+2.8 	&	\\ 
CL$+$   	& 1.069& 1.072& $-$0.649& $-$0.640& 1.+5.6 	&	\\ 
\hline
CL  		& 1.062& 1.066& $-$0.647& $-$0.639& 2.+22.	&$3.\times 10^{-3}$\\ 
CLD    	& 1.069& 1.073& $-$0.649& $-$0.640& 2.+22. 	&	\\ 
CL$+$    	& 1.069& 1.073& $-$0.649& $-$0.640& 2.+22. 	&	\\ 
CL$-$   	&$-$1.139&$-$1.117& $-$0.106& $-$0.181& 2.+22. 	&	\\ 
\hline
exact 	& 1.069& 1.073& $-$0.649& $-$0.640&	 	&	\\  
ex.$+$ 	& 1.069& 1.073& $-$0.645& $-$0.640&	 	&	\\  
ex.$-$ 	&$-$0.798&$-$1.606& $-$0.211&  0.070&	 	&$0.75\times10^{-4}$\\ 
ex.pq	& 1.071& 1.071& $-$0.644& $-$0.644&	 	&	\\ 
\hline 
\end{tabular}
\caption{Simulation results at $\beta=0.25, \kappa=0.12$, 
$N_\tau=8$ and $\mu=1.375,1.425$, in the ordered case,
from all trajectories (CL), 
from trajectories  with $d_{\rm min}>d_c$ (CLD), 
from all trajectories after dropping the points with $\Re D < 0$ (CL+),
using 100 or 50 trajectories with 
varying length of Langevin time $t= t_{\rm therm.}+t_{\rm meas.}$. Errors 
are not indicated but are at the permille level. Imaginary parts are zero 
within the error.
Also indicated are exact results: full, restricted to $\Re D\gtrless 0$, 
and phase quenched (pq).  
$w_-$  is the relative weight of the 
$\Rep D <0$ region in the partition function, for the simulation $p_-$
is given instead, estimated via the proportion of $\Re D <0$ points.
}
\label{table3}
\end{center}
\end{table}

\begin{figure}[t]
\begin{center}
\epsfig{file=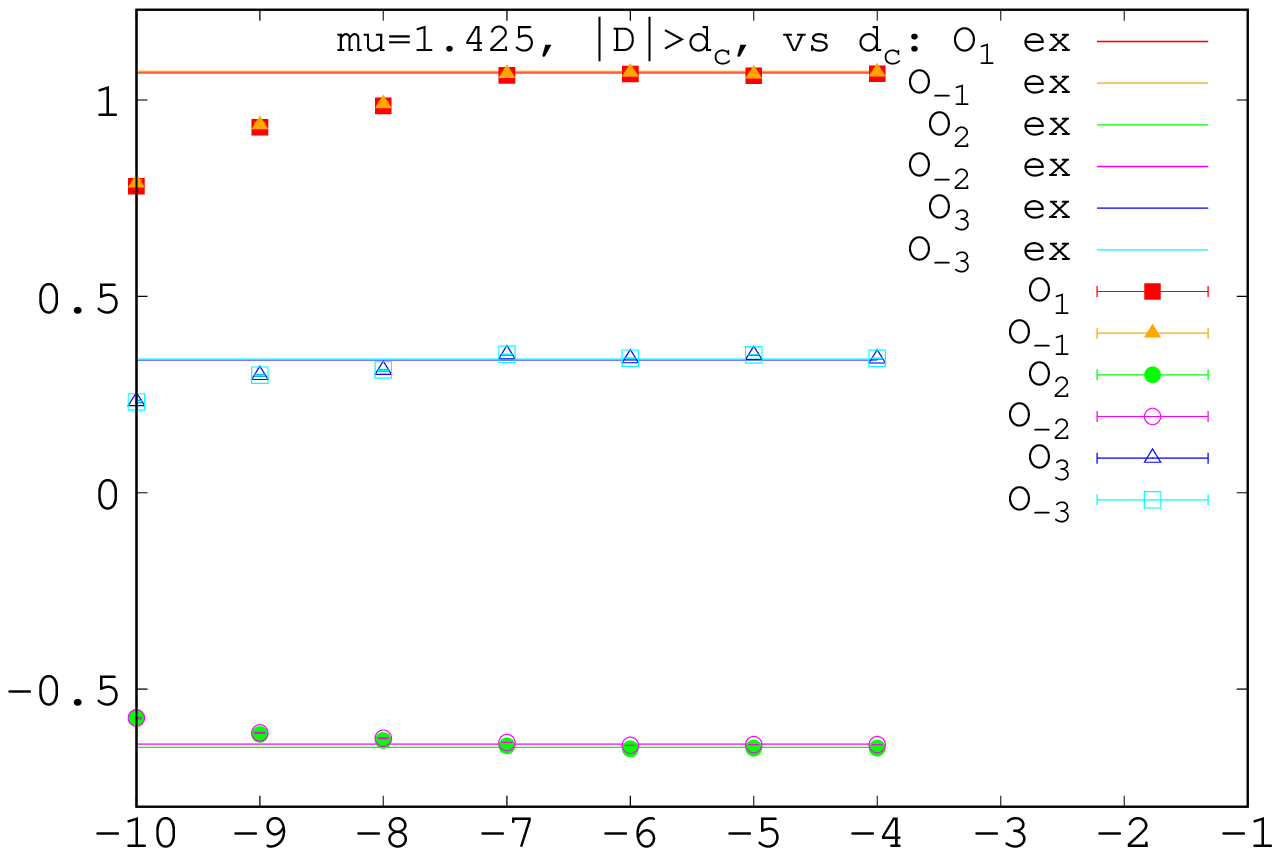, width=0.48\textwidth}
\epsfig{file=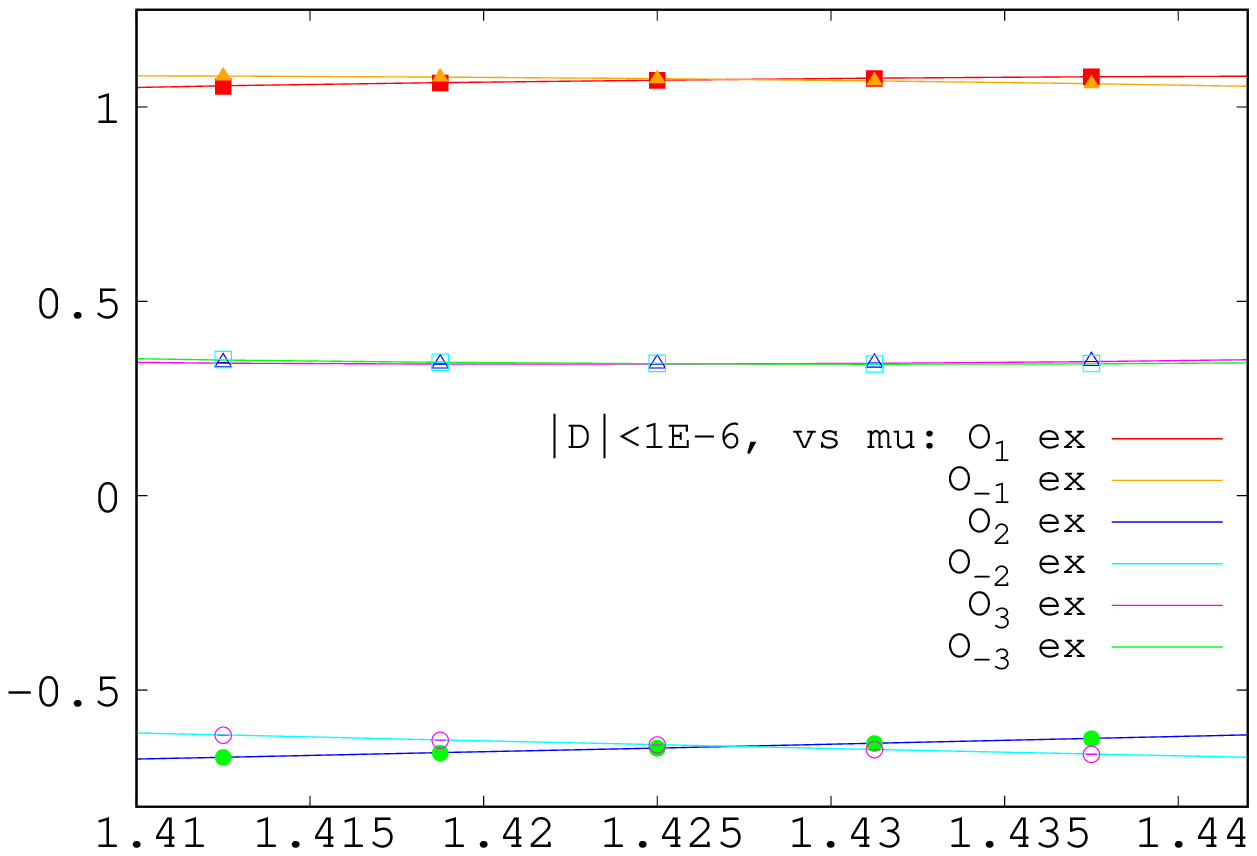, width=0.48\textwidth}
\caption{
Left: dependence of observables on the cutoff $\log(d_c)$ at $\mu=1.425$. 
Right: observables vs $\mu$ from all trajectories after dropping configurations with 
$ \Rep D < 0$, cf.\ Fig.\ \ref{fig:fsmu_E2u3} (right). Parameters as in 
Fig.\ \ref{fig:fsmu_E2u3}.
}
\label{f.cutdep}
\end{center}
\end{figure}

The exact relative weight  of the region $\Re D < 0$ is easily found and is 
$O\left(10^{-4}\right)$.
 We find that the process, once arrived in the positive region, only rarely 
 visits the negative region and typically only very briefly. Hence random starts
  with $\Re D<0$
 give that region an artificially large weight and a considerate choice 
 of the start configuration will reduce the necessity of long thermalisation times. 
  These findings suggest it might be useful to discard trajectories with
   $d_{\rm min} < d_c$ and 
   thus sample  the $\Re D >0$ region only, to ensure nearly 
   correct convergence.
We find that the value of $d_c$ does not need tuning, in the above case any value
between $10^{-5} - 10^{-8}$ is acceptable, see  Fig.\ \ref{f.cutdep} (left).  
Alternatively one can keep all trajectories but drop contributions from 
configurations with  $\Re D < 0$, as not to lose statistics. 
This is demonstrated in Fig.\ \ref{f.cutdep} (right). 
Keeping only trajectories with $d_{\rm min}>d_c$ leads to the similar results.

\begin{figure}[t]
\begin{center}
\epsfig{file=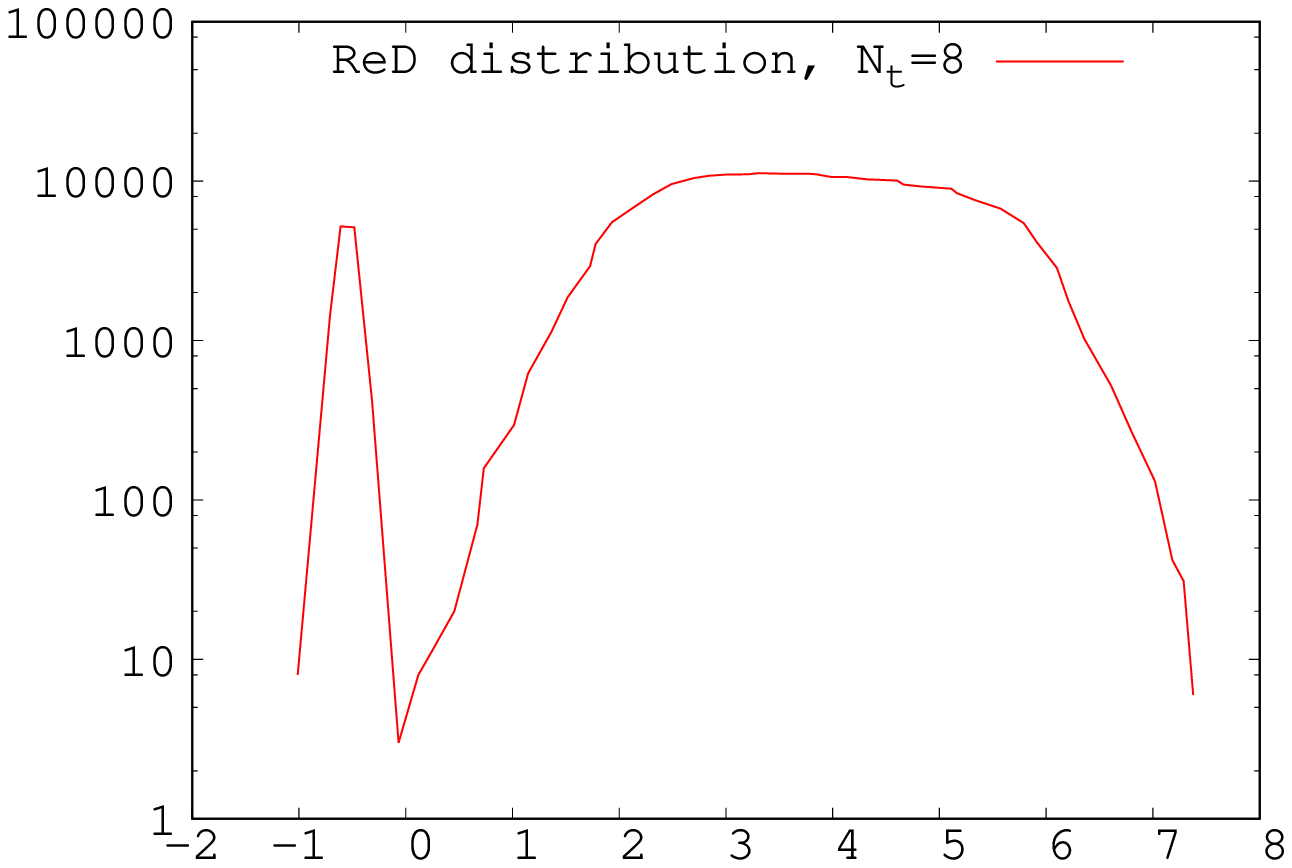, width=0.48\textwidth} 
\epsfig{file=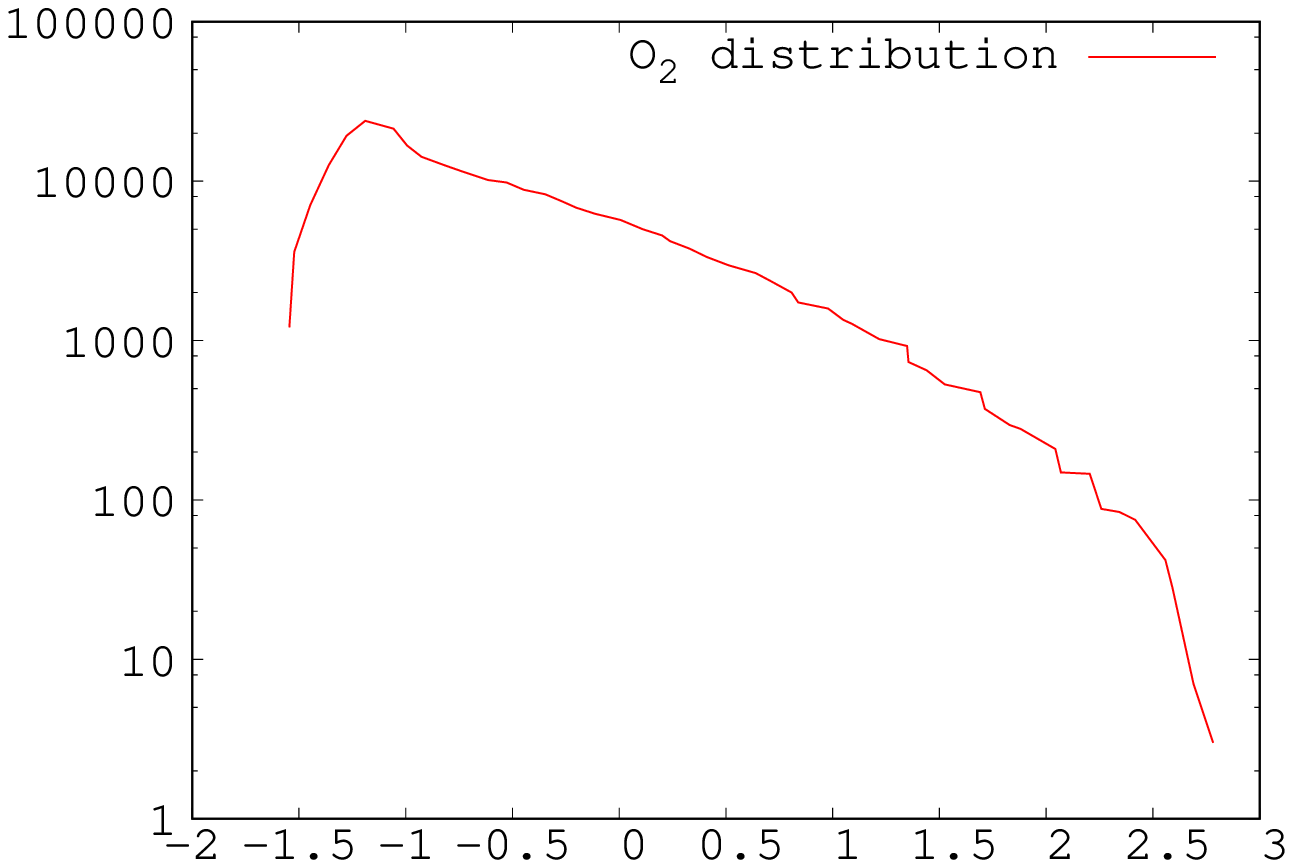, width=0.48\textwidth}
\vspace{0.1cm}
\epsfig{file=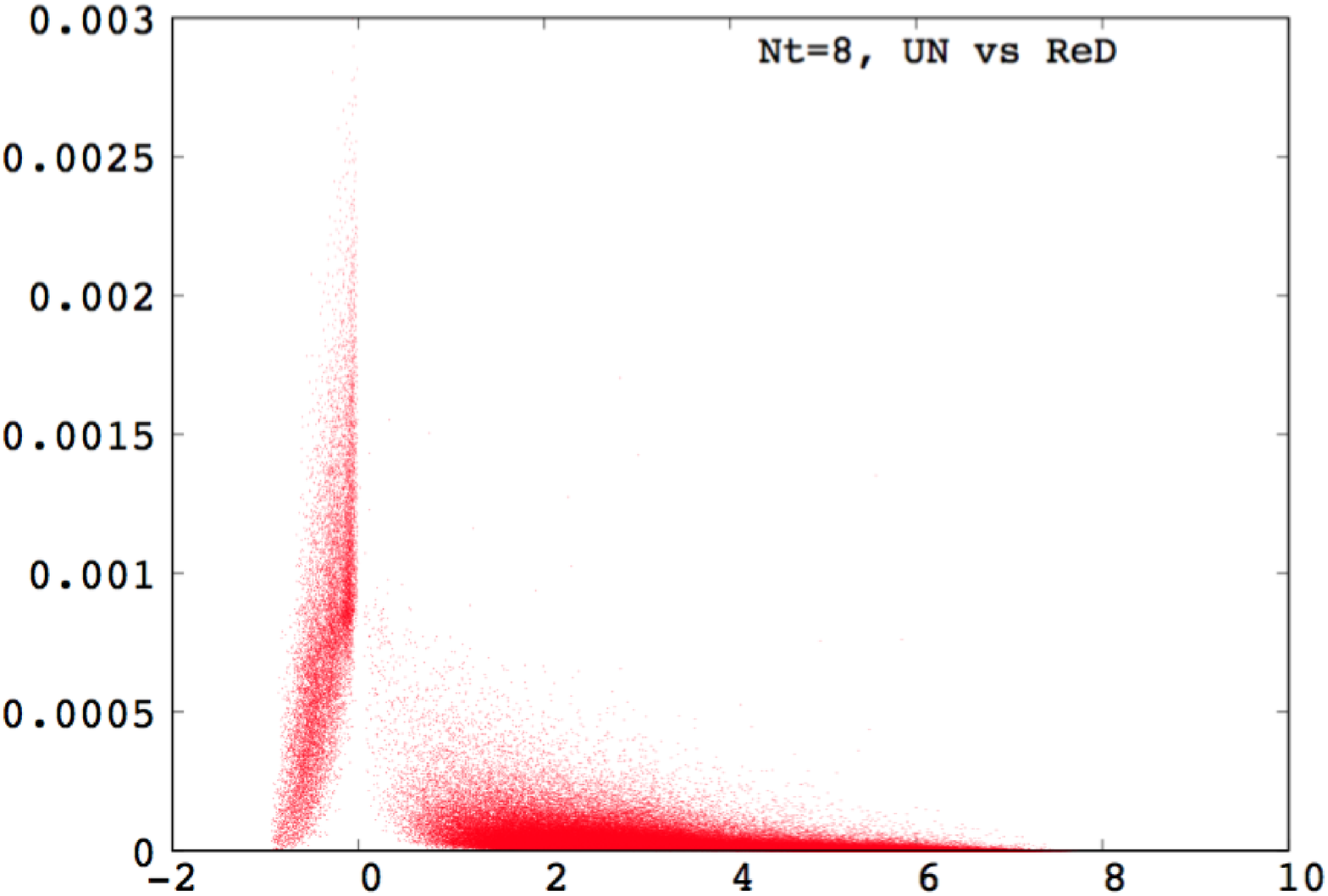, width=0.48\textwidth} 
\epsfig{file=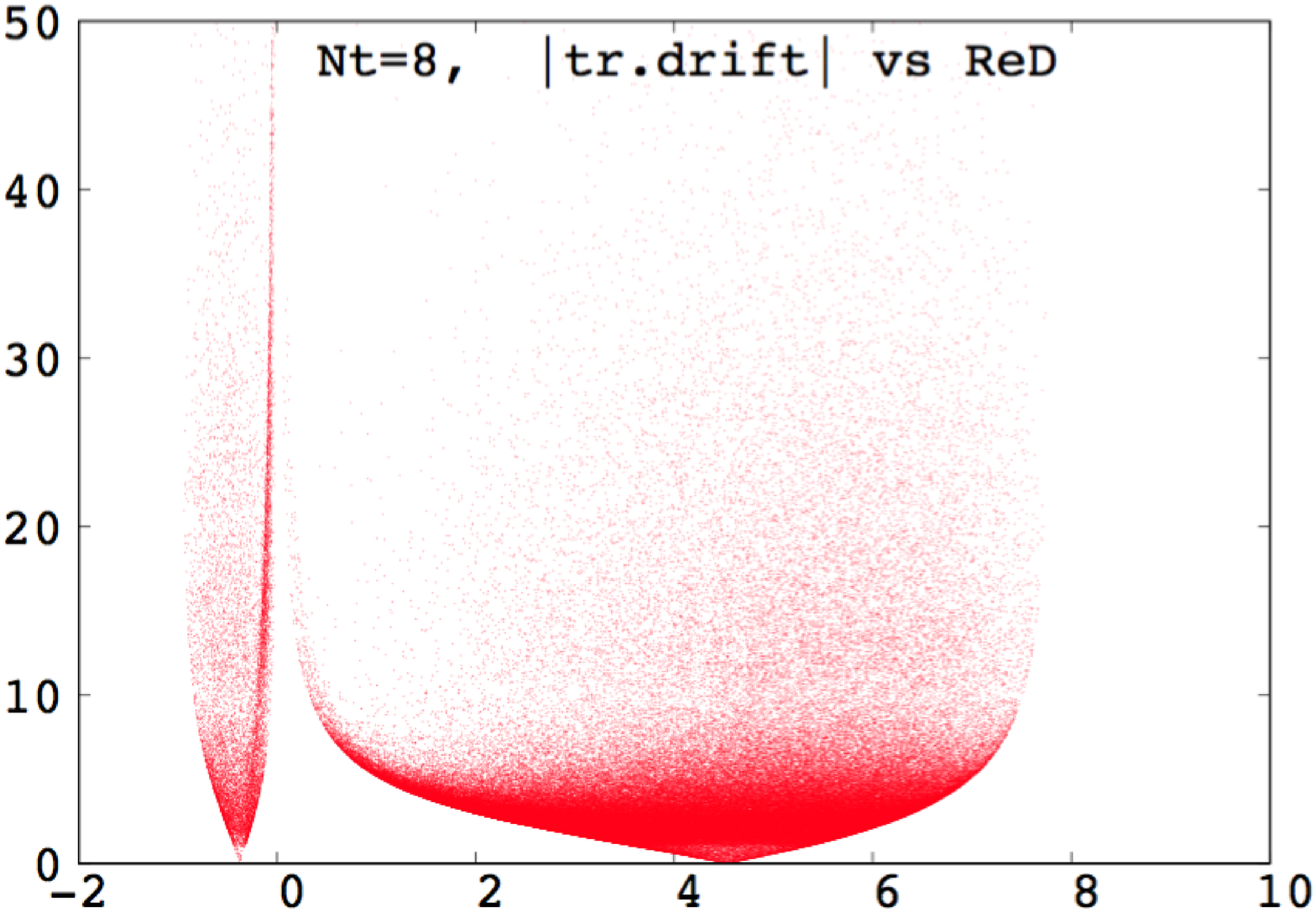, width=0.48\textwidth}
\vspace{0.1cm}
\epsfig{file=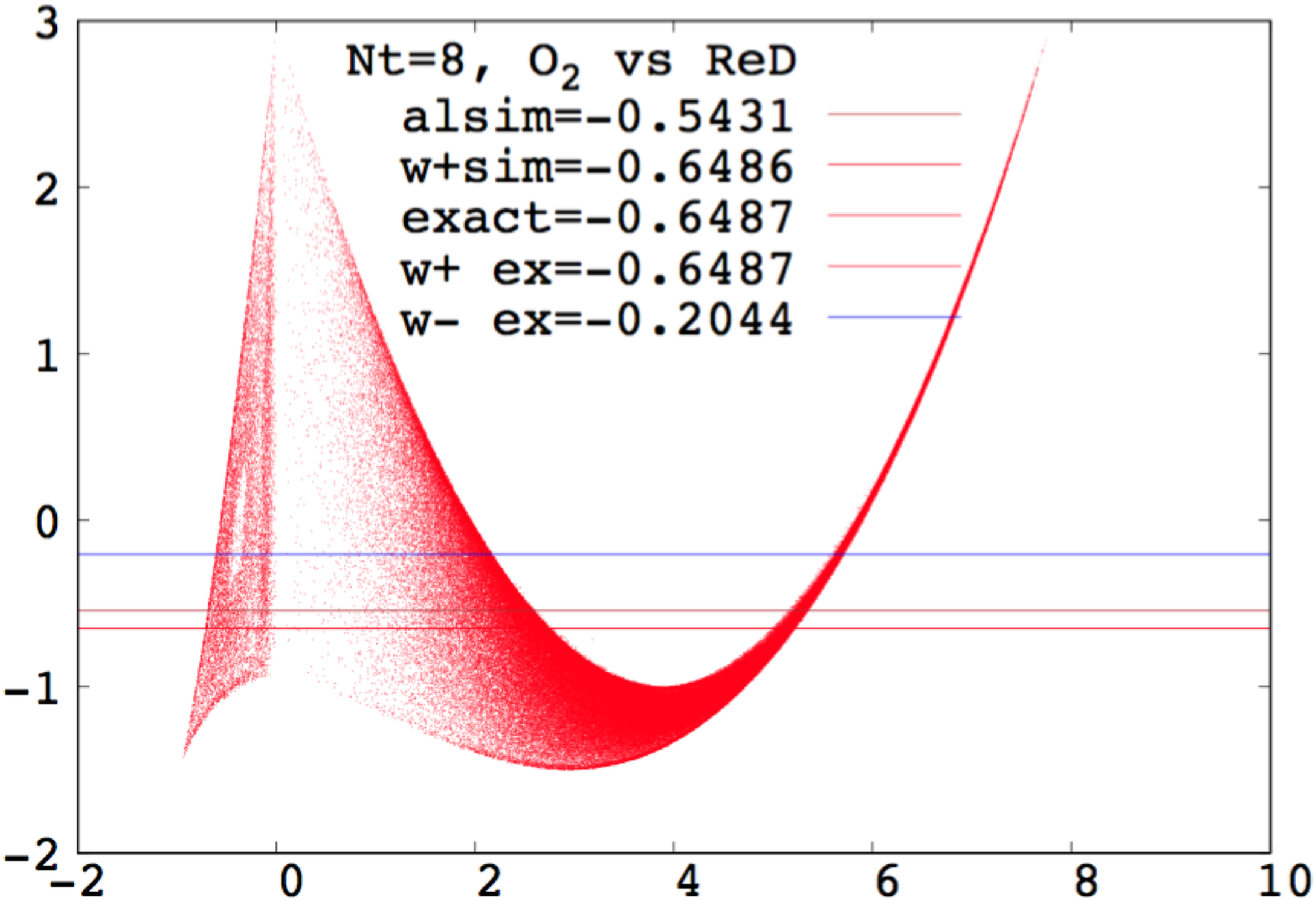, width=0.48\textwidth} 
\epsfig{file=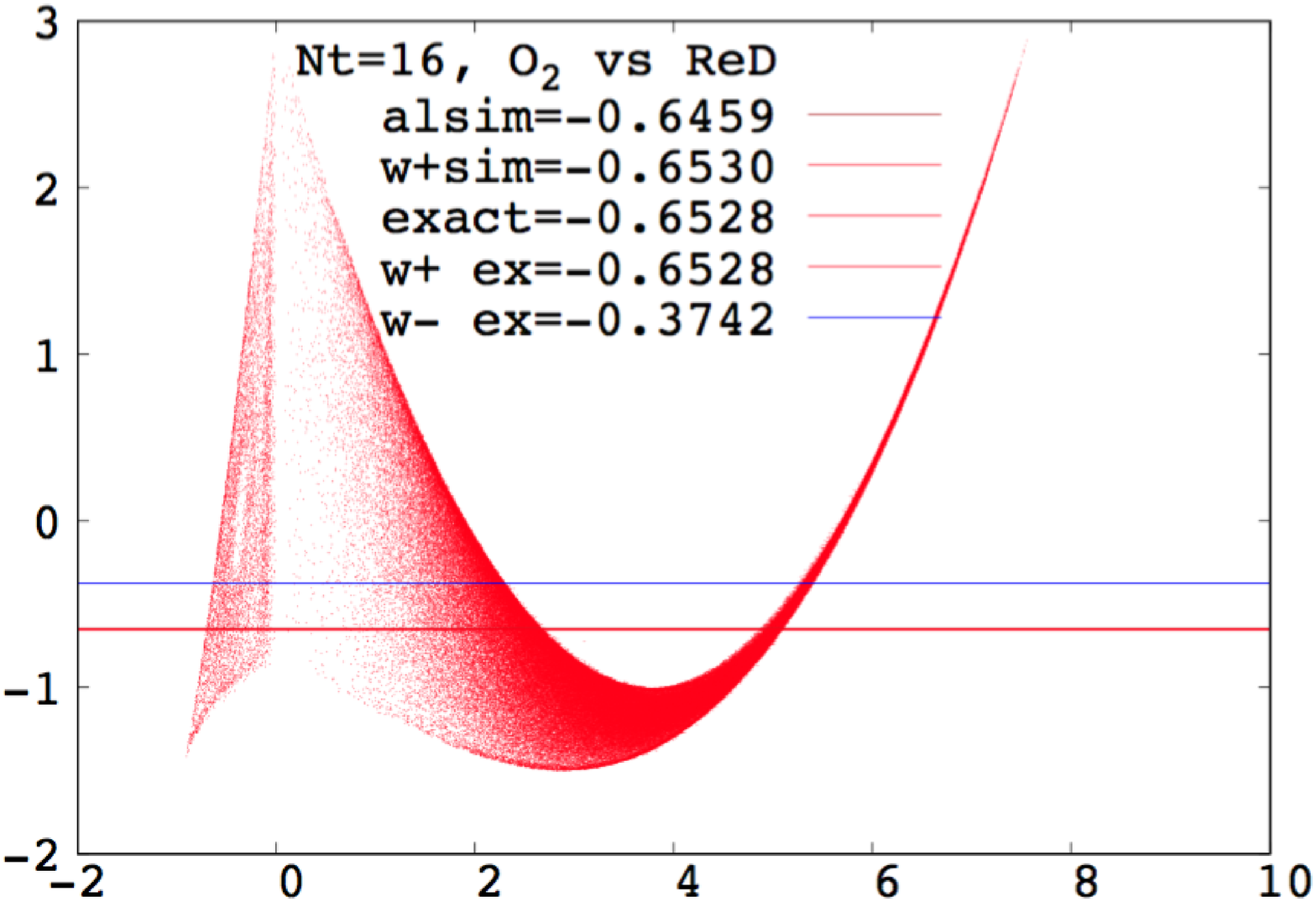, width=0.48\textwidth}
\caption{
Top: histograms of $\Re D$ (left) and $O_2=\tr\, U^2$ (right).
Middle: correlation between $\Re D$ and the unitarity norm $\tr(U^\dagger U-\id)$ (left),
 and between $\Re D$ and the 
 norm of the instantaneous drift force.
Bottom: correlation between $\Re D$ and $O_2$ for $N_t=8$ (left) and $16$ (right).
 Parameters as above, with $\mu=1.425$.
  }
\label{f.dcorr}
\end{center}
\end{figure}

As already suggested by Fig.\ \ref{fig:fcor_C13} the signal for deviant contributions 
also shows up in observables, as can be seen in Fig.\ \ref{f.ftr_NAL_2} (right).
In this long run and rather representative case only three 
trajectories out of fifty contain configurations with $\Re D<0$, which lead to 
outlying contributions in the observables. These few contributions
falsify the averages by nearly 1\%, while the average 
over the other trajectories reproduces the exact result within the error ($\lesssim$ 0.1 \%). 
The irregular signal is clear in the scatter plot of the observables, which also suggests that they are related to certain initial configurations, such that their weight will diminish in long runs.
 
The above effects become evident by plotting the correlation between
the determinant and various other quantities. In Fig.\ \ref{f.dcorr}
we show the histograms of the probability distributions of $\Re D$
and of one selected observable, $O_2 = \tr\,U^2$ (top). 
The scatter plots (middle, bottom) show a clear correlation between $\Re D$ 
and the unitarity norm $\tr(U^\dagger U-\id)$, the drift, and $O_2$.

\subsection{Disordered lattice}

\begin{figure}[t]
\begin{center}
\epsfig{file=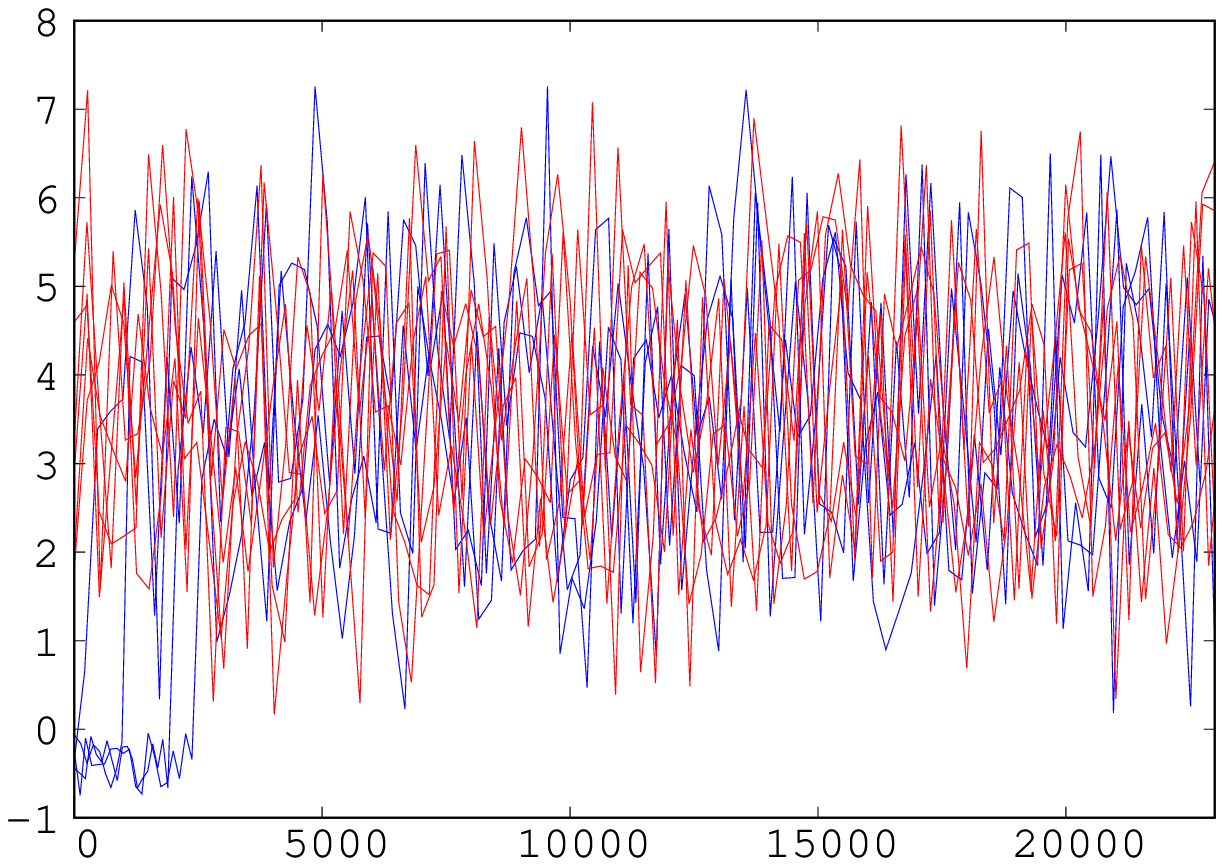, width=0.48\textwidth}\\
 \epsfig{file=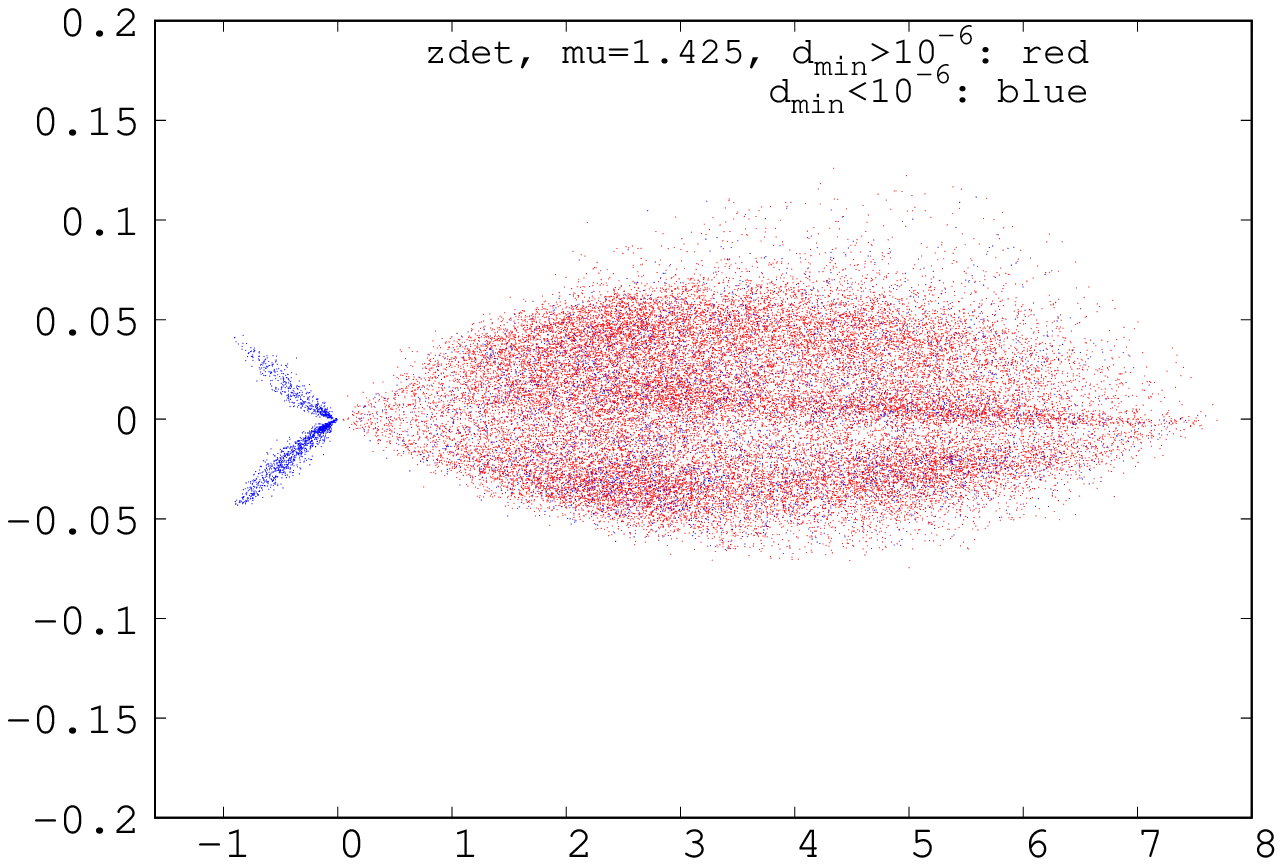, width=0.48\textwidth}
 \epsfig{file=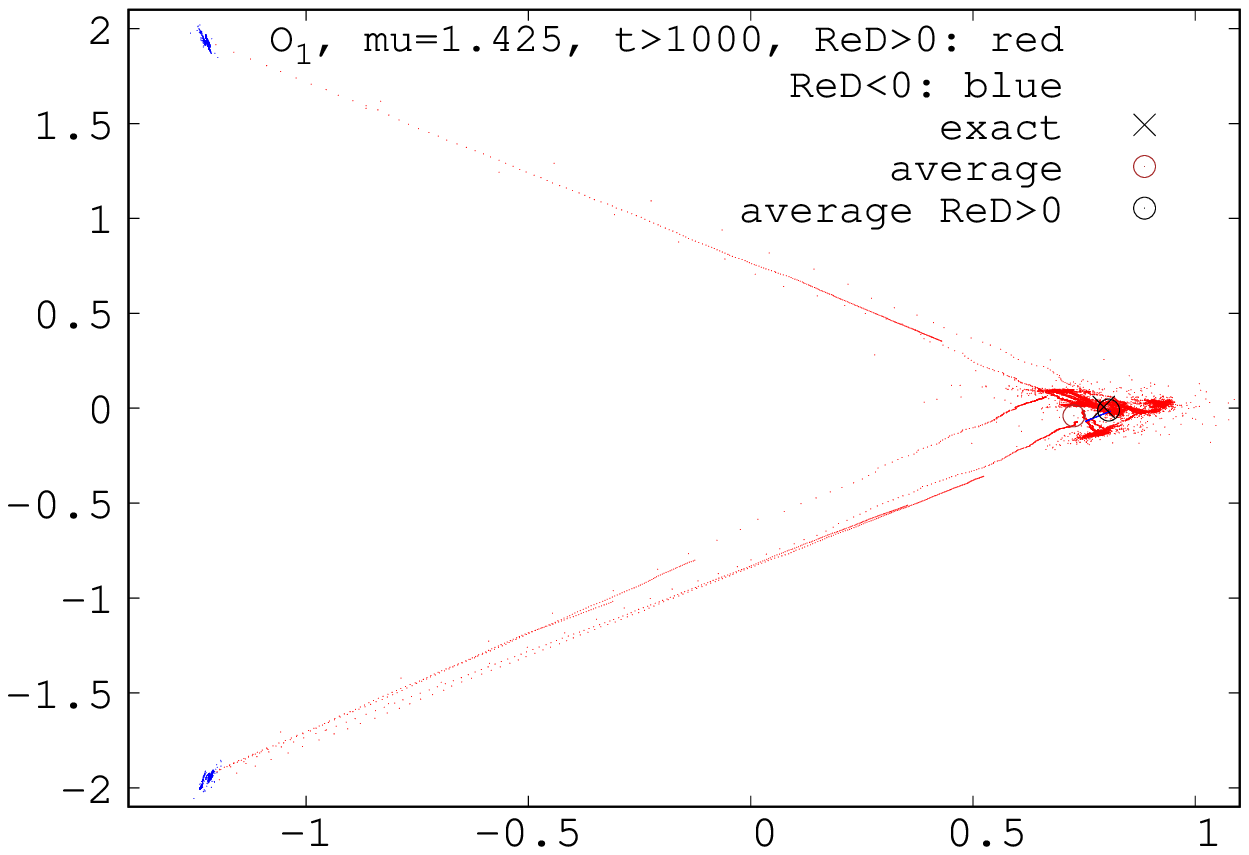, width=0.48\textwidth}
\caption{
As in Fig.\ \ref{f.ftr_NAL_2} for the strongly disordered lattice
 }
\label{f.trjsc1}
\end{center} 
\end{figure}

Next we consider a disordered lattice, i.e.\ we take into account the nontrivial 
effect of the neighbours when the lattice theory is reduced to a one-link model, 
see Eq.\  (\ref{eq:Ssu}). We take $\alpha_k \ne 0$ and choose the values given 
at the end of Sec.\ \ref{sec:subhd}.
Fig.\ \ref{f.trjsc1} demonstrates the behaviour of the unsquared determinant 
and for $O_1$, see also Table \ref{table4}, which is very similar as for the
 ordered case. 
We find that  trajectories starting with $\Re D < 0$ (``blue whiskers") need 
much more time to  switch to the region with $\Re D > 0$ (``red fish"). 
Nevertheless the weight of the former 
 is only about $0.001 - 0.05$, and discarding the contribution with $\Re D<0$ 
 decisively improves the results.  
The results, however, deteriorate with increasing lattice disorder, which may 
indicate the
 effect of large excursions in the noncompact directions, for which the adaptive 
 stepsize and gauge cooling become essential. Since this was studied in previous 
 papers, we do not analyse this problem any further here.

\begin{table}[t]
\begin{center} 
\begin{tabular}{l|r|r|r|r}
1.375	& $\bra\tr\,U\ket\qquad$ & $\bra\tr\,U^{-1}\ket\quad$ & $\bra\tr\,U^2\ket\qquad$ 
		& $\bra\tr\,U^{-2}\ket\qquad$   \\ 
\hline
CL 		& 0.666$-$0.005$i$ & 0.829+0.021$i$ & $-$0.779$-$0.037$i$ & $-$0.490+0.032$i$  \\ 
CLD   	& 0.653$-$0.025$i$ & 0.817+0.050$i$ & $-$0.774$-$0.105$i$ & $-$0.489+0.098$i$  \\ 
CL$+$  	& 0.670$-$0.005$i$ & 0.832+0.021$i$ & $-$0.780$-$0.002$i$ & $-$0.490+0.031$i$  \\ 
\hline
exact 	& 0.660+0.011$i$ & 0.823+0.014$i$ & $-$0.774$-$0.010$i$ &$-$0.488$-$0.004$i$ 	  \\ 
ex.$+$ 	& 0.659+0.011$i$ & 0.823+0.014$i$ & $-$0.774$-$0.010$i$ &$-$0.489$-$0.004$i$ 	  \\ 
ex.$-$ 	&$-$2.170+0.294$i$ &$-$0.309+0.298$i$ &  0.184+0.159$i$ &$-$0.848$-$0.140$i$ 	 \\ 
ex.pq	& 0.704+0.007$i$ & 0.717+0.015$i$ & $-$0.625$-$0.011$i$  &$-$0.605$-$0.020$i$ 	  \\ 
\hline 
 &&&&\\
1.425	& $\bra\tr\,U\ket\qquad$ & $\bra\tr\,U^{-1}\ket\quad$ & $\bra\tr\,U^2\ket\qquad$ 
		& $\bra\tr \,U^{-2}\ket\qquad$ \\ 
\hline
CL 		& 0.727$-$0.039$i$ & 0.748+0.044$i$ & $-$0.666$-$0.020$i$ & $-$0.626+0.019$i$  \\ 
CLD   	& 0.805$-$0.010$i$ & 0.825+0.016$i$ & $-$0.693$-$0.034$i$ & $-$0.650+0.071$i$  \\ 
CL$+$  	& 0.808$-$0.020$i$ & 0.827+0.024$i$ & $-$0.692$-$0.060$i$ & $-$0.653+0.059$i$  \\ 
CL$-$    	& $-$2.5 +1.1$i$    & $-$2.5$-$1.1$i$  & 0.4	+1.2$i$	& 0.2$-$1.8$i$     \\ 
\hline
exact 	& 0.794+0.007$i$ &  0.809+0.012$i$ & $-$0.684$-$0.007$i$ & $-$0.654$-$0.001$i$   \\ 
ex.$+$ 	& 0.795+0.007$i$ &  0.810+0.012$i$ & $-$0.684$-$0.007$i$ & $-$0.654+0.001$i$   \\  
ex.$-$ 	&$-$1.021$-$0.219$i$ & $-$1.385+0.219$i$ & $-$0.111+0.097$i$ & $-$0.026$-$0.101$i$ \\  
ex.pq	& 0.797+0.007$i$ & 0.806+0.012$i$ & $-$0.678$-$0.007$i$ & $-$0.660+0.001$i$ 	  \\ 
\hline 
\end{tabular}
\caption{As in the previous Table, for the disordered case. The Langevin time is 
$t \simeq (1.+5.6)\times 10^3$. 
For $\mu=1.375$, $w^-/w=(-2.9+0.3i)\times 10^{-4}$  and  for $\mu=1.425$,  
$w^-/w=(2.2-0.07i)\times 10^{-4}$.
}
\label{table4}
\end{center}
\end{table}

\subsection{Expansion}

\begin{figure}[t]
\begin{center}
\epsfig{file=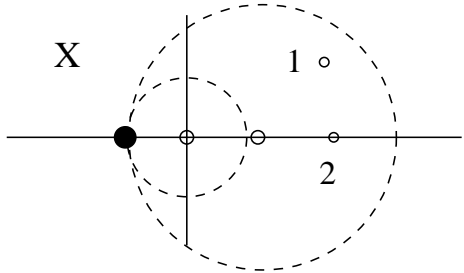, width=0.4\textwidth}
 \caption{Complex $X$ plane, with the pole at $X=-1$. The expansion around 
 $X=\lambda$ has a larger radius of convergence than around $X=0$.
  }
\label{fig:ancon}
\end{center}
\end{figure}

Finally we study the  possibility to ameliorate the dynamics using an 
expansion of the determinant, which is also discussed in
 App.\ \ref{sec:expansions} for the simple models.
We restrict ourselves here to the ordered case.
The fermionic part of the drift is of the form
\be
K = D^{-1} \partial D  +  {\tilde D}^{-1}  \partial {\tilde D}.
\ee
Neglecting the factor $1+C^3$, which cancels in the drift, we write 
\bear
D=1+X, \qquad\qquad X=  a P + b P',
\eear
and similar for $\tilde D$. The pole is at $X=-1$.
We then write a Taylor expansion centred at the shifted point $\lambda$, such that
\be
\frac{1}{D} = \frac{1}{\lambda+1}\sum_{n=0}^{\infty}\left(\frac{\lambda - X}
{\lambda +1}\right)^n,
\label{e.dexp2}
\ee
and again similar for $\tilde D$, with parameter $\tilde \lambda$.
Notice that the $\lambda$ used here differs by one unit from $D_0$ used in 
App.\ \ref{sec:expansions}. 
Since the pole is at $X=-1$, the expansion around $X=\lambda$ with conveniently
chosen $\lambda$  has an increased 
radius of convergence compared to the expansion centred at $X=0$, see
Fig.\ \ref{fig:ancon}. 

 The ``regularisation parameters'' $\lambda, \tilde \lambda$ can be 
 chosen conveniently, and can also be adapted during the simulation 
 (dynamical analytic continuation, see App.\ \ref{sec:expansions}).
We note here that this procedure can also be used for inverting matrices 
$\id + {\bf X}$, where
a simple choice for the regularisation term is $\lambda\id$. 
In lattice QCD, this can e.g.\ be applied to the fermion matrix.
Notice that this shift is an exact procedure and does not represent 
an approximation for which a subsequent extrapolation is needed. 
In practice the expansions are of course truncated and their effectiveness 
depends on the radius of convergence, which is however improved
by the $\lambda$-shift.

\begin{figure}[t]
\begin{center}
 \epsfig{file=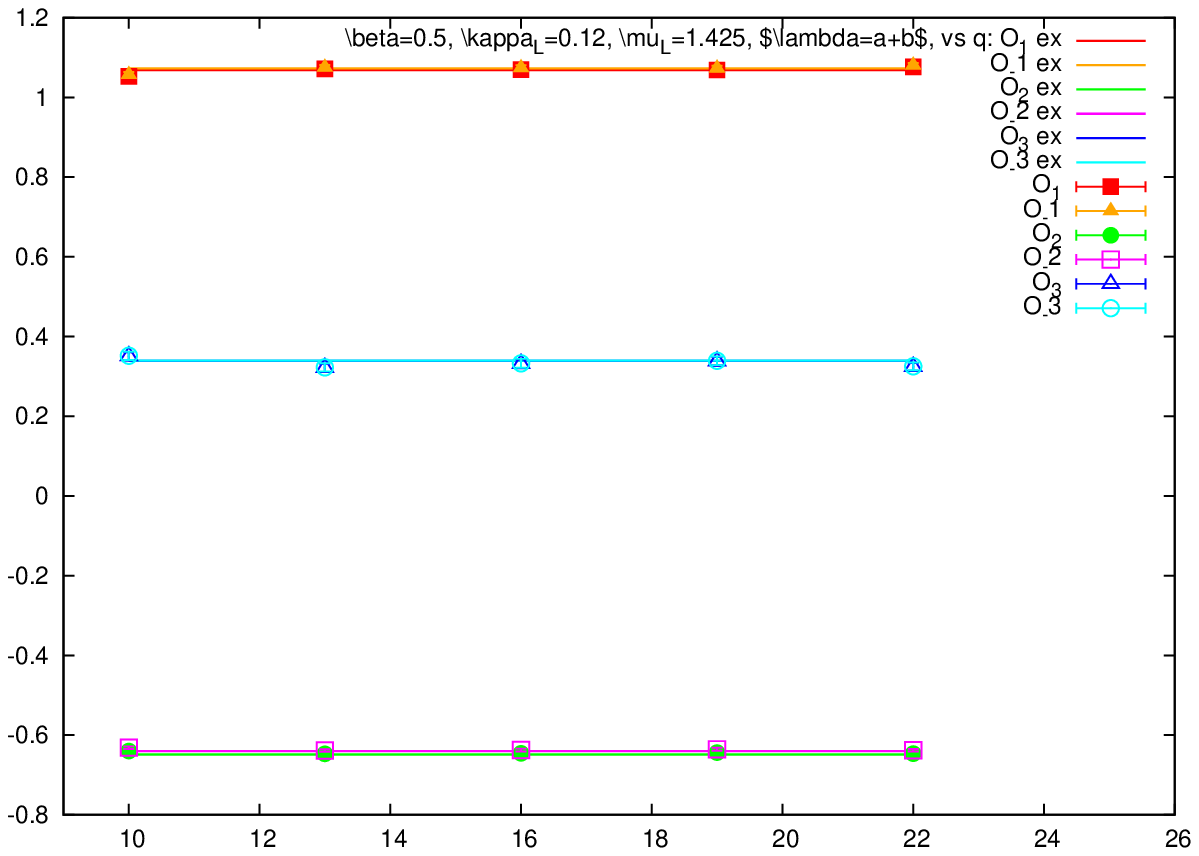, width=0.48\textwidth}
 \epsfig{file=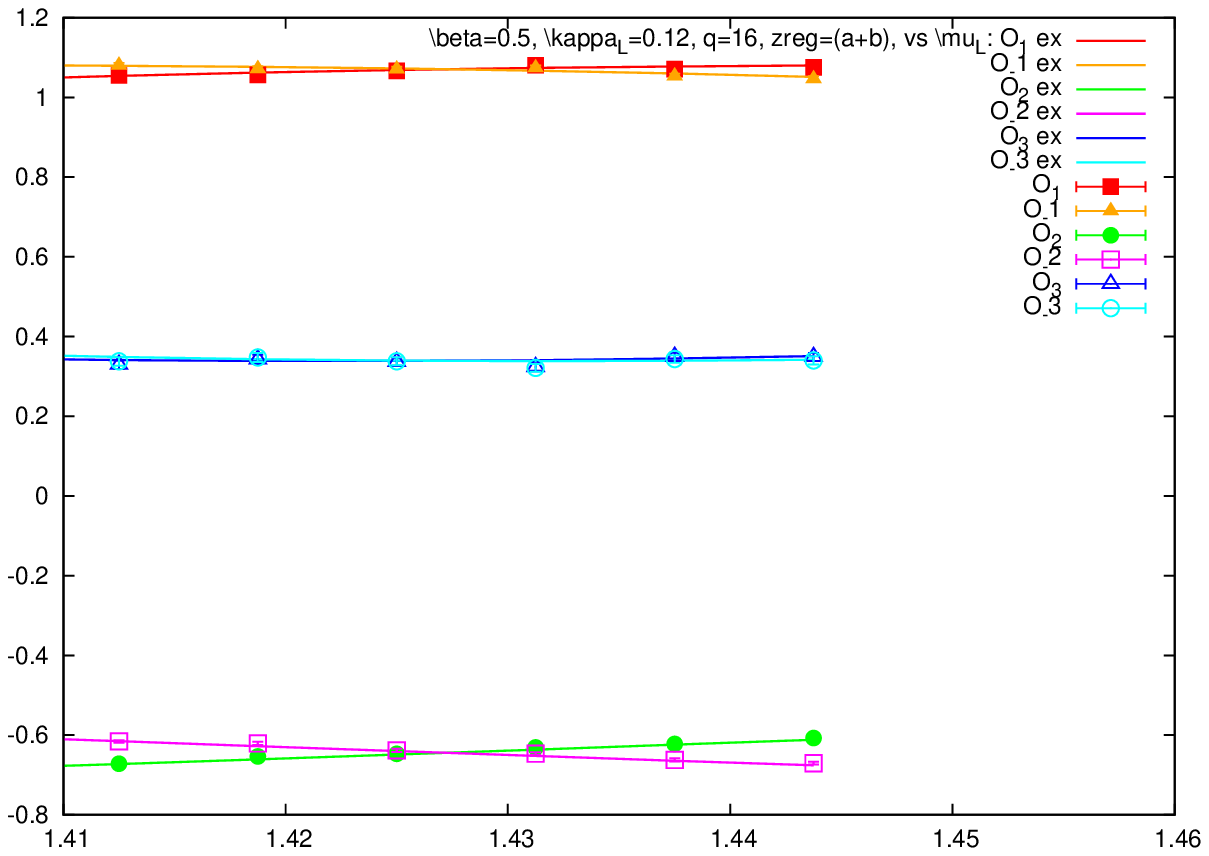, width=0.48\textwidth}
 \caption{
 Left: observables vs expansion order $N$ using $\lambda= (a+b)$. 
 Right:  observables vs $\mu$ for $\lambda= (a+b)$ at fixed $N=16$. 
 Parameters are $\beta=0.25$, $\kappa=0.12$, $\mu =1.425$, $N_\tau=8$, 
 for the ordered lattice, using short runs with Langevin time $t \sim 40$.
  }
\label{fig:fsq_F2u3}
\end{center}
\end{figure}

 In Fig.\ \ref{fig:fsq_F2u3} results for this procedure are shown.
 We have tested $\lambda$ real, imaginary and 0 (no regularisation). For the latter,  
 the expansion shows runaways and does not converge. With imaginary $\lambda=i(a+b)$,
  the convergence was found to be not very good. On the other hand,  
  for real $\lambda=a+b$
   the convergence and results are excellent. In Fig.\ \ref{fig:fsq_F2u3} (left) 
   convergence of the expansion is shown at $\mu=1.425$, i.e.\ the $\mu$ value where 
   the ``whiskers'' affect the result. The expansion is truncated, $n \leq N$, 
   and the dependence on $N$ is shown. Excellent convergence is observed. 
  In Fig.\ \ref{fig:fsq_F2u3} (right), observables are shown as a function of $\mu$, 
  using a fixed $N=16$ , leading to good agreement with the exact results 
  (cf.\ Fig.\ \ref{fig:fsmu_E2u3}).

We find therefore that meaningfully regularised expansions converge to the correct results, 
even at $\mu$ values for which the standard procedure does not work.
This can be understood in the following way:
Choosing the shift such that the expansion point lies in the ``fish" 
region of the scatter 
plots, e.g.\  by choosing $\lambda =a+b$,  
the trajectories explore this region and never enter the ``blue whiskers" region, 
due to the bottleneck. 
Hence it 
 has the same effect as avoiding the $\Re D<0$ region altogether.
A dynamical expansion which adapts $\lambda$ when approaching the edge 
of the domain of
 convergence
is easy to implement as well and will cover the full analyticity domain. 
In this case, 
however, it will also collect data from the ``blue whiskers", leading to 
the same wrong 
results as when all trajectories are used.

\subsection{Discussion and tentative conclusions}

As long as the parameters $a, b, \tilde a, \tilde b$ are below 1, the Langevin process 
is found always to converge 
to the correct results, within the error. Significant discrepancies appear for  $C\simeq 1$ 
where $a, b \simeq 1.5$. 

 Although the measure is $\det M = (D\tilde D)^{2N_f} \simeq D^{2N_f}$,  
the relevant factor for the analysis is $ D\tilde D \simeq D$.  The sign of
 $\Re D$ appears 
to identify two separate regions with two different contributions to 
expectation values. 
This conspicuous situation appears close to onset. The determinant also
becomes squeezed in the imaginary direction. 
These regions show up in the scatter plots  as ``red fish'' ($\Re D>0$) and 
``blue whiskers'' ($\Re D<0$), the former producing good expectation
values for the observable but the latter leading to strongly deviant 
contributions. This outlying behaviour is also visible in the scatter plots of 
the observables directly sensitive to the pole or the fermionic degrees of freedom, 
which may provide a practical test in realistic lattice
simulations at no cost, as the gauge invariant observables 
are calculated anyway. The clear correlation
between  the $\Re D< 0$ region and the outlying contributions to various 
quantities is shown 
in Fig.\ \ref{f.dcorr}, which also illustrates the small weight of these regions.  

 \begin{figure}[t]
\begin{center}
\epsfig{file=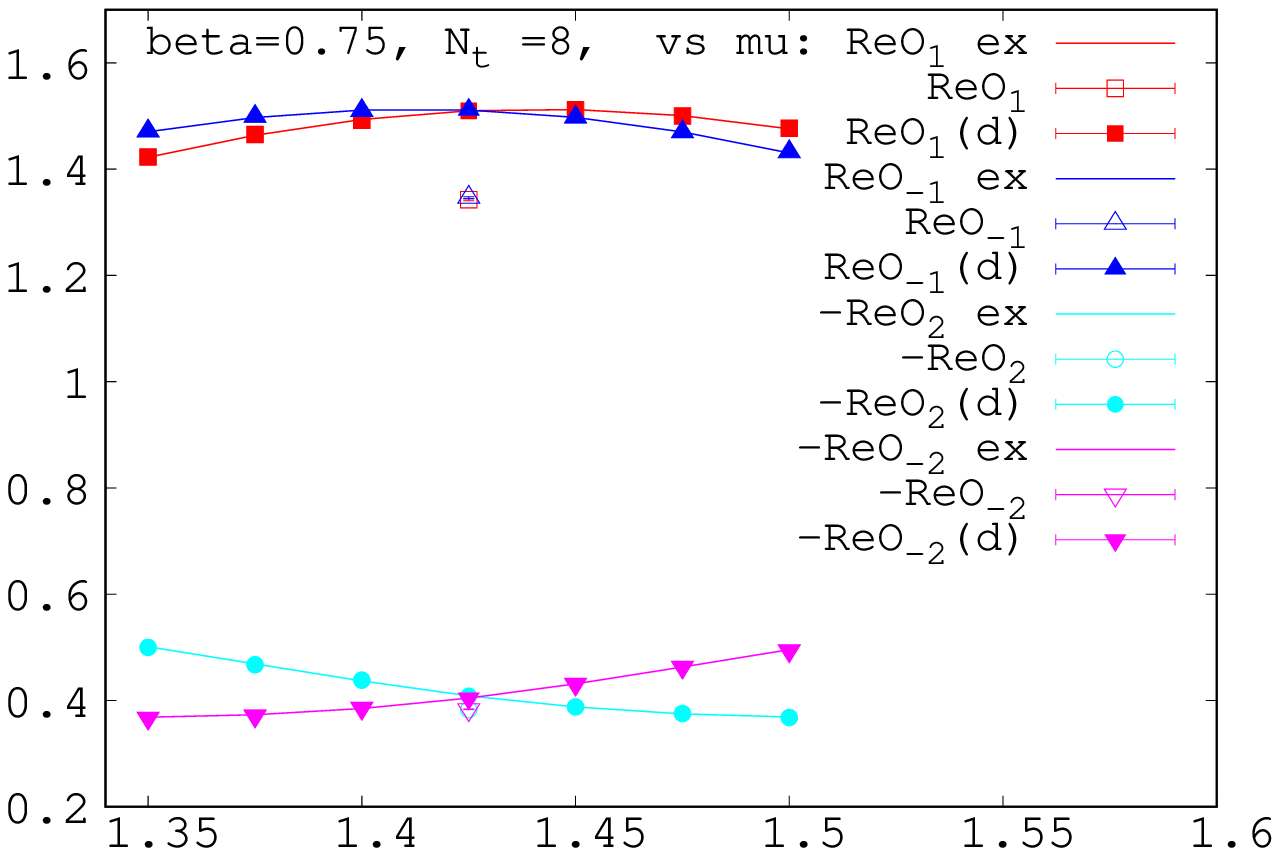, width=0.48\textwidth}
\epsfig{file=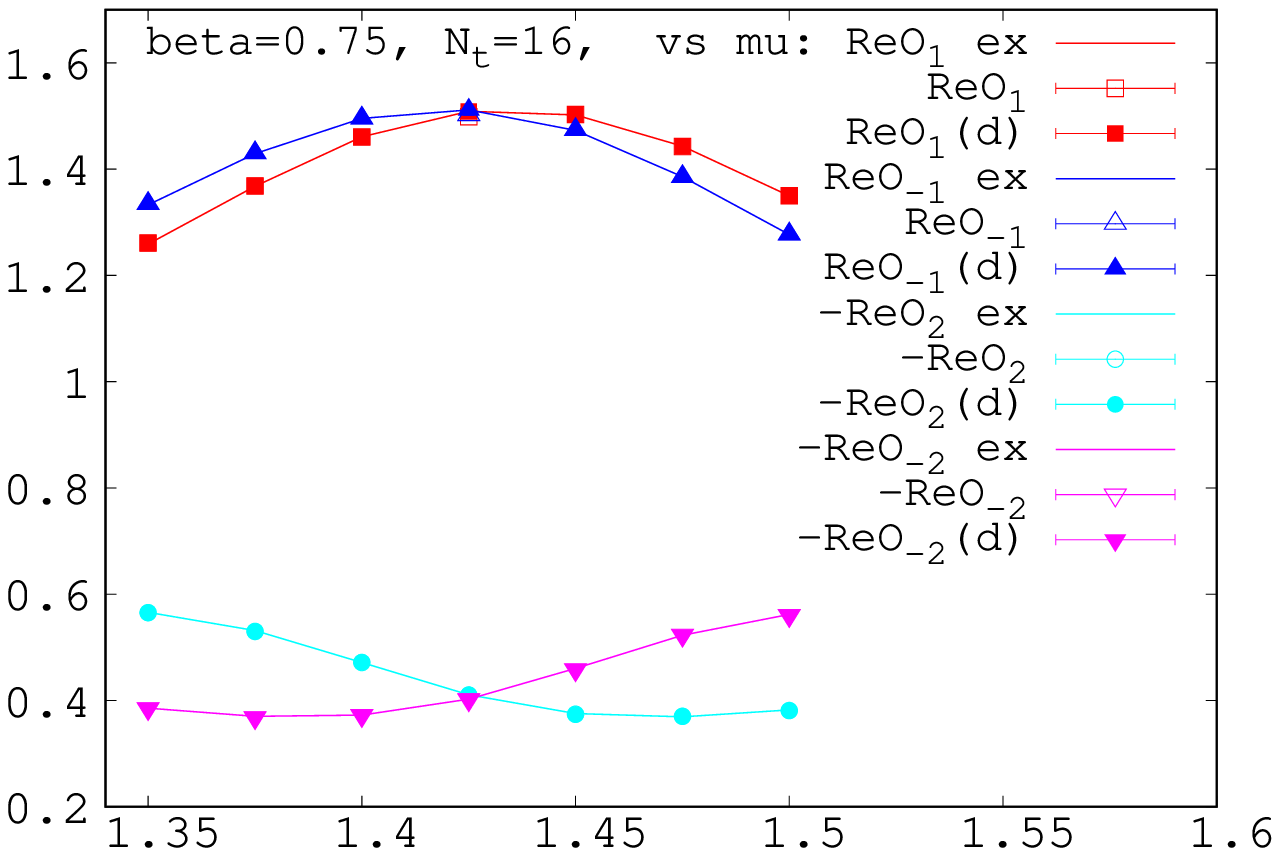, width=0.48\textwidth}
 \caption{
 Observables vs $\mu$ using all trajectories (CL) and trajectories 
 with $d_{\rm min}>d_c$ (CLD)
 for  $\beta=0.75$,  $\kappa=0.12$, $N_\tau=8$ (left) and 16 (right). 
 The Langevin time is $t \sim 2500$.
 }
\label{f.largebN}
\end{center}
\end{figure}

At least in the examples analysed here the two regions appear separated by a bottleneck  
which can only be crossed by trajectories approaching $|D|=0$ below a 
certain threshold. 
The approach to the pole can therefore signal the possible
 sampling of ``deviant'' contributions from the  $\Re D < 0$ region.
The latter  has a significantly smaller weight,
which may only appear as
a quasi-transient whose contribution for very long Langevin trajectories is extremely small. 
This suggests that discarding the contributions from the region with $\Re D <0$ will 
automatically lead to good results. Long enough thermalisation times, or adequate starting 
points in the 
$\Re D > 0$  region can help by inhibiting the development of these regions.
The expansion, on the other hand, seems to do just that if we set the expansion 
centre in the region of $\Re D > 0$, leading to good results via this approach.
We note that the appearance of the bottleneck, separating the complex configuration space 
into two regions, with $w^-\ll w^+$, is consistent with the findings in the U(1) model.

Larger $\beta$ and/or larger $N_\tau$ show a picture consistent with the
above one, supporting these conclusions,  see Fig.\ \ref{f.largebN}.
Larger $N_\tau$ seems to increase the transient character of
the ``blue whiskers'' region; for $N_\tau=16$ only one out of 50 trajectories enters
this region and it yields only a very small contribution.

We conclude that the development of $\Re D <0$ regions signals 
failure in the simulation: although the weight of these regions is small,
their contributions deviate strongly from the ones with $\Re D>0$ and hence they may affect
the results by many standard deviations. Dropping, one way or the other, those contributions
leads to results agreeing with the exact ones, 
at the level of the statistical errors (at the permille level in these runs). 
The fact that the process fails to account correctly for the region with $\Re D<0$ at
 certain parameter values 
suggests, however, that we should attribute a possible systematic error 
proportional with the weight of this region, which is  
${\cal O}\left(10^{-2} - 10^{-4}\right)$ in the examples studied here.

\begin{figure}[t]
\begin{center}
\epsfig{file=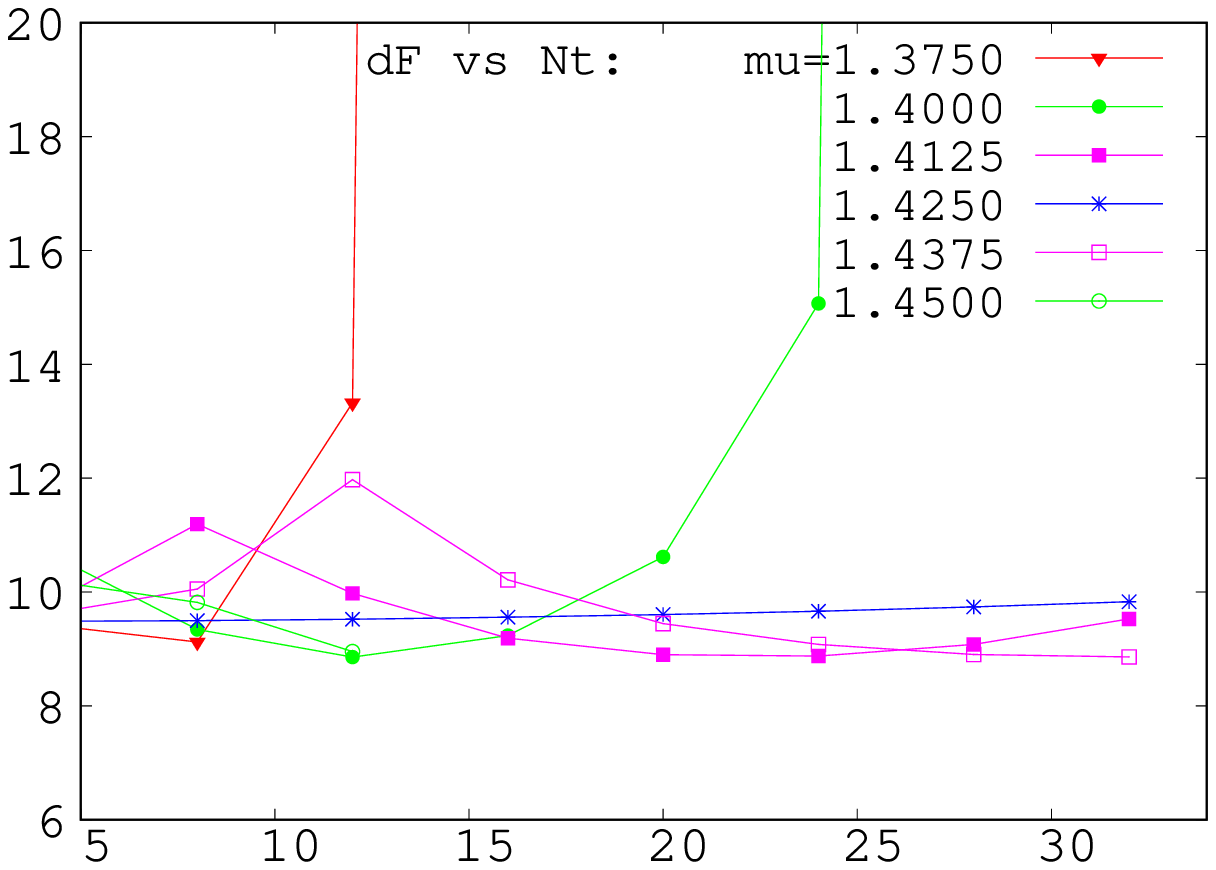, width=0.48\textwidth}
\epsfig{file=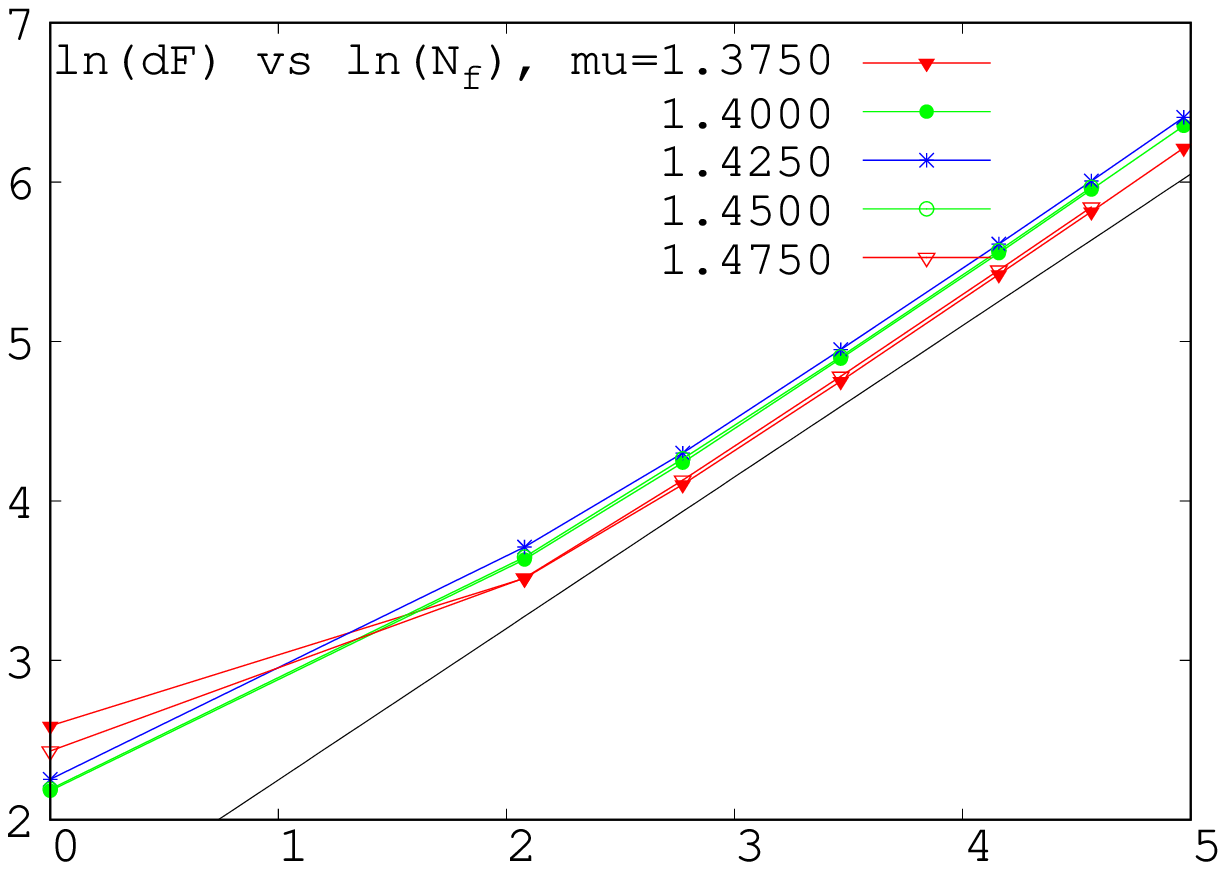, width=0.48\textwidth}
 \caption{
Left: Difference $\Delta F_{\rm eff}$ between the free energy associated 
with the $\Re D <0$ region 
 and the full one vs $N_\tau$.  
Right: $\Delta F_{\rm eff}(N_f)$ vs $N_f$ for $N_f $ flavours
  (double logarithmic scale).
  The straight line suggests a scaling $\Delta F_{\rm eff}(N_f)= \Delta F_{\rm eff}(1) N_f^{p}$ with $p \simeq 0.95$.
 Parameters are  $\beta=0.25$,  $\kappa=0.12$, ordered lattice, various $\mu$.
 }
\label{f.ntdep}
\end{center}
\end{figure}

We may try to quantify the relevance of the region $\Re D <0$
by calculating the logarithm of its relative weight, 
 $\Delta F_{\rm eff} = - \ln (w_-/w_+)$, using the exact integral expressions.  
 As can be seen in
 Fig.\ \ref{f.ntdep} (left),  
  $\Delta F_{\rm eff}$ appears bounded from below, and even increasing with
 $N_{\tau}$ for some $\mu$ values far from $\mu_c^0$. The bound is about $10\%$
 lower 
 on the disordered lattice but shows similar behaviour.
 With increasing number of fermion species, the difference in free energy scales approximately with the number of flavours, 
 $\Delta F_{\rm eff} (N_f) \simeq N_f \Delta F_{\rm eff} (1)$, see Fig.\ \ref{f.ntdep} (right).  
   Since in HDQCD the lattice determinant  
 is a  product of factors of the form $D^2$ over the spatial lattice, we may ask 
 how the spatial volume would manifest itself in $\Delta F$.
 Fully correlated Polyakov loops would then behave as if in the presence of many flavours, 
 but the general case is not trivial and  will be discussed in the next section.

Extrapolating the lesson from this discussion to realistic QCD lattice calculations, 
we conclude that 
one has to monitor the appearance of disconnected regions with a bottleneck at
 $|D| \simeq 0$. 
 This can be done  by monitoring various quantities such as some selected 
observables, the drift or the lowest determinant modes avoiding time consuming 
procedures. Dropping {\em by  hand} the occasional contributions of regions of
type ``blue whiskers''  ($\Re D < 0$)
should 
already produce good results, while an estimate of the relative 
impact of such contributions  
would suggest a measure for possible  systematic errors.

\section{Lattice QCD}
\label{sec:lattice}

In the following section we aim to apply the lessons found above to the case of QCD at nonzero quark density, first in the case of heavy quarks (HDQCD) and then for full QCD, using the staggered fermion formulation. 

In QCD the partition function is, after integrating out the quarks fields, written as
\be
Z = \int DU\, e^{-S_{\rm YM}} \det M \equiv \int DU\, e^{-S},
\qquad S=S_{\rm YM} - \ln\det M,
\ee
where $S_{\rm YM}$ is the Yang-Mills action, $U$ are the gauge links, and $M$ is the fermion matrix.
The Langevin update for the gauge links $U$ reads \cite{PhysRevD.32.2736}, in a first-order discretised scheme, with Langevin time $t=n\eps$,
\be
U_{x,\nu}(t+\eps) = \exp\left[ i\lambda^a\left(\eps K^a_{x,\nu}+\sqrt{\eps}\eta^a_{x,\nu}\right) \right]  U_{x,\nu}(t), 
\ee
where $\lambda^a$ are the Gell-Mann matrices, normalised as $\Tr \lambda^a\lambda^b=2\delta^{ab}$,  and the sum over $a=1,\ldots,8$, is not written explicitly. More details can be found in Refs.\ \cite{Aarts:2008rr,Seiler:2012wz,Sexty:2013ica}.
The drift is generated by the action $S$ and reads
\be
\label{eq:driftSU}
K^a_{x,\nu} = - D^a_{x,\nu}S = -D^a_{x,\nu}S_{\rm YM} + \Tr\left[ M^{-1}D^a_{x,\nu}M\right].
\ee
Hence the zeroes of the determinant show up as poles in the drift.

\subsection{Heavy dense QCD}
\label{sec:hdqcd}

To assess the importance of these poles, we need to specify the fermion matrix. We consider first heavy dense QCD. This approximation to full QCD can be obtained by a systematic hopping-parameter expansion of the fermion determinant, preserving terms that cannot be ignored for large chemical potential, as well as the terms related by symmetry under $\mu\to-\mu$. For Wilson quarks, this amounts to an expansion in terms of $\kappa$, keeping $\kappa e^\mu$ fixed and preserving terms that go as $\kappa e^{-\mu}$. A detailed discussion can be found in Refs.\ \cite{Bender:1992gn,DePietri:2007ak,Aarts:2014bwa}, see also Refs.\ \cite{Fromm:2011qi,Fromm:2012eb} for combined hopping and strong-coupling expansions.

Here we consider the resulting theory at leading order, using $N_f$ degenerate quark flavours, for which the fermion determinant reads \cite{Aarts:2008rr} (see also Sec.\ \ref{sec:spin})
\be
 \det M = \prod_\xv 
		\det\left(1 + h e^{\mu/T}  \mathcal{P}_\xv \right)^{2N_f} 
		\det\left( 1 + h e^{-\mu/T}  \mathcal{P}^{-1}_\xv \right)^{2N_f}.
\end{equation}
The remaining determinant is in colour space only. The parameter $h=(2\kappa)^{N_{\tau}}$ arises from the hopping expansion ($N_\tau$ the number of time slices in the temporal direction) and $\mathcal{P}_\xv^{(-1)}$ are the (inverse) Polyakov loops, see Eq.\ (\ref{eq:P}).
Note that the gluon dynamics is included in Eq.\ (\ref{eq:driftSU}) via the usual Wilson Yang-Mills lattice action, with gauge coupling $\beta$.

In order to study the zeroes of the determinant, we identify the basic building block of determinant, defined such that the full determinant in the path integral weight is written as 
\be
\det M = \prod_\xv  \left[ \det \widetilde M_\xv \right]^{2N_f}.
\ee
Here the {\em local} determinants are
\bea
\det \widetilde M_\xv &=&
	\det\left(1 + h e^{\mu/T}  \mathcal{P}_\xv \right) \det\left( 1 + h e^{-\mu/T}  \mathcal{P}^{-1}_\xv \right) 
	\nn\\
	&=& \left(1+3z P_\xv+3z^2P_\xv^{-1}+z^3\right) \left(1+3\bar z P_\xv^{-1} + 3\bar z^2P_\xv+\bar z^3\right),	 
\label{eq:detloc}
\eea
 where $z=he^{\mu/T}, \bar z=he^{-\mu/T}$ and 
 \be
 P_\xv = \frac{1}{3}\Tr\,  \mathcal{P}_\xv, \qquad\qquad P_\xv^{-1} = \frac{1}{3}\Tr\,  \mathcal{P}_\xv^{-1}.
 \ee  
 We study the zeroes of the local determinants rather than the full determinant, since this is what is closest to the analysis carried out above and will allow us to focus on individual factors getting small. 
 
HDQCD has been used extensively to justify the results obtained with CL, e.g.\ via reweighting \cite{Seiler:2012wz,Aarts:2013uxa,Aarts:2016qrv}. Since reweighting and CL have very different  systematic uncertainties, 
 the agreement of the results obtained by both methods is a strong argument for
the correctness of either approach. In particular, since reweighting does not suffer from potential problems caused by zeroes of the determinant, agreement indicates that the latter do not cause problems for CL.

We note here that HDQCD has also been used to test and compare variations of the hopping parameter expansion to higher order \cite{Aarts:2014bwa}, in particular with regard  to the standard hopping expansion, obtained via \cite{Seiler:2015uwe}
\be
\det M \equiv \det(1-\kappa Q) = \exp\sum_{n=1}^\infty -\frac{\kappa^n}{n}\Tr\, Q^n,
\ee
for which zeroes of the determinant do not appear. An expansion in spatial hopping terms only, for which HDQCD is the leading-order term, has been discussed and assessed in Ref.~\cite{Aarts:2014bwa}.

\begin{figure}[t]
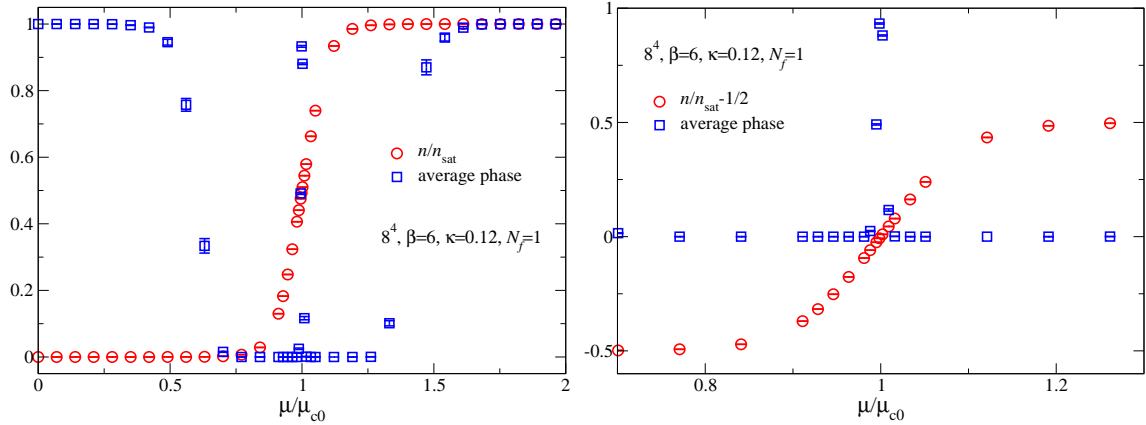

\begin{center}
\includegraphics[width=0.49\columnwidth]{figs-QCD/signdens-xm.eps}  
\includegraphics[width=0.49\columnwidth]{figs-QCD/signdens-zoom-xm.eps}
\caption{Left:
 fermionic density in units of the saturation density, $n/n_{\rm sat}$, and average phase factor, see Eq.\ (\ref{signaver}),
as a function of the chemical potential. 
Right: a blow-up around onset. Parameters as in Eq.\ (\ref{eq:para}) on a $8^4$ lattice.
}
\label{HDsigndens}
\end{center}
\end{figure}

We now discuss some results in HDQCD, focussing on potential zeroes of the determinant.  We mostly use the following choice of parameters 
\be
\label{eq:para}
\beta =6, \qquad \kappa=0.12, \qquad N_f=1, \qquad N_s^3 = 8^3, \qquad N_\tau=8,  16.
\ee
Note that the lattice spacing (or gauge coupling $\beta$) and the spatial volume ($N_s^3$) are fixed, but we consider two temperatures ($N_\tau=8,16$). 
In this theory, the quark number at zero temperature changes from 0 below onset  to saturation above onset, with $n_{\rm sat}=N_{\rm spin}\times N_{\rm colour}\times N_f=6N_f$,  and the critical chemical potential is given by 
\be
 \mu_c^0  =-\ln (2\kappa) = 1.427.
\ee
At nonzero temperature, this transition is smoothed and the critical chemical potential $\mu_c(T)<\mu_c^0$, eventually connecting to the thermal confinement-deconfinement transition line at higher temperature and lower chemical potential.
A study of the phase diagram at fixed lattice spacing can be found in Ref.\ \cite{Aarts:2016qrv}.
In Fig.\ \ref{HDsigndens} we show the density, in units of saturation density, for the parameters in Eq.\ (\ref{eq:para}) on the $8^4$ lattice. The rapid rise around $\mu=\mu_c^0$ is indeed observed.

Writing the determinant as a product of its absolute value and phase,
\be
\det M = |\det M | e^{i\varphi},
\ee
and using the symmetry
\be
[\det M(\mu)]^* = \det M(-\mu^*),
\ee
we can extract the average phase factor in the full (i.e.\  not in the phase-quenched) theory via a computation of \cite{Aarts:2008rr}
\be 
 \label{signaver}
 \langle e^{2i\varphi} \rangle = \left\langle \frac{\det M(\mu)}{\det M(-\mu)} \right\rangle,
\ee
which is accessible using CL dynamics. The result is shown in Fig.\ \ref{HDsigndens} as well. The sign problem is severe in the onset region. At the critical chemical potential, the fermions are at `half-filling', i.e.\ 
half of the available fermionic states are filled, and the theory
becomes particle-hole symmetric \cite{Rindlisbacher:2015pea}. 
Exactly at $\mu_c$, the sign problem becomes very mild, as the first dominating factor in Eq.\ (\ref{eq:detloc}) becomes real ($z=1$).
The sign problem due to the second factor is very small, since $\bar z = (2\kappa)^{2N_\tau}\ll 1$. 

\begin{figure}[t]
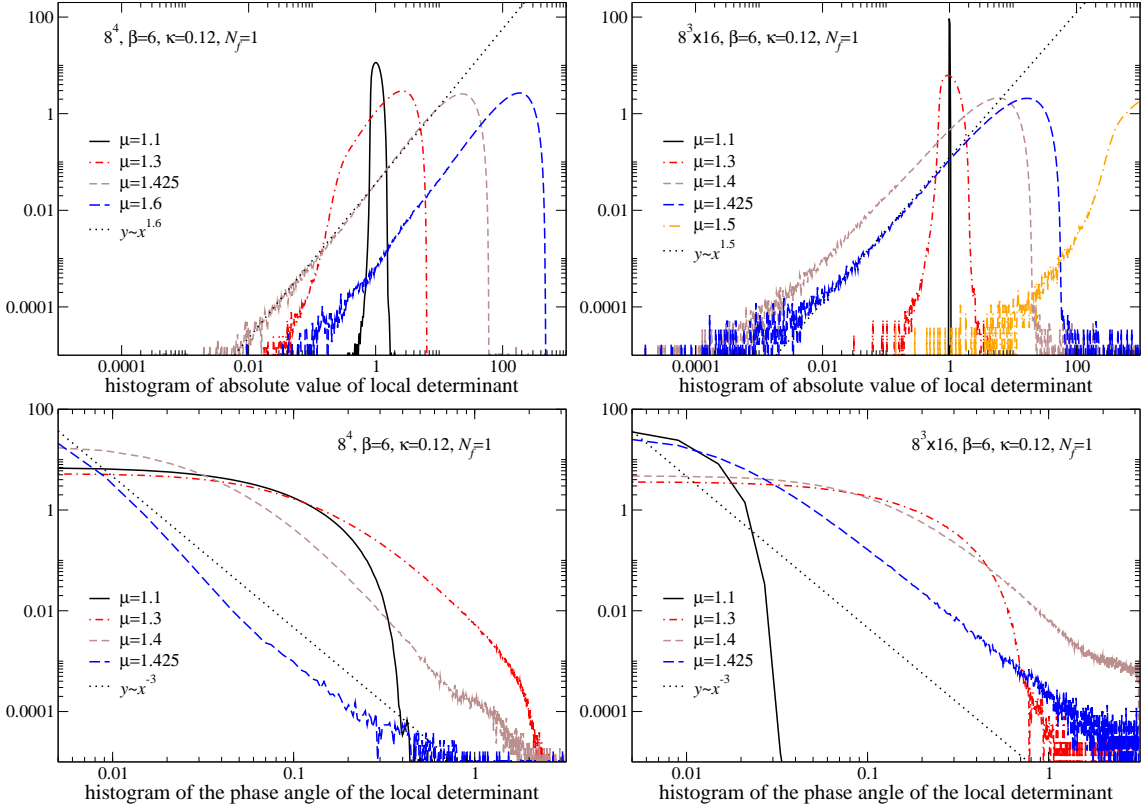

\begin{center}
\includegraphics[width=0.49\columnwidth]{figs-QCD/rhist1-xm.eps}  
\includegraphics[width=0.49\columnwidth]{figs-QCD/rhist2-xm.eps}  
\includegraphics[width=0.49\columnwidth]{figs-QCD/anglehist1-xm.eps}  
\includegraphics[width=0.49\columnwidth]{figs-QCD/anglehist2-xm.eps}  
\caption{Histogram of the absolute value (top) and the  phase angle (bottom) of the local determinant $\det \widetilde M_\xv$ 
for several chemical potentials, on a $8^4$ (left) and a $8^3\times 16$ (right) lattice.
}
\label{HDhist}
\end{center}
\end{figure}

To investigate the zeroes of the measure in HDQCD, we have analysed $\det \widetilde M_\xv$, see Eq.\ (\ref{eq:detloc}).
Note that the corresponding factor  in the effective SU(3) model was discussed in Sec.~\ref{sec:spin}.
We find that the simulations largely avoid the zeroes of the determinant, except in the vicinity of the critical chemical potential.
This is illustrated in Fig.~\ref{HDhist} (top), where a histogram of the absolute value of the local determinant $\det \widetilde M_\xv$ is shown, on a double-logarithmic scale for two lattice sizes. For small chemical potential, the absolute value of the determinant is close to 1, as $z,\bar z\ll 1$. As $\mu$ is increased, the distribution widens and its maximum shifts towards larger values. However, we also observe that the distribution is nonzero for smaller values, with an apparent power decay towards zero. Based on these simulations, we find the following behaviour
\be
\mbox{probability}\left(\left|\det\widetilde M\right|\right) \sim \left|\det\widetilde M\right|^\alpha, \qquad\qquad \alpha\sim1.5-1.6.
\ee
Values close to zero are more likely when  the chemical potential is close to the critical value, but remain suppressed. 
This behaviour is seen for both lattice sizes, $8^4$ and $8^3 \times 16$, with a broader distribution on the larger lattice.

Also shown in Fig.~\ref{HDhist}  are the distributions for the phase angle $\Phi$, defined via
\be
\det \widetilde M = \left| \det \widetilde M \right| e^{i\Phi}.
\ee
Note that $-\pi<\Phi<\pi$ and that the distributions are symmetric around zero. Away from the critical chemical potential, the distribution drop to zero rapidly; around $\mu_c$ we observe a decay $\sim 1/\Phi^3$.
The relation between this phase and the phase of the full determinant $\varphi$ is not immediate. However, we note that in general it is expected that the latter will vary rapidly, as the full determinant is a product of $2N_fN_s^3$ local determinants.

\begin{figure}[t]
\begin{center}
\includegraphics[width=0.49\columnwidth]{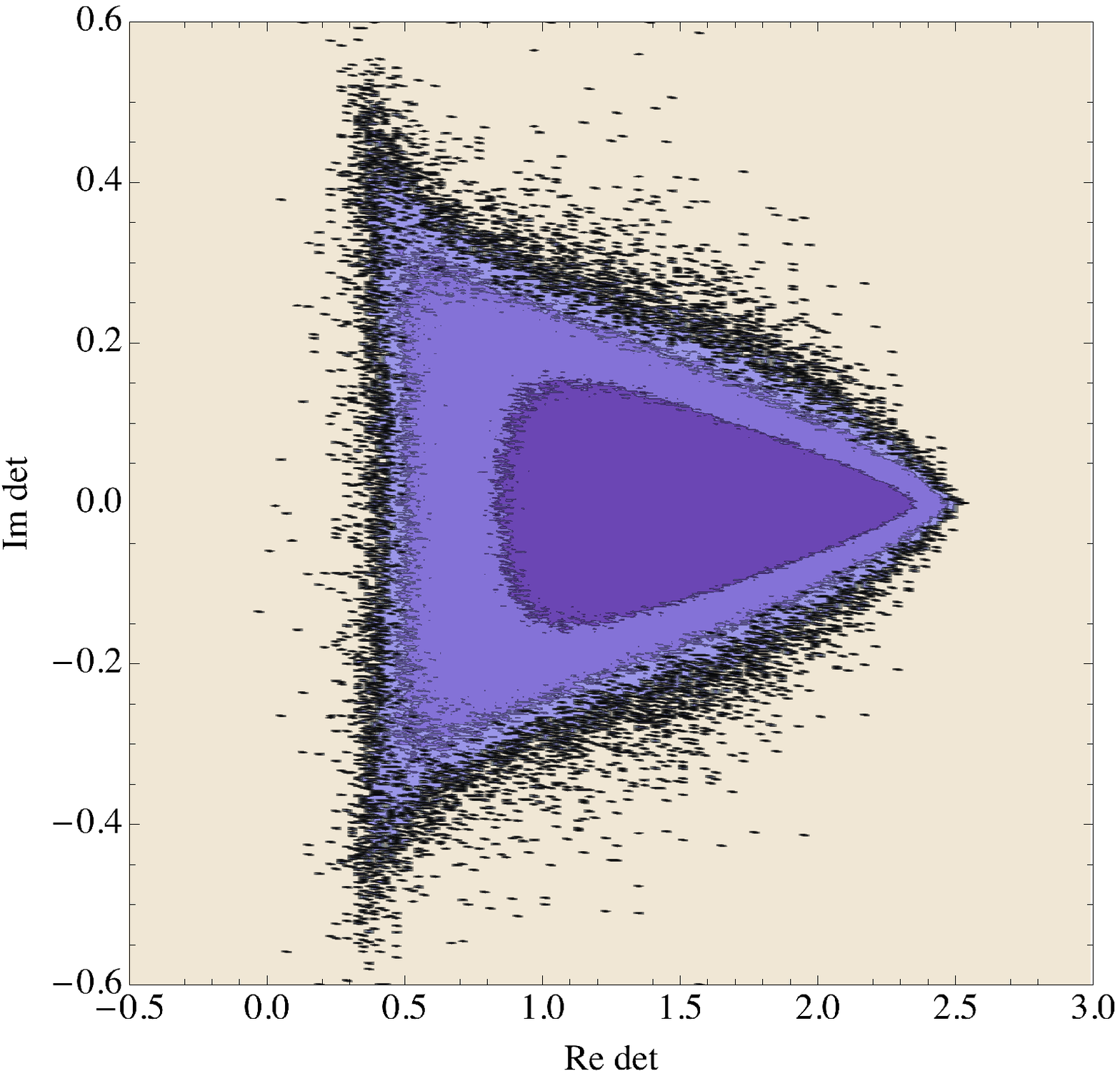}  
\includegraphics[width=0.49\columnwidth]{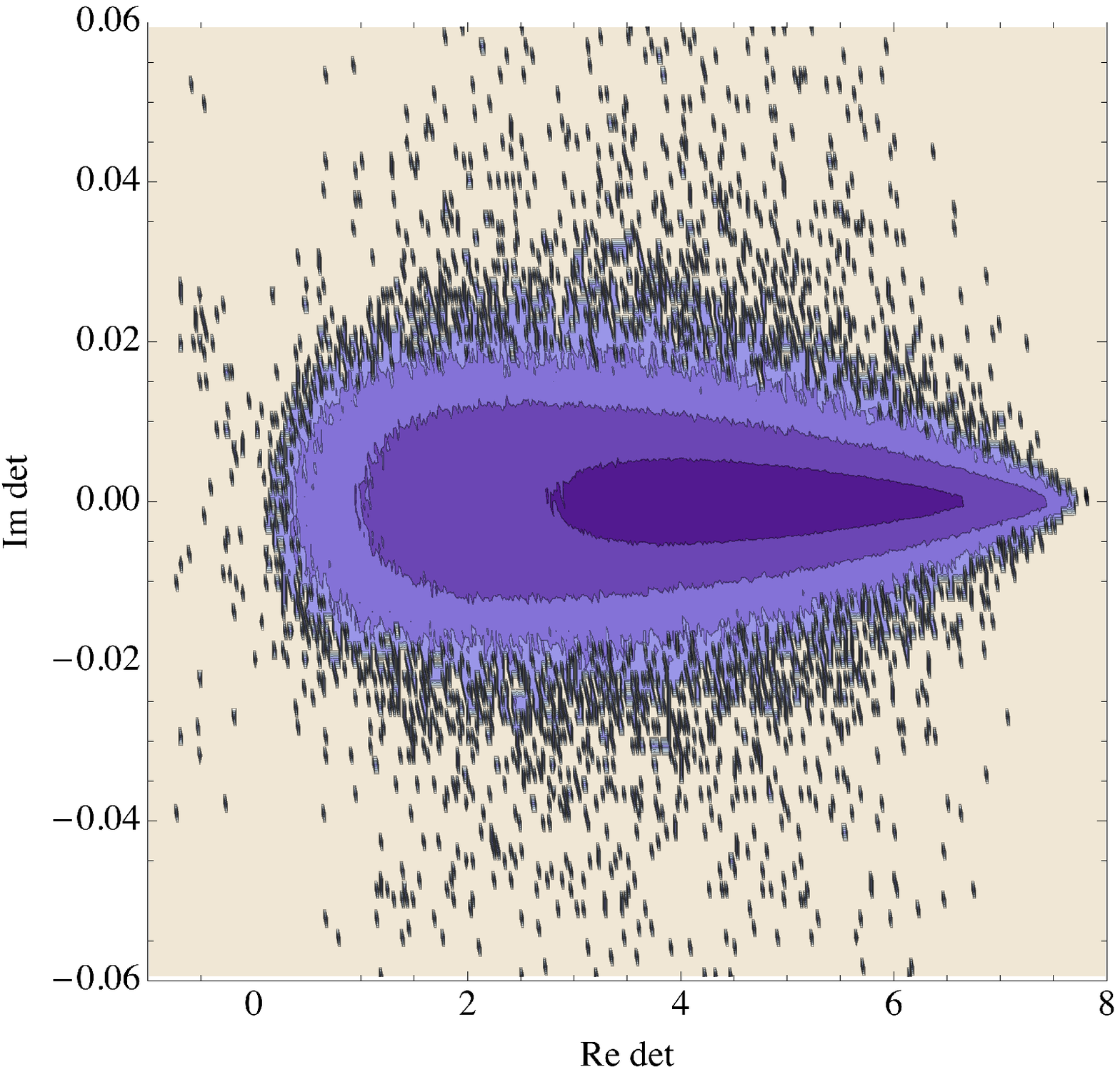}  
\caption{Probability density of the local determinant $\det\widetilde M$ on a logarithmic scale, for 
$\mu=1.3$ (left) and $\mu=1.425$ (right), on an $8^4$ lattice with $\beta=6.0, \kappa=0.12, N_f=1$. Note the different vertical and horizontal scales. 
}
\label{HDscatter}
\end{center}
\end{figure}

To compare with the results presented in the previous sections, we show in Fig.~\ref{HDscatter} the probability density of the local determinants on a logarithmic scale, for two values of the chemical potential close to $\mu_c^0=1.427$, namely $\mu=1.3$ (left) and 1.425 (right). Note the very different horizontal and vertical scales.  These distributions look remarkably similar to those encountered in the simpler cases, although with only a very  thin presence of the `whiskers', if at all. 
We find that at the critical point the distribution is highly elongated in the positive real direction, and that it shrinks again for $\mu>\mu_c$.
Exactly at $\mu_c$,  the distribution shrinks in the imaginary direction, leading to a much smaller typical phase 
of the determinant, and thus a milder sign problem. This explains the milder sign problem at $\mu=\mu_c$, as observed via $\bra e^{2i\varphi}\ket$  in Fig.~\ref{HDsigndens}. There are some configurations where Re $\det \widetilde M<0$, but these appear very infrequent and do not carry substantial weight.

\begin{figure}[t]
\begin{center}
\includegraphics[width=0.49\columnwidth]{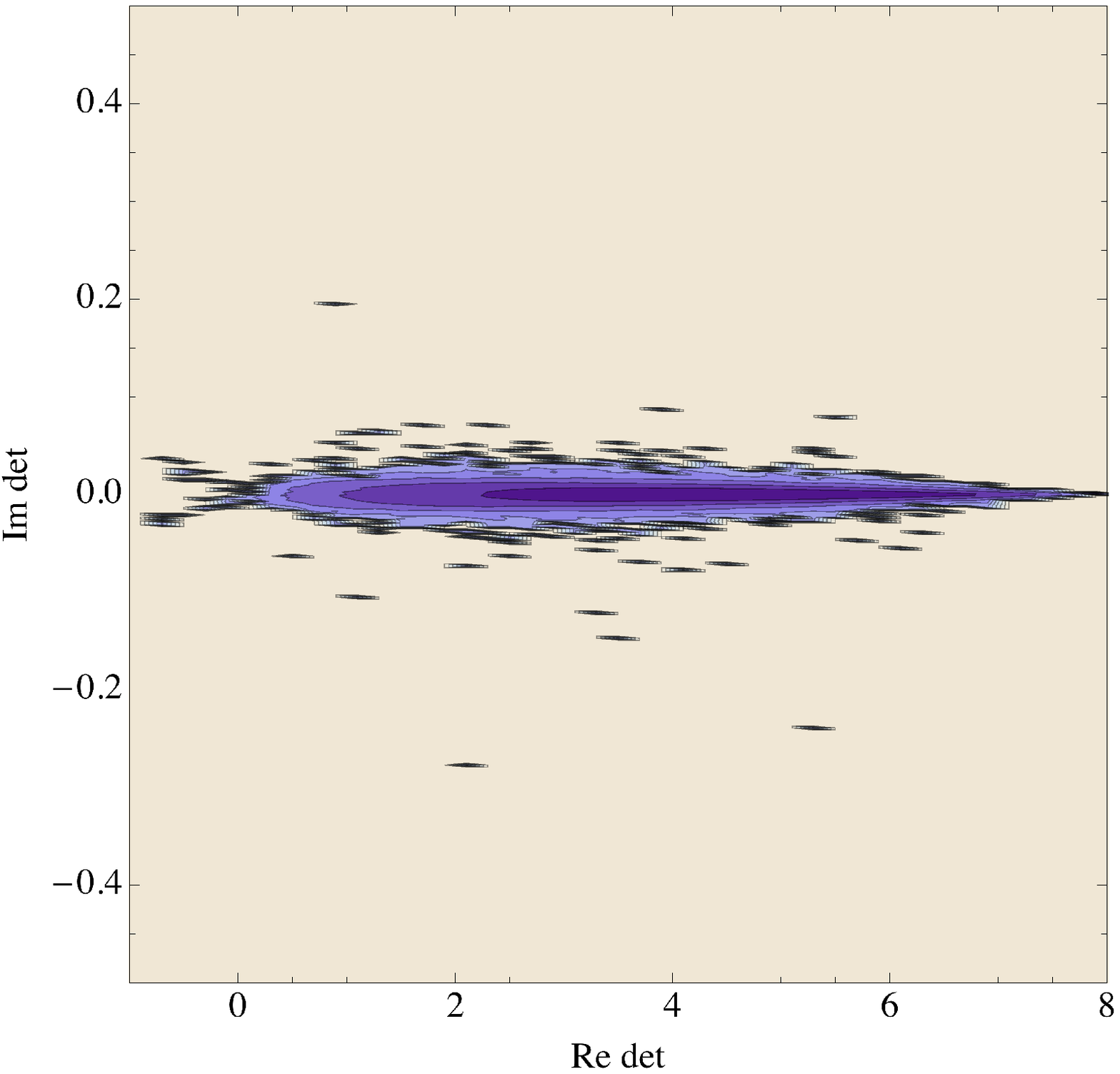}  
\includegraphics[width=0.49\columnwidth]{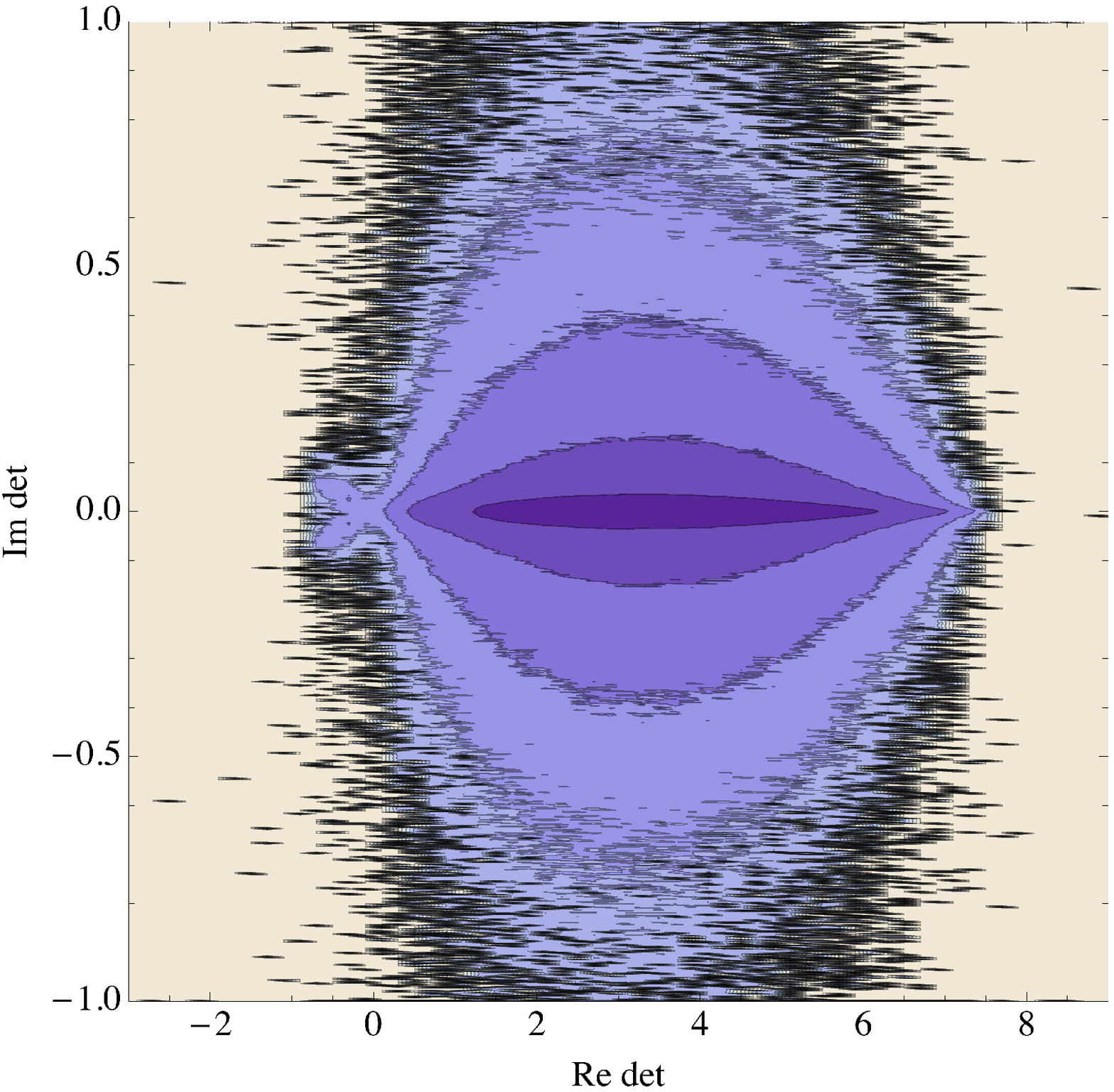}  
\caption{As in Fig.\ \ref{HDscatter}, on a $8^4$ (left) and $16^4$ (right) lattice at $\beta=5.9, \kappa=0.12, \mu=1.425, N_f=1$.  Note the different scale.}
\label{HDbiglat}
\end{center}
\end{figure}

\begin{figure}[t]
\begin{center}
\includegraphics[width=0.55\columnwidth]{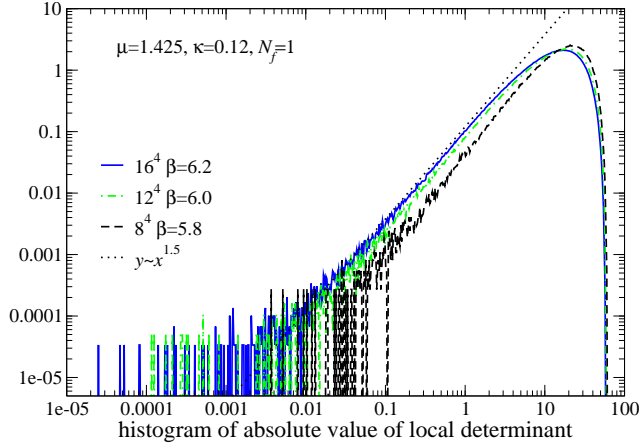}  
\caption{As in Fig.\ \ref{HDhist}, with a decreasing lattice spacing ($\beta=5.8, 6.0, 6.2$), while keeping the physical volume 
approximately constant (with lattice volume $8^4, 12^4, 16^4$), using $\kappa=0.12, \mu=1.425, N_f=1$. }
\label{HDcontlim}
\end{center}
\end{figure}

However, at lower temperatures a clear sign of the whiskers appears, which indicates the possibility of contamination from configurations with Re $\det\widetilde M<0$. This is illustrated in Fig.~\ref{HDbiglat}, where we show results at a fixed lattice spacing, on a $8^4$ and $16^4$ lattice, such that the temperature is twice as low on the $16^4$ lattice. Close to $\mu_c$, the weight of the region with Re $\det\widetilde M<0$ is approx.\ 0.005\% on the $8^4$ lattice and  0.08\% on a $16^4$  lattice, indicating a growing importance as the temperature is lowered. At the lower temperature, the ``whiskers'' are remarkably similar to those encountered in the SU(3) one-link model.
In Fig.~\ref{HDcontlim} we aim to reduce the lattice spacing, while keeping the physical volume and the temperature constant. Here we see that the power decay towards zero remains approximately the same and hence the role of configurations with a small absolute value of the local determinant does not change when going (somewhat) closer to the continuum limit.
We also investigated whether changing $N_f$ influences the appearance of the 
whiskers. While using a larger $N_f$ is beneficial, configurations with determinants in the whiskers do appear at low temperatures.
Finally we have  studied the volume dependence of the weights $w_\pm$, signifying
the importance of the regions $G_\pm$, as suggested by the analysis of the one-link models, but did not find a clear suppression of the region
with $\Re D<0$.

We hence conclude that the zeroes of the determinant for HDQCD appear to have no effect except close to the critical chemical 
potential. Since for those chemical potentials the sign problem is quite mild, this observation is not related to the severeness of the sign problem but to details of the CL process. Although the zeroes might influence the results in this region, the relative weight of configurations with Re $\det\widetilde M<0$ is quite small and their influence is therefore suppressed.

\subsection{Full QCD}
\label{sec:qcd}

Finally, we consider full QCD, with dynamical fermions. We note that full QCD is numerically much more costly than HDQCD, as the inversion of the fermion matrix has to be carried out numerically  for every update. Similarly, an assessment of the zeroes of the determinant is harder, since it requires the computation of the full determinant. It should be noted that the determinant itself is not required for the Langevin update, only $M^{-1}$ evaluated on a fixed vector \cite{Sexty:2013ica}.

 Here we show results obtained using staggered fermions, with the (unimproved) staggered fermion matrix 
 \be 
 M_{xy} = m\delta_{xy} +\sum_\nu \half\eta_{\nu x} \left[ e^{\mu\delta_{\nu 4}}U_{\nu,x}\delta_{x+a_\nu y} -  e^{-\mu\delta_{\nu 4}}U_{\nu,y}^{-1}\delta_{x-a_\nu y} \right],
\ee
 where $\eta_{\nu x}$ are the staggered sign functions, $\eta_{1x}=1$, $\eta_{2x}=(-1)^{x_1}$, $\eta_{3x}=(-1)^{x_1+x_2}$, $\eta_{4x}=(-1)^{x_1+x_2+x_3}$. Note that this formulation describes four tastes, due to the fermion doubling. 
 
\begin{figure}[t]
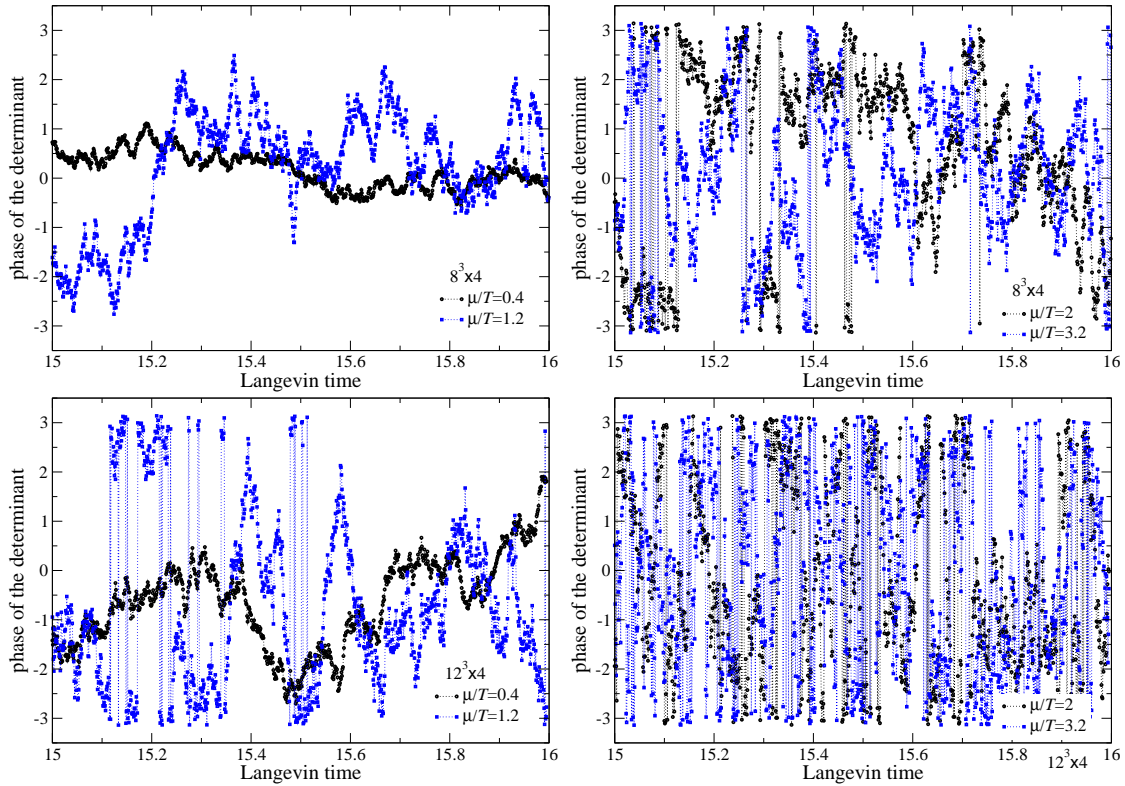

\begin{center}
\includegraphics[width=0.48\columnwidth]{figs-QCD/fullphase8a-xm.eps}  
\includegraphics[width=0.48\columnwidth]{figs-QCD/fullphase8b-xm.eps}  
\includegraphics[width=0.48\columnwidth]{figs-QCD/fullphase12a-xm.eps}  
\includegraphics[width=0.48\columnwidth]{figs-QCD/fullphase12b-xm.eps}  
\caption{Typical Langevin evolution of the phase of the fermion determinant, on a $8^3\times 4$ (top) and $12^3 \times 4$ (bottom) lattice at $\beta=5.3$, $m=0.05$ and $N_f=4$ staggered fermion flavours, for $\mu/T=0.4, 1.2$ (left) and $\mu/T=2, 3.2$ (right).
}
\label{fullphase12}
\end{center}
\end{figure}

There are several indications for full QCD that,  at least at high temperatures  and for the quark masses considered, the CL simulations are unaffected by poles in the drift: these include comparisons 
to systematic hopping-parameter expansions,  which have holomorphic actions \cite{Aarts:2014bwa},
comparisons to reweighting, which does not depend on the action being holomorphic \cite{Fodor:2015doa}, and by observing spectral properties of the fermion matrix \cite{Sexty:2014dxa} (see also below).
At low temperature, it is at present  not known whether poles affect the CL results, partly because 
simulations are more expensive due to the larger values of $N_\tau$ required, or are hindered by the ineffectiveness of gauge cooling on coarse lattices.

\begin{figure}[t]
\begin{center}
\includegraphics[width=0.48\columnwidth]{figs-QCD/phasehist2-xm.eps}  
\includegraphics[width=0.48\columnwidth]{figs-QCD/phasehist4-xm.eps}  
\caption{Histogram of the phase of the determinant for three spatial volumes for $\mu/T=0.4$ (left) and $\mu/T=1.2$ (right). Other parameters  as in Fig.\ \ref{fullphase12}.
 }
\label{phasehist}
\end{center}
\begin{center}
\includegraphics[width=0.5\columnwidth]{figs-QCD/rhist-full.eps}  
\caption{Histogram of the absolute value of eigenvalues for full QCD, gained averaging the spectrum from 100 configurations.}
\label{fullrhist}
\end{center}
\end{figure}

\begin{figure}[t]
\begin{center}
\includegraphics[width=0.45\columnwidth]{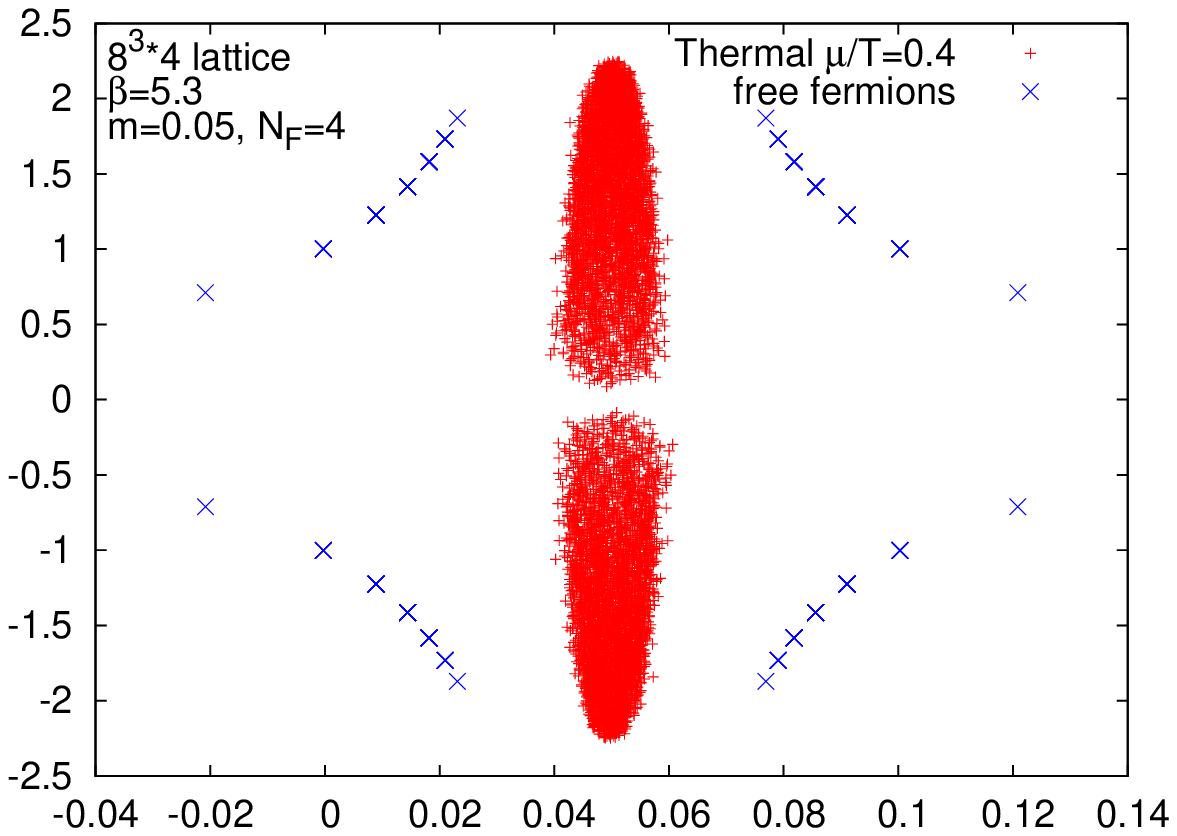}  
\includegraphics[width=0.45\columnwidth]{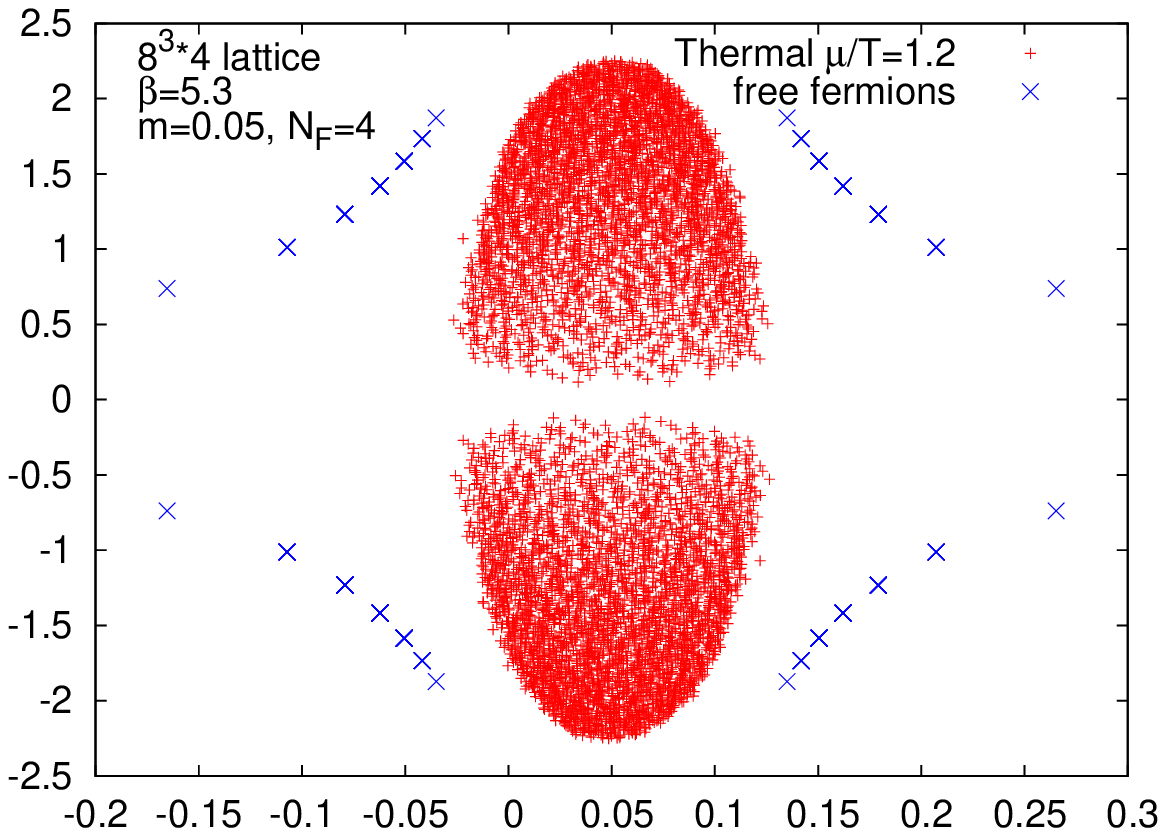}  
\includegraphics[width=0.45\columnwidth]{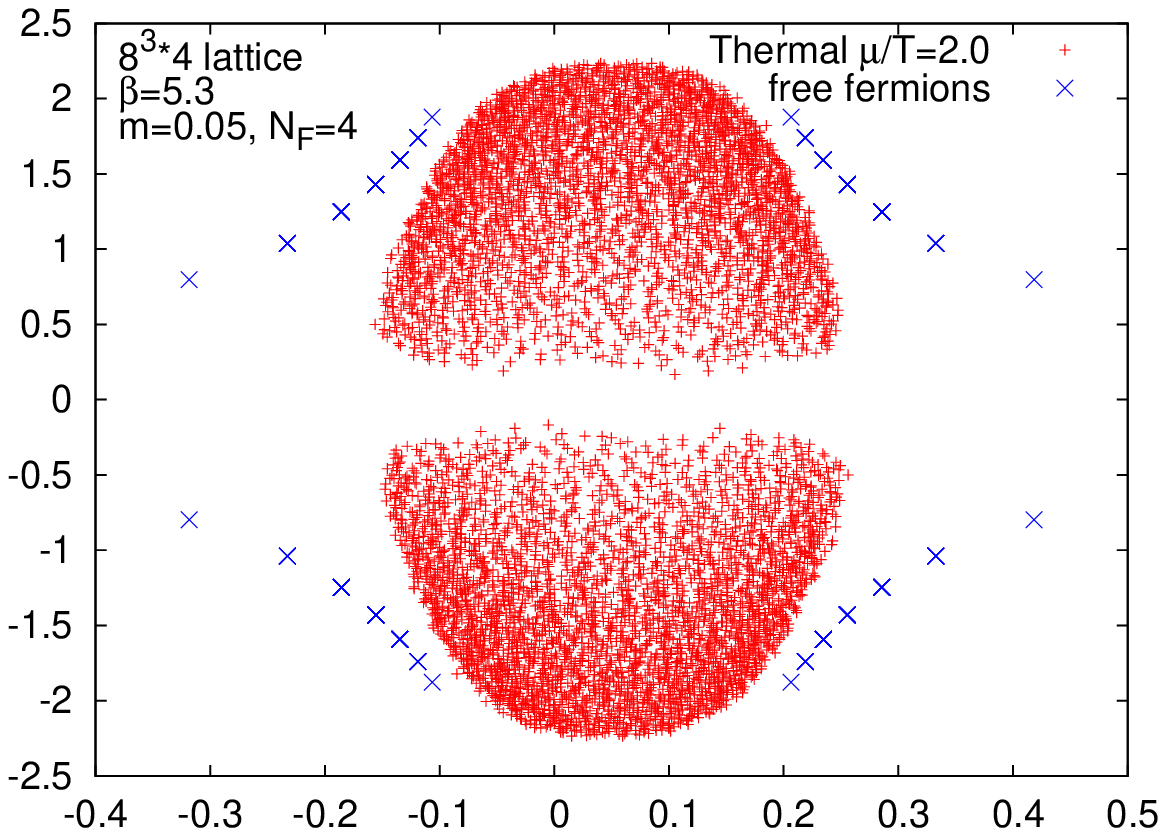}  
\includegraphics[width=0.45\columnwidth]{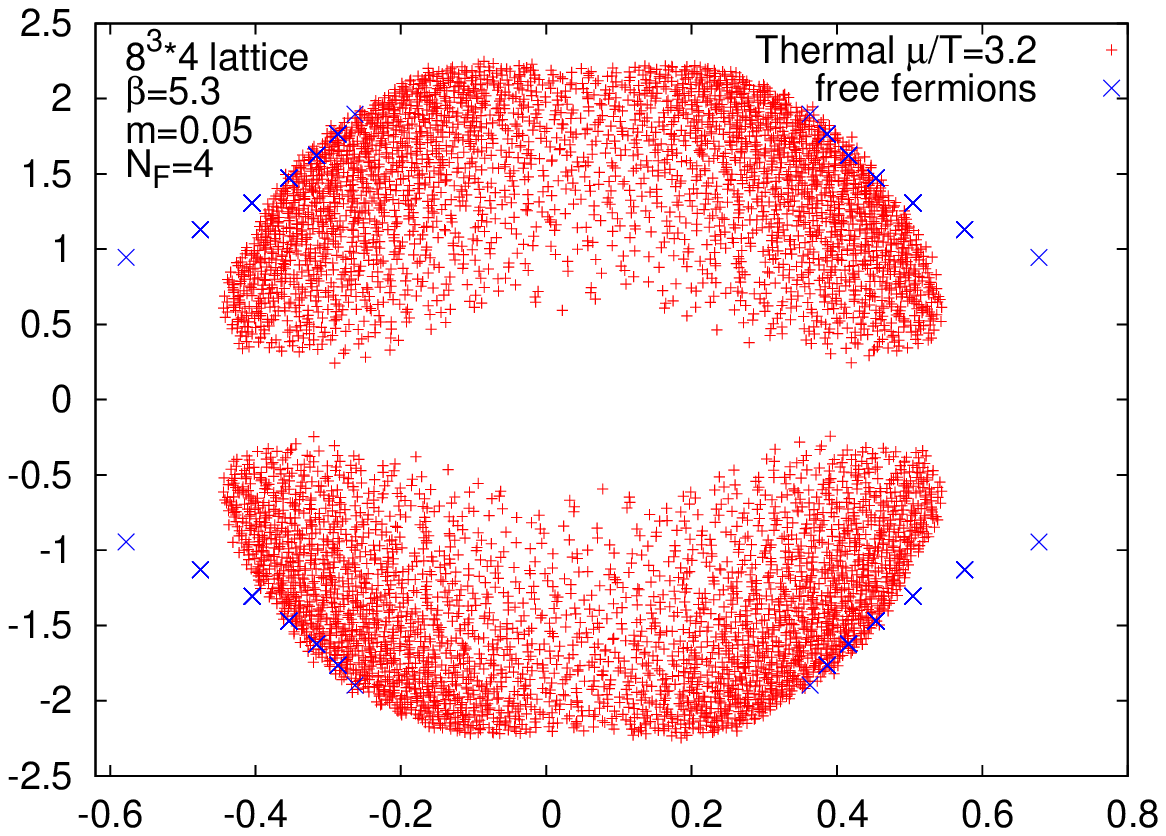}  
\label{spectra8}
\includegraphics[width=0.45\columnwidth]{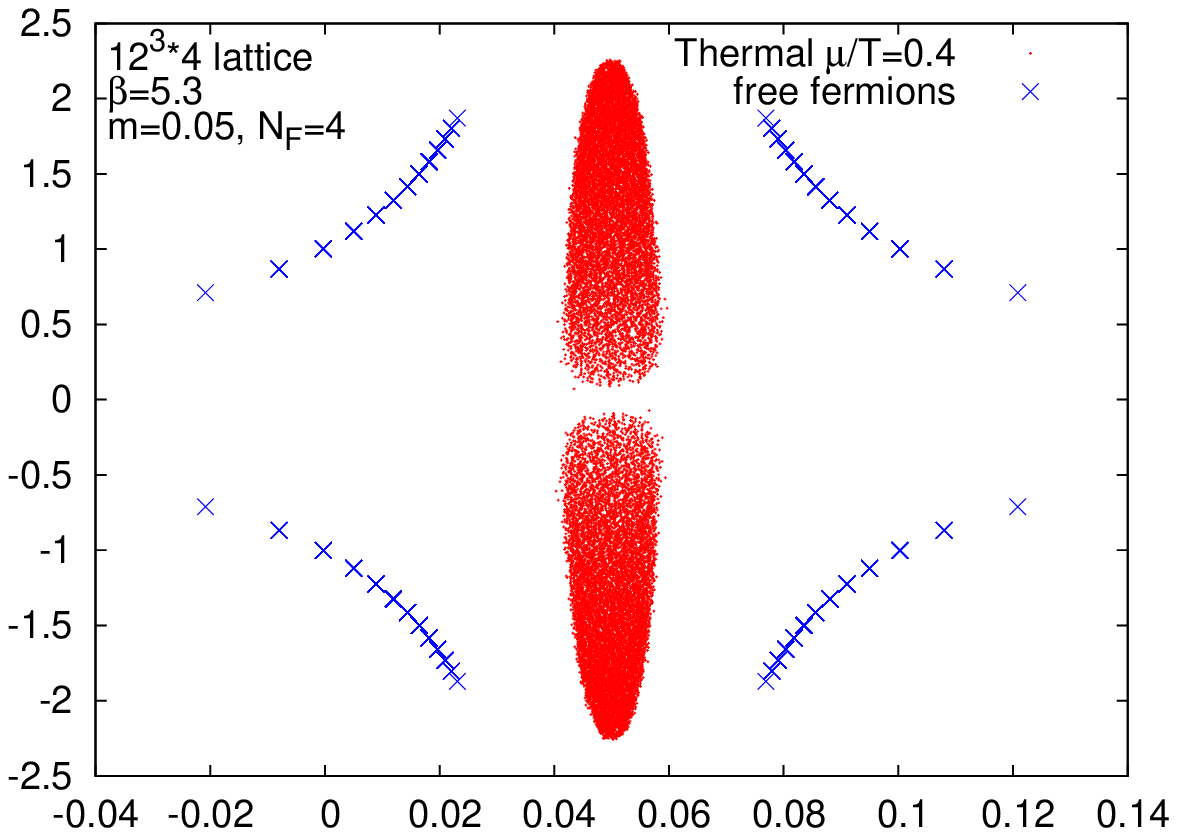}  
\includegraphics[width=0.45\columnwidth]{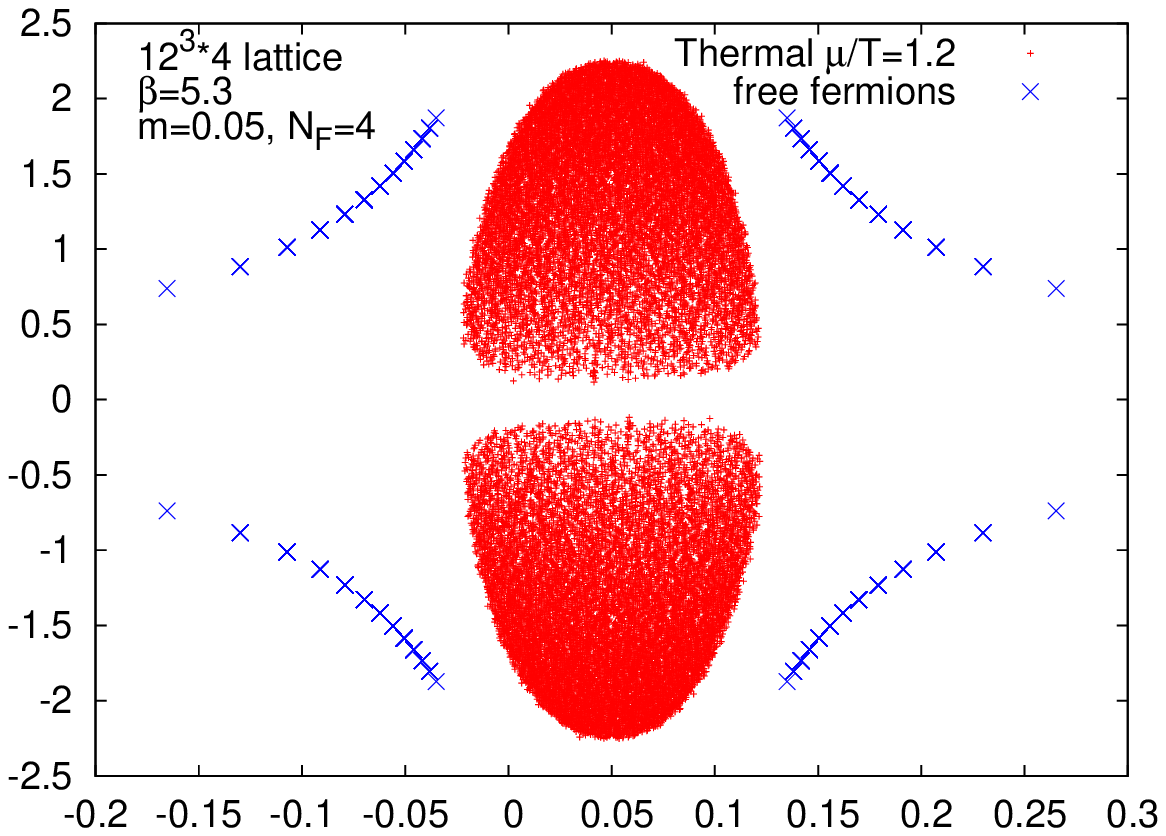}  
\includegraphics[width=0.45\columnwidth]{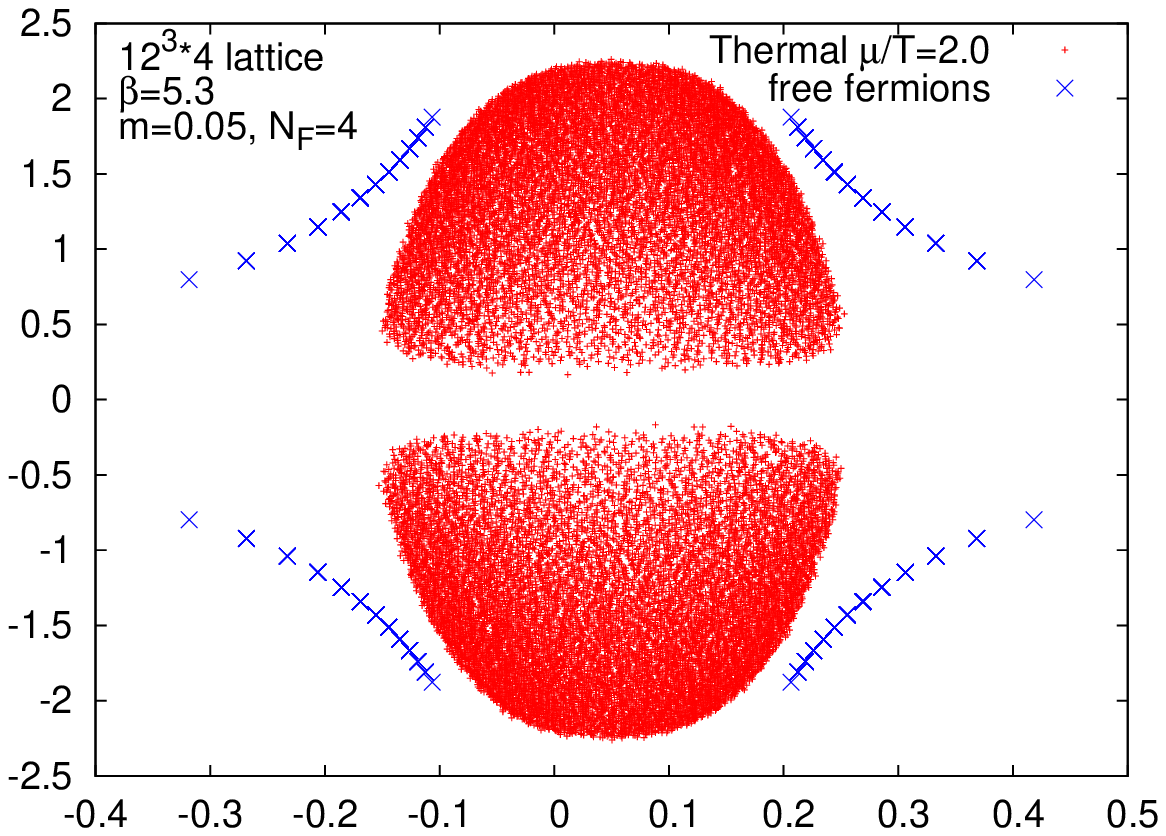}  
\includegraphics[width=0.45\columnwidth]{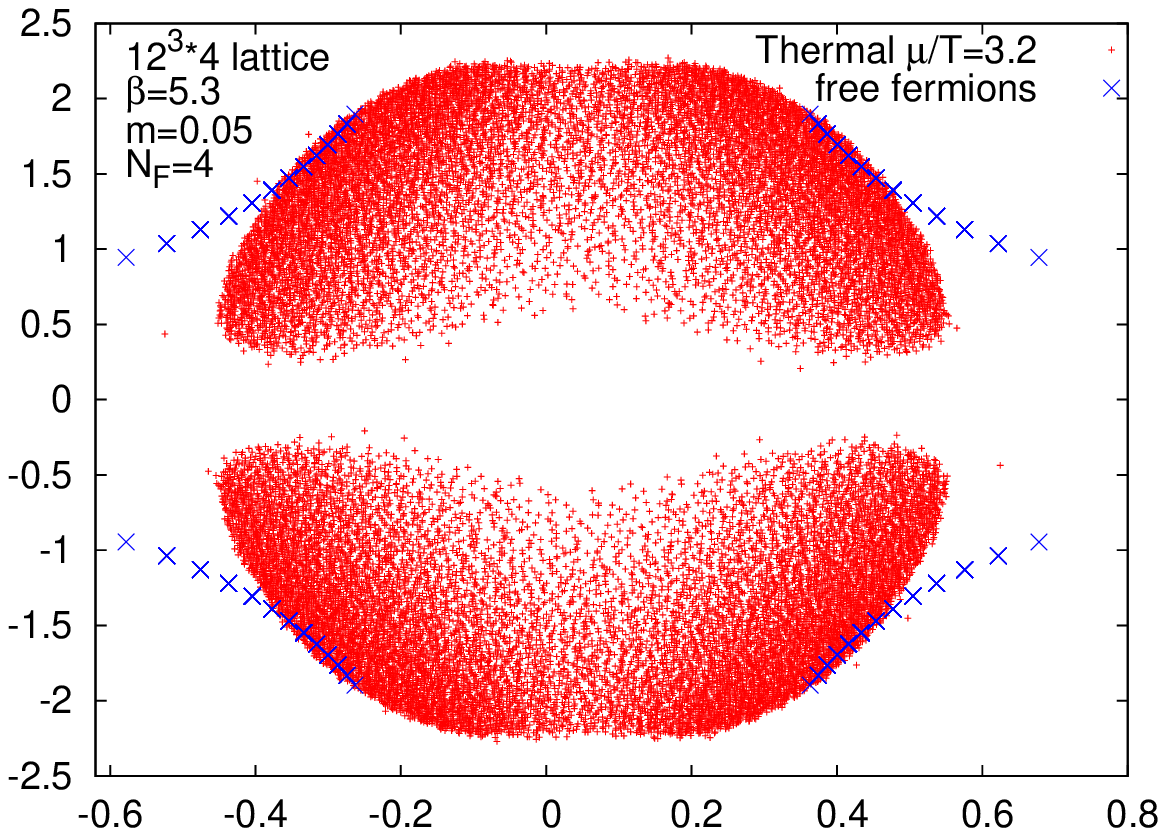}  
\caption{Spectrum of the staggered fermion operator on a $8^3\times 4$ lattice (top four panels) and  $12^3\times 4$ lattice (bottom four panels), for $\mu/T=0.4, 1.2, 2.0, 3.2$. The free spectrum is shown as well. Other parameters as in Fig.~\ref{fullphase12}.
}
\label{spectra12}
\end{center}
\end{figure}

In order to study the phase of the determinant in full QCD we start from an initial configuration on
the SU(3) submanifold. After thermalisation we then follow the evolution 
of the phase. The results are shown in Fig.~\ref{fullphase12}, for a $8^3\times 4$ and a $12^3\times 4$ lattice. On the smaller volume and for the smaller chemical potentials $\mu/T=0.4$ and 1.2, one observes very mild time dependence with no winding of the phase around the origin. For the larger chemical potentials 
$\mu/T=2.0$ and 3.2, however, we see frequent crossings of the 
negative real axis. This signals that there is a sign problem 
in the theory which is hard to counter with reweighting, as the average 
phase factor gets close to zero. As expected, this behaviour gets worse as the volume is increased.

The corresponding histograms are shown in Fig.\ \ref{phasehist}, for three different spatial volumes, namely $8^3, 12^3, 16^3$, with fixed $N_\tau=4$. 
 As expected, the distribution is localised on the smallest volume, but gets increasingly wider as the volume and/or the chemical potential are increased.

To relate this behaviour to possible zeroes of the determinant, we have computed the eigenvalue spectrum of the Dirac operator for typical configurations in the ensemble, using the same setup as in Fig.~\ref{fullphase12}. Fig.~\ref{spectra12} contains the spectra, while Fig.~\ref{fullrhist} contains histograms of the 
absolute values of the eigenvalues, obtained by averaging over  100 configurations.
We note that in spite of the frequent circlings of the origin by the fermionic determinant there are typically no eigenvalues close to zero, suggesting 
that the probability density of configurations around the singularities of the drift is very small, as in the simpler models discussed above. 
The change of the total phase is given by the sum of the changes over all the eigenvalues, so in contrast to toy models the frequent crossing of the negative real axis does not suggest that the poles are affecting the CL dynamics. 
We note that the increase of the volume, from $8^3$ to $12^3$, leaves the shape of the spectrum very similar, but increases the density of eigenvalues.

To summarize our findings, in full QCD at high temperatures the singularities of the drift appears to be outside
the support of the probability density of configurations. We conclude therefore that in this situation complex Langevin dynamics  provides correct results, in line with the formal arguments and the lessons from the simple models.
What happens at lower temperatures remains an open question.

\section{Discussion}
\label{sec:disc}

In this section we summarise the key findings and discuss them in the context of the various models.
The main objective was to understand the role of zeroes in the path integral measure after analytic continuation in the context of complex Langevin dynamics, in which the zeroes show up as poles in the Langevin drift. Since the derivation of the justification of the complex Langevin approach for complex measures relies on holomorphicity of the drift, the presence of poles makes a re-analysis of this derivation necessary.

We have shown that a crucial role is again played by the behaviour of  the observables considered and the (real and nonnegative) distribution, which is  a solution of the associated Fokker-Planck equation and effectively sampled during the Langevin process. While for holomorphic drifts it is the behaviour at large imaginary directions in the complex configuration space that matters, for meromorphic drifts we have shown that an additional constraint arises from the behaviour close to the poles: to justify the method it is necessary to be able to perform partial integration, without picking up finite boundary terms, both around the poles and for  large imaginary directions. This condition gives a requirement of fast decay of the distribution in those regions. In simple cases it is possible to verify this requirement analytically, for instance when it can be shown that the distribution is strictly zero in those regions, but for most models this has to be verified a posteriori by diagnosing the output from numerical simulations.

Besides this, we also found that time-evolved observables typically have an essential singularity at a pole. However, this is counteracted by the distribution going to zero at this pole, with nontrivial angular behaviour. This ensures that the contribution to expectation values from this region is finite, but not that boundary terms are necessarily absent.

In order to further understand and support these analytical considerations, we have subsequently analysed a number of models and theories with increasing complexity, from the one-pole model with one degree of freedom to full QCD at nonzero baryon density, with the aim of extracting common features.
Logically a pole can be outside, on the edge of, or inside the distribution, and we have given examples of each of these. As expected, when the pole is outside, it does not interfere with the Langevin process. The possibility of a pole on the edge is important, since it indicates that it is not the winding around the pole that matters, but the decay of the distribution towards the pole. Indeed, we have encountered both correct and incorrect convergence in this case, and this can be traced back to the fast decay of the distribution, or lack thereof. 

As a side remark, we note that further support for our analytical understanding comes from the observed interplay  
between observables, drift and the distribution:  if the observable is naturally suppressed (enhanced) near the pole, it is possible to obtain correct (incorrect) results. This explains why in one analysis one can encounter both correctly and incorrectly determined or nonconverging expectation values.

When the pole is inside the distribution, we have found that it typically leads to a bottleneck, i.e.\ a region in configuration space which is difficult to pass and effectively divides the configuration space in two, nearly disjoint regions. 
For QCD and QCD-like models, it is zeroes of the determinant that correspond to poles and determine the location of the bottleneck. Hence the complex-valued determinant, preferably not raised to any powers (such as $N_f$), or the real part of the determinant,  
provide useful diagnostic observables to analyse the dynamics. We have studied a number of models in this way, namely U(1) and SU(3) one-link models and heavy dense QCD (HDQCD) in four dimensions. In all of these, we indeed observed similar behaviour: the emergence of two regions, which were denoted with $G_\pm$ and are identified by the sign of the real part of the determinant (before raising it to a power).
We found that it is typically the region with positive real part that dominates the dynamics, but that excursions to $G_-$ can upset expectation values, even when their relative weight is suppressed.
In some cases the bottleneck is particularly difficult to pass, which may occur when the order of the zero is increased. 
It is possible to analyse each region $G_\pm$ separately. 
The error made by restricting the simulation to $G_+$ can then be estimated and was typically seen to be small, depending on the parameters used.
 In the SU(3) one-link model and HDQCD, the process is affected by zeroes close to the half filling point, where the sign problem is milder. 
In this case the region $G_-$ took the form of characteristic ``whiskers'': even though the relative weight of these regions was small, the contribution to expectation values could be large. Hence an exclusion of this region improves the results, but with a systematic uncertainty. 
In HDQCD we found indications that the role of zeroes is unchanged when the lattice spacing is decreased, while keeping the physical volume approximately constant.
 Finally, we considered QCD with dynamical staggered quarks at high temperature
and analysed the eigenvalues of the Dirac operator. Here we noted that there are typically no eigenvalues close to zero, which suggests that complex Langevin dynamics is applicable in this part of the phase diagram.

\section{Summary and Outlook}
\label{sec:concl}

We have given a detailed analysis of the role of poles in the Langevin drift,  in the case of complex Langevin dynamics for theories with a sign problem. 
Since the standard derivation of the formal justification relies on holomorphicity of the drift, we have revisited the derivation and shown that, besides the requirement of a fast decay of the probability distribution at large imaginary directions, an additional requirement of fast decay near the pole(s) is present. The probability distribution is typically not known a priori, but its decay can be analysed a posteriori. 

We then studied a number of models, from simple integrals to QCD at nonzero baryon chemical potential, and found support for the analytical considerations. In the cases when the simulation is affected by the pole(s), we found that typically the configuration space is divided into two regions, connected via a bottleneck. For theories with a complex fermion determinant, such as QCD, the bottleneck is determined by the zeroes of the determinant.
In the simple models, and even for QCD in  the presence of heavy (static) quarks, this understanding is sufficient to analyse the reliability of the Langevin simulation. 

In full QCD, with dynamical quarks, ideally it requires knowledge of the (small) eigenvalues of the fermion matrix throughout the simulation, which is nontrivial.
At high temperature, it was shown that the eigenvalues are typically not close to zero.
Hence the most important outstanding question for QCD refers to the low-temperature region in the phase diagram. Here a number of hurdles remains to be taken. 
When eigenvalues become very small, the conjugate gradient algorithm used in the fermion matrix inversion becomes ineffective. This is common to many problems at nonzero density, even in the absence of a sign problem, and requires e.g.\ a regulator. A successful approach in this case has not yet been developed.
Besides this, on coarse lattices  gauge cooling, which stabilises the Langevin process, is ineffective. This situation is improved on finer lattices, which are however more expensive due to the larger values of $N_\tau$ required. 
A definite statement on the applicability of complex Langevin dynamics throughout the QCD phase diagram can only be made once further analysis of theses issues has been brought to a conclusion.

\section*{Acknowledgements} 

\noindent
IOS would like to thank Prof.\ Reimer K\"uhn from King's College London for interesting discussions.
GA is supported by STFC (grant ST/L000369/1), the Royal Society and the Wolfson 
Foundation.  
DS, ES and IOS thankfully acknowledge support from the DFG via projects STA 283/16-2 (EHS and IOS)  and SE 2466/1-1 (DS).
The authors gratefully acknowledge the Gauss Centre for Supercomputing 
e.V. (www.gauss-centre.eu) for funding this project by providing computing 
time on the GCS Supercomputer SuperMUC at Leibniz Supercomputing 
Centre (LRZ, www.lrz.de), as well as for providing computing time 
through the John von Neumann Institute for 
Computing (NIC) on the GCS share of the supercomputer JURECA and 
JUQUEEN \cite{juqueen} at J\"ulich Supercomputing Centre (JSC). 
Some parts of the numerical 
calculation were done on the GPU cluster at E\"otv\"os and Wuppertal Universities.

\appendix

\section{Second-order poles: a solvable real example}
\label{sec:2nd}

In this appendix, we discuss the special case of a second-order pole in a solvable example.
{\bf Statement:}  When there is a pole with residue 2 in the drift 
(corresponding to a second-order zero in the density) and the Laurent 
expansion around the pole has no constant term, then $\cO(x+iy;t)$ has no 
essential singularity, only a simple pole. 
   
A simple example is the following: consider the action 
\be
-S(z) =-\frac{\omega z^2}{2} + \ln(z^n),
\ee
leading to
\be
K(z)=\frac{n}{z}-\omega z,
\quad\qquad
  \tilde L= \frac{d^2}{dz^2} + \left(\frac{n}{z}-\omega z\right)\frac{d}{dz}.
\ee
A simple computation then gives
\be
\tilde L z = \frac{n}{z}-\omega z,
\quad\quad
 \tilde L^2z-\frac{n(n-2)}{z^3}+\omega^2 z,
\quad\quad 
\tilde L \frac{1}{z}=\frac{2-n}{z^3}+\omega \frac{1}{z},
\label{langtilde}
\ee
so that for $n=2$ no higher singularities are produced by $\tilde L^n$ 
to $z$. The more general statement follows easily by computation.

This model (which is a special case of the models considered in Sec.\
2.1), has some interesting features, in particular the complete spectrum of $\tilde L$ and the evolution of 
holomorphic observables can be determined.
Because the model is real we now write $x$ instead of $z$ and drop the 
tilde, i.e.\ we consider
\be
L= \frac{d^2}{dx^2}+\left(\frac{n}{x}-\omega x\right)\frac{d}{dx}
\ee
and its dual
\be
L^T=\frac{d^2}{dx^2}-\frac{d}{dx}\left(\frac{n}{x}-\omega x\right).
\ee
For $n=2$ one can actually find the complete spectrum of $L, L^T$: because 
the model is real, a well-known fact \cite{Damgaard:1987rr} is that $L$ is conjugate to a 
self-adjoint operator $-H$,
\be
H=-\exp(-S/2)L \exp(S/2)=-\frac{d^2}{dx^2} -\frac{1}{4}\omega^2 
x^2 -\frac{3\omega}{2},
\ee
which is, up to constant, the Hamiltonian of a harmonic oscillator.  It 
has apparently a negative eigenvalue; this is, however, deceptive: so 
far we have been sloppy about the boundary conditions at $x=0$. The drift 
is strongly repulsive away from the origin along the real axis, the 
probability density $\rho$ is vanishing quadratically there and the 
Langevin process does not cross the origin. 

This means that mathematically we have to consider $H$ with 0-Dirichlet 
boundary conditions at the origin. To avoid confusion, we call the corresponding 
Hamiltonian $H_D$. All functions in the domain of definition of $H_D$ have 
to have at least a square integrable second derivative; this means that 
they are continuous and vanish at the origin. The ground state wave 
function of $H$ (ignoring the b.c. at 0), which superficially seems to 
belong to a negative eigenvalue of $H_D$, is not in the domain of 
definition (neither are all even eigenfunctions of $H$). Actually, because 
there is no communication between the two half-lines, we may as well 
consider the problem only on one of the half-lines $\R^\pm$. From now on 
we choose $\R^+$.

The eigenfunctions of $H_D$ are thus the odd eigenfunctions of $H$
\be
\psi_{2n+1}(x)=N_{2n+1}\exp\left(-\frac{\omega x^2}{4}\right) 
H_{2n+1}\left(\sqrt{\frac{\omega}{2}}x\right),  \qquad n=0,1,\ldots,
\ee
with eigenvalues $(2n+1)\omega$ and $N_{2n+1}$ such that they are normalised on 
$\R^+$; $H_n$ are the Hermite polynomials. The eigenfunctions of $L$ are 
then found as
\be
\phi_n=\psi_{2n+1}\exp(S/2)= \psi_{2n+1} \exp(\omega x^2/4)/x,\qquad n=0,1,\ldots,
\ee
while those of $L^T$ are
\be
\hat\phi_n=\psi_{2n+1}\exp(-S/2)= \psi_{2n+1} \exp(-\omega x^2/4)x,\qquad n=0,1,\ldots.
\ee
The ground state of $L$ is thus a constant $c$ and the ground state of  
$L^T$ is
\be
\hat\phi_0(x)=\rho(x)= c x^2 \exp(-\omega x^2/4),
\ee
as it has to be. 

The first excited state of $L$ is $\phi_2=N_3(x^2-3/\omega)$; it belongs 
to the  eigenvalue $-2\omega$. We thus find
\be
e^{t L}x^2=e^{-2\omega t} x^2+\frac{3}{\omega}\left(1-e^{-2\omega
t}\right)\,,
\label{x2evol}
\ee
which converges to the correct expectation value
\be
\bra x^2\ket =\frac{3}{\omega}
\ee
for $t\to\infty$. Similarly convergence to the correct value is found for
all even functions.

How about the superficially unstable mode $1/x$, see Eq.\ (\ref{langtilde})? 
It corresponds to the ground state of $H$ on $L^2(\R)$, which is not in 
the domain of definition of $H_D$. The odd eigenfunctions of $H$ are 
complete in the subspace of odd functions; hence the odd eigenfunctions 
restricted to $\R^+$ are complete in $L^2(\R^+)$ and so the apparent 
eigenvector of $H_D$ with a negative eigenvalue $\psi_0$ should be 
considered as a $L^2(\R^+)$ convergent series 
\be
\psi_0(x) =\sum_{n=0}^\infty a_n \psi_{2n+1}(x)  
\ee
with
\be
a_n= \int_0^\infty dx\, \psi_0(x)\psi_{2n+1}(x). 
\ee
Instead of considering $-H_D$ as a self-adjoint operator on $L^2(\R^+)$
we may consider $L$ itself as a self-adjoint operator on the Hilbert space 
$\cH$ obtained as (the completion of) the set of functions $\phi$ on 
$\R^+$ with the scalar product 
\be
(\phi,\phi')\equiv\int_0^\infty dx\, e^{-S} \phi(x)^*\phi'(x).
\label{scalar}
\ee
The eigenfunctions $\phi_n,\, n=0,1,\ldots$, are orthogonal with respect to 
this scalar product. So we can write equivalently 
\be
\frac{1}{x}=\sum_{n=0}^\infty a_n \phi_{n}(x),
\ee
where the convergence is now to be understood in the sense of $\cH$.
$\frac{1}{x}$ is in $\cH$, but not in the domain of definition of $L$.

\section{Solutions to the sign problem for real models with poles}
\label{sec:real}

In this appendix we consider real models with a zero in the density and hence a pole in the Langevin drift, with a weight of the form motivated by the U(1) one-link model in Sec.\ \ref{sec:poles}, i.e.
\be
 \rho(x) = \left[ 1+\kappa \cos(x) \right]^{n_p}. 
\ee

\subsection{One pole, $n_p$ odd}

Since the models are real, one might attempt to treat them by the real 
Langevin method. But a simple consideration shows that for a real model 
with a sign problem this cannot produce correct results. The reason is 
that the real Langevin equation will have a {\it positive} equilibrium 
measure on the real axis and thus cannot reproduce all the averages which 
would be obtained with a signed measure. When we modify the process to 
allow it move out into the complex plane, the story changes: it seems then 
a priori not impossible that a {\it positive} measure on $\C$ reproduces 
correctly the averages of {\it holomorphic} observables with a {\it 
signed} measure on $\R$ and there are examples that bear this out. A 
simple example is given by
\be
\rho(x)=1+\kappa \cos(x),   \qquad \kappa>1.
\ee
It is easy to verify that the correct expectation values are reproduced by 
the positive density in $\C$, 
\be
P(x,y)=
(1+\cos(x))\,\frac{1}{\sqrt{2\pi\sigma}}
\exp\left[-\frac{y^2}{2\sigma}\right]\,
\ee
with 
\be
\sigma=2\log\kappa\,.
\ee
This simple solution is, however, unrelated to the CL method. 

\subsection{One pole, $n_p$ even}

For $n_p>0$ and even, there is no sign problem, but the lack of ergodicity 
exists as well. In this case, because of the stronger repulsion away from 
the pole, our simulations typically do {\it not} cross the pole, and so 
produces incorrect results when started on one side of the pole. (If there is a symmetry $x \to -x$, this defect can 
easily be remedied by starting the process with equal probability on 
either side of the pole).
Another way to facilitate the crossing and achieve correct results is by 
adding a small imaginary noise term.

 A simple cure consists in the reweighting with the sign factors, as follows.
Replace the observable $\cO(x)$ by
\be
\cO(x)\,{\rm sgn\,\rm Re} \rho(x)
\ee
and compute by real or complex Langevin
\be
\bra \cO\ket  \equiv \frac{\bra \cO(x+iy)\,{\rm sgn\, Re}\, \rho(x)\ket} {\bra {\rm sgn\, Re}\, \rho(x)\ket},
\ee
where the symbol $\bra \cdot \ket$ stands for the ordinary real or complex 
Langevin long time average. But this cure, like any reweighting method, 
while it works for one-variable models, it is not very useful for lattice 
models. So we will not pursue it any further.

\subsection{A cure for compact real models} 

The final cure we consider is for a real but nonpositive weight $\rho$. Let $c$ be a constant such that
\be
\rho+c>0
\ee
and define
\be
\sigma\equiv \rho+c.
\ee
Then for any observable satisfying $\int \cO dx=0 $ we can rewrite $\bra \cO\ket$ as
\be
\bra \cO\ket_\rho =  \frac{\bra\cO\ket_{\sigma}}{\left\bra \rho/\sigma\right\ket_{\sigma}},
\ee
because
\be
\bra \cO\ket =\frac{\int\sigma \cO dx}{\int \sigma dx} 
\frac{\int \sigma dx} {\int \rho dx}\,.
\ee
So a correct procedure is to run real Langevin with the drift derived from
the positive density $\sigma$ and correct the normalisation as shown 
above.

We take the U(1) model with $n_p=1$, such that
\be
\sigma(x) = c+ \left[ 1+\kappa \cos (x) \right] e^{\beta\cos (x)}.
\ee
The drift for the modified Langevin process is now
\be
K_{\sigma} = - \frac{\kappa\sin (x) +  \beta\sin (x) \left[1+\kappa \cos (x)\right]} {c e^{-\beta \cos (x)} +1 + \kappa\cos (x)}.
\ee
A full set of observables satisfying the condition  $\int \cO dx=0 $ are 
the exponentials $\exp(ikx), k\neq 0$. For the normalisation factor we have
\be
\bra n(x)\ket_{\sigma} 
\equiv \left\bra \rho/\sigma\right\ket_{\sigma} =
\left\bra \frac{1+\kappa \cos(x)}{c e^{-\beta\cos(x)} +1 +\kappa\cos(x)}\right\ket_{\sigma}.
\ee

\begin{figure}[t]
\begin{center}
\includegraphics[width=0.75\columnwidth]{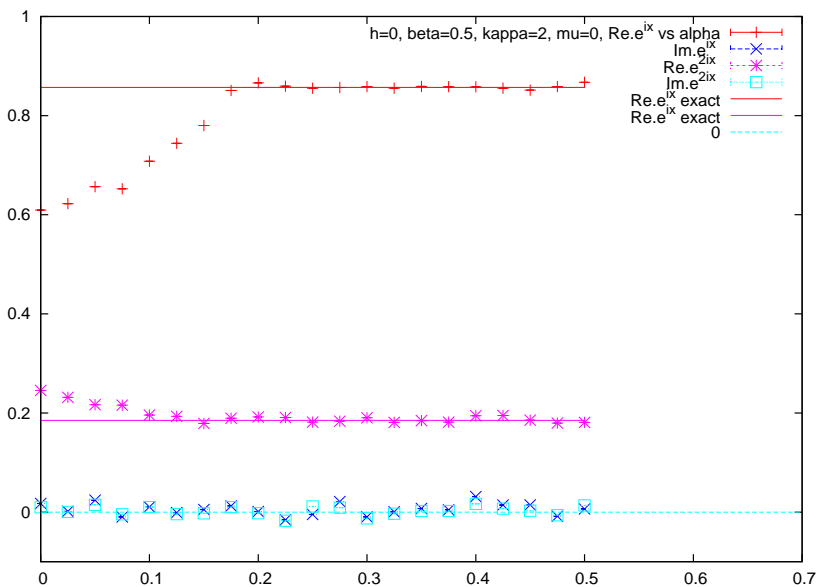}
\caption{$U(1)$ one-link model for $\kappa=2,\mu=0,\beta=0.5$; data
points: cured CLE vs $\alpha=c e^{-\beta}/\kappa$,  solid lines: exact 
results.}
\label{cure2b}
\end{center}
\end{figure}

This procedure works very well, as shown in Fig.\ \ref{cure2b}. Note that in 
the figure, the horizontal axis is 
\be 
\alpha =\frac{c e^{-\beta}}{\kappa}, 
\ee 
which increases as $c$ is increased ($c=0$ is the original process). The observables are Re/Im $e^{ikx}$ with $k=1,2$.

As expected, the numerical results start agreeing with the exact results 
as soon as $c$ is large enough to make $\sigma$ nonnegative. Therefore a 
plot like this can also serve to determine the minimal $c$ for which correct results are obtained
and a priori knowledge about the zeroes of the density $\rho$ is not required.

The advantage of this cure is that it can be easily generalised to the 
complex case $\mu\neq 0$; it turns out that it does work reasonably well, 
but not perfectly, provided $\mu$ is not very large. But again,  since the cure 
involves some reweighting, it is not very useful for lattice systems.

\section{Zeroes of the HDQCD determinant}
\label{app-zeroes}

In this Appendix, we further consider the HDQCD determinant for gauge group SU(3) or 
SL(3,$\C$), reduced to a single link $U$. We will demonstrate that zeroes of the determinant do not come as isolated points.

As discussed in Sec.\ \ref{sec:spin}, the determinant contains the factors
\be
D = \det(1+CU), \qquad\qquad \tilde D = \det(1+\tilde C U^{-1}), 
\ee
such that $\det M = (D\tilde D)^{2N_f}$.
For a discussion of its zeroes we may look at each factor separately. Let us first consider $D$. The eigenvalues of $U$ can be parametrised as $z_1,z_2,1/(z_1z_2)$; in terms of these
\be
D=(1+Cz_1)(1+Cz_2)(1+\frac{C}{z_1z_2}).
\ee
In $\C^2$, parametrised by $z_1,z_2$, the determinant vanishes on the
three submanifolds given by
\be
\label{eq:123}
(1):\;z_1=-\frac{1}{C},\qquad (2):\; z_2=-\frac{1}{C},\qquad (3):\; z_1z_2=-C.
\ee
But there is a different way to think about this: define
\be
 u\equiv \tr\, U=z_1+z_2+\frac{1}{z_1 z_2},
 \qquad\qquad
 v\equiv \tr\, U^{-1}= \frac{1}{z_1}+\frac{1}{z_2}+z_1z_2.
\ee
$u$ and $v$ are algebraically independent and can be used instead of $z_1$
and $z_2$ to parametrise conjugacy classes of SL(3, $\C$).  The map from
$z_1,z_2$ to $u,v$ is not one-to-one, since interchanging $z_1$ and $z_2$
will leave $u,v$ unchanged. The inverse map from $u,v$ to $z_1,z_2$ will
have branch points where any two eigenvalues coincide.

The fact that $u,v$ ignore permutations of the eigenvalues is actually an
advantage because $D$ too remains unchanged. The zeroes of $D$ are then  
determined by
\be
1+C^3+Cu+C^2 v=0\qquad {\rm or}\qquad u=-Cv-C^{-1}-C^2.
\label{eq0}
\ee
The three manifolds (1,2,3) in Eq.\ (\ref{eq:123}) are thus mapped into a single manifold 
given by Eq.\ (\ref{eq0}). This manifold is an affine complex plane in the 
space $\C^2$ (affine means that it does not go through the origin).

$\tr\,U^{-1}$ does not have to be computed by taking the inverse of $U$;
one can instead use the identity, valid for $U\in$ SL(3, $\C$),
\be
\tr\, U^{-1}= \frac{1}{2} \left((\tr\, U)^2 -\tr\, U^2\right).
\ee
So we get,  using $u_2=\tr\, U^2$,
\be
D= 1+ C^3+ Cu +\frac{C^2}{2}\left(u^2-u_2 \right),
\ee
which vanishes on a complex parabola.

The determinant factor $\tilde D$ is quite similar; proceeding as before 
we find 
\be
\tilde D=\tilde C+\tilde C^2+u+\tilde C^{-1} v= 
\tilde C+\tilde C^2+u+\frac{1}{2}\tilde C^{-1}(u^2-u_2),
\ee
so its zeroes are again described either by a complex affine plane in $u,v$ 
or a complex parabola in $u,u_2$. The main point is that they are real 
manifolds of codimension two, so there are no isolated zeroes.

\section{Expansion methods}
\label{sec:expansions}

One possibility to deal with the pole is to use power series 
expansions in order to approximate the meromorphic drift by polynomials. 
In QCD the pole in the drift is always due to a zero of the determinant. 
In the one-pole model the role of the determinant is played by the factor
$D(x)\equiv x-z_p$.
 
 \subsection{One-pole model}

We explain the approach in the one-pole model. We study two basic procedures:

\begin{itemize}
\item[(1)] Fixed expansion:
Let $D^{n_p}$ be the `determinant' causing problems due to its zeroes. 
Consider the drift caused by $D$,
\be
K_D(z)=n_p\frac{D'(z)}{D(z)}.
\ee
In order to obtain a holomorphic approximation to $K_D$, we choose a point 
$(x_0,y_0)$ not too far from the peak of the distribution but far enough 
from the pole(s). We then expand $1/D$ around this point to order $N$ 
as follows: let $D_0\equiv D(x_0,y_0)$,  then replace $1/D$ by 
\be
\frac{1}{D_N} \equiv\frac{1}{D_0}\sum_{n=0}^N 
\left(\frac{D_0-D}{D_0}\right)^n=\frac{1-(1-D/D_0)^{N+1}}{D}.
\label{expans}
\ee
Since this is a polynomial in $D$, there is no indeterminacy when $D=0$.
The difference between the exact value $1/D$ and $1/D_N$ is 
$(1-D/D_0)^{N+1}$, which converges to 0 if and only if $|D/D_0|<0$. There 
could be problems if the process goes outside the region of convergence, 
but experience shows that typically the drift will tend to keep the 
process inside the region of convergence. An illustrative example is 
shown in Fig.\ \ref{flowexp}. Because the expansion point $(x_0,y_0)$ is 
chosen once and for all, we call this the fixed expansion. 
In any case, by varying $N$ one can check whether this is the case.
Numerical studies using this fixed expansion are presented below for the 
U(1) one-link model and for the SU(3) one-link model in Sec.\ \ref{sec:spin}.

\begin{figure}[t]
\includegraphics[width=.48\columnwidth]{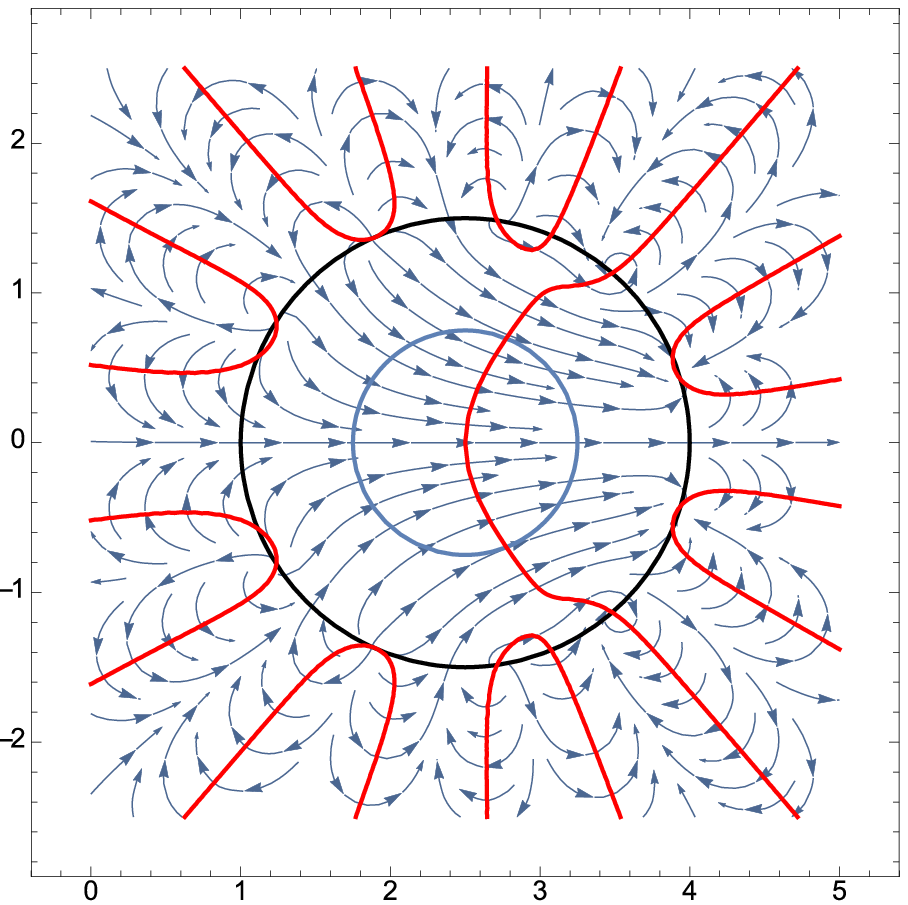}
\includegraphics[width=.48\columnwidth]{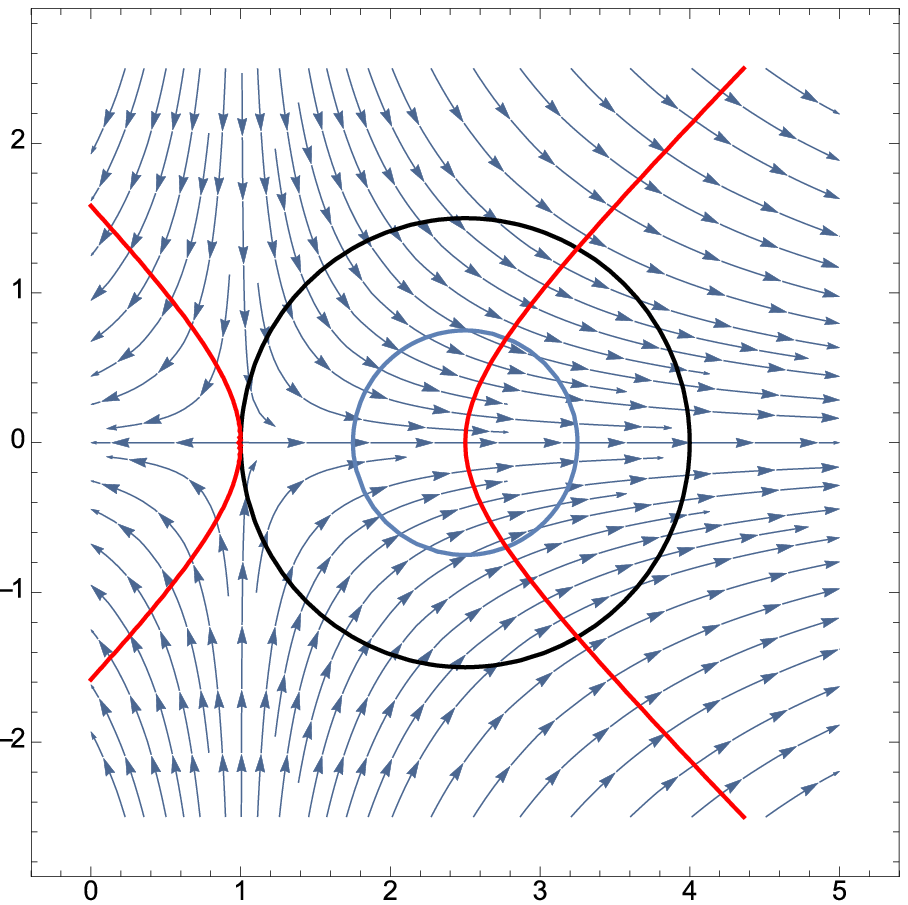}
\caption{Flow pattern for the one-pole model at $\beta=0$, $z_p=1$, using
the Taylor expansion to order $N=10$ with $D_0=1$ (left) and the full expression (right). 
The black circle indicates the radius of  convergence, while the red lines show where the radial component 
of the flow changes sign.} 
\label{flowexp}
\end{figure}

\item[(2)] Dynamic expansion: one may choose different expansion points $(x_i,y_i)$, with 
$D(x_i,y_i)$, in such a way that the domain of analyticity is covered, while always staying 
well inside the domain of convergence. The quality of the 
expansion can be fixed by changing $(x_i,y_i)$ to the actual configuration point 
whenever $(1-D/D_0)^{n+1}>\epsilon$ with some pre-chosen $\epsilon$. If 
$\epsilon$ is chosen small enough we should not find any appreciable 
difference between the results using $D$ and $D_N$.
By studying various situations we find that, as expected, the dynamically 
expanded drift generally performs just like the unexpanded one: it works 
where the latter works and it fails where the latter fails.

\end{itemize}

\subsection{U(1) one-link model}

\begin{figure}[t]
\includegraphics[width=.48\columnwidth]{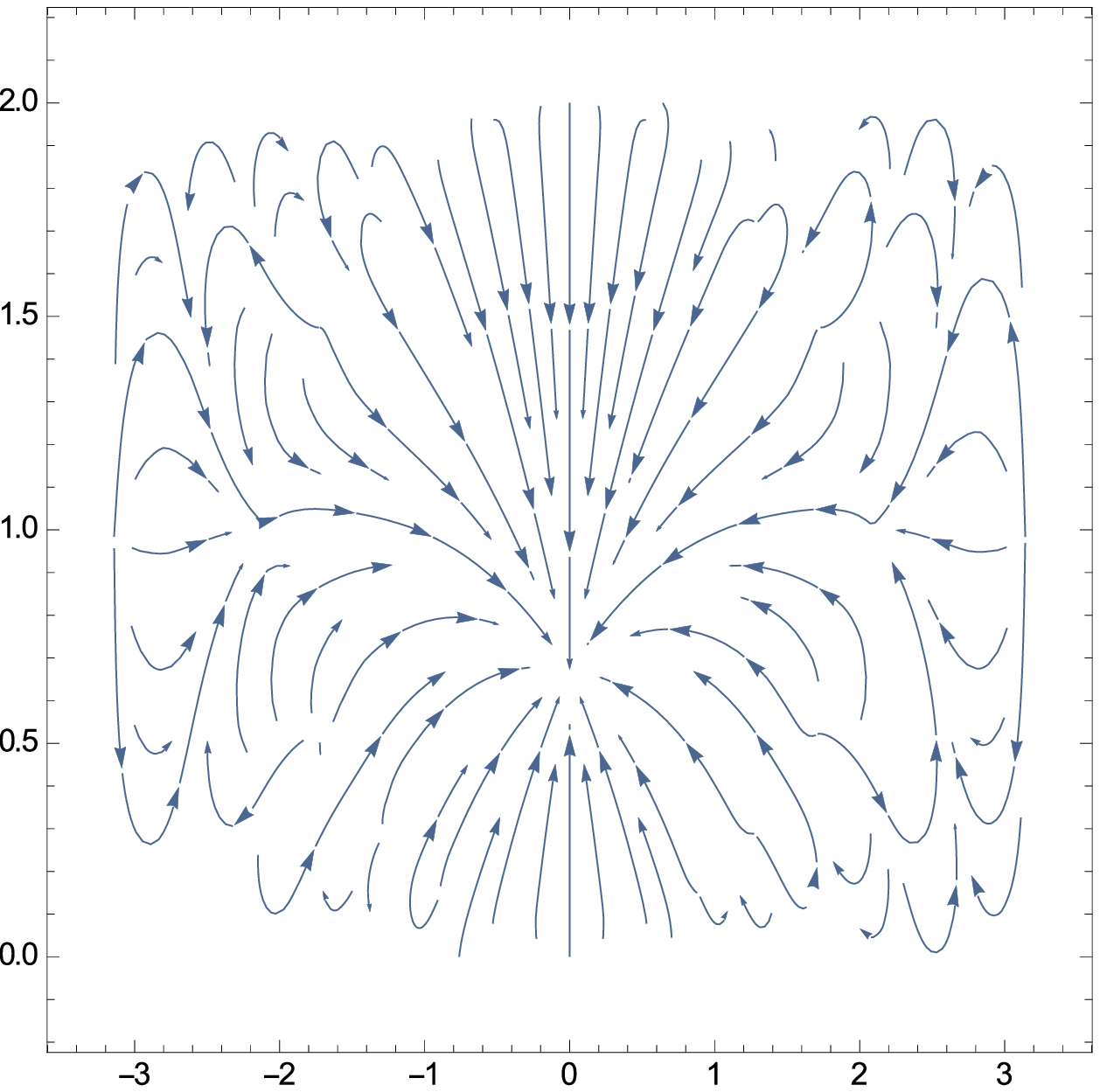}
\includegraphics[width=.48\columnwidth]{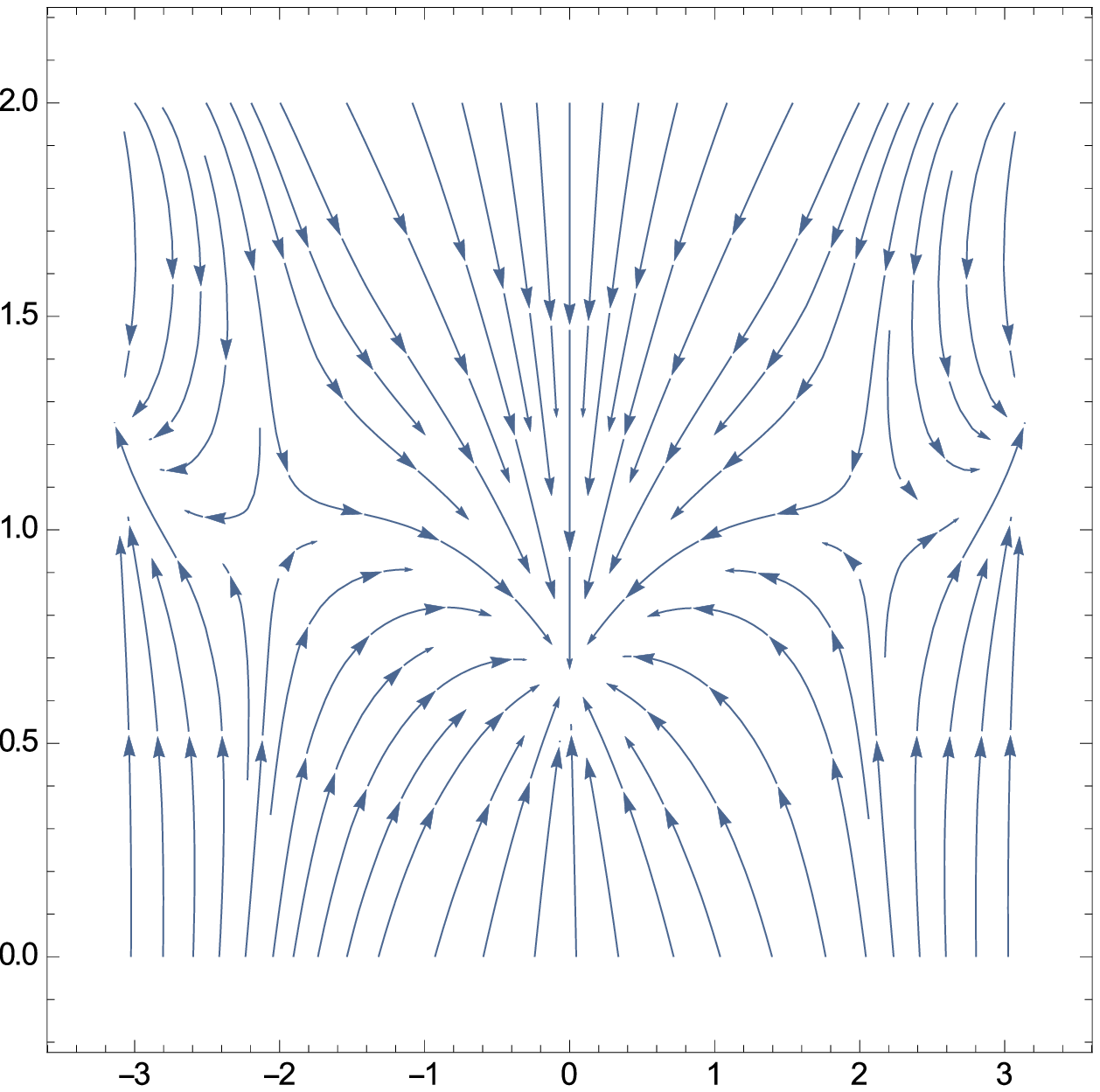}
\caption{Flow pattern for the U(1) one-link model at $\beta=0.3$, 
$n_p=1, \mu=1,\kappa=2$;, using the Taylor expansion to order $N=10$ with 
$D_0=1+\kappa\cos(0.8i)$ (left) and the full expression (right).}
\label{flowu1exp}
\end{figure}

We now apply the expansion method to the U(1) one-link model and focus on the fixed expansion.
Consider the factor
\be
D^{n_p}(x)=\left(1+\kappa \cos(x-i\mu)\right)^{n_p},
\ee
appearing in $\rho(x)$. The drift caused by this,
\be
K_D(z)=n_p\frac{D'(z)}{D(z)}\,
\ee
has two poles for $\kappa>1$. As described for the one-pole model we 
obtain a holomorphic approximation to $K_D$ by choosing a point 
$z_0=x_0+iy_0$ somewhere near the center of the equilibrium distribution; here 
the natural choice is $z_0=i\mu$, i.e.\ $D(z_0)=1+\kappa$.  The Taylor 
expansion of $1/D$ around this point to order $N$ then looks as in 
Eq.\ (\ref{expans}). Again the drift from the expansion tends to keep the 
process inside the region of convergence, as shown in Fig.\ \ref{flowu1exp}.

We have tested this approach numerically, using expansions to order 10 and 20, making sure that the centre of the expansion was chosen reasonably far away from the poles and near the maximum of the 
distribution $P(x,y)$ in the region with positive $\Re D$, i.e.\ $G_+$. 
We found the results to be more or less comparable to the ones obtained by restricting the process to $G_+$, and not a great difference between orders 10 and 20.

Hence we conclude that the fixed expansion is a potential cure of the ills of 
meromorphic drift in the cases where restricting the process to $G_+$ 
works as well. It should be noted though that potentially significant, 
though much reduced deviations from the exact results remain.

\providecommand{\href}[2]{#2}\begingroup\raggedright\endgroup

\end{document}